\pdfoutput=1
\documentclass[11pt,twoside,a4paper,cmspaper,final,collab]{cms-tdr}
\def\svnVersion{85ffdc3}\def\svnDate{2024/09/17}\def\cmsCernNoTag{CERN-EP-2024-016}\def\cmsCernDate{\today}\def\cmsMessage{Published in Physical Review D as \href{http://dx.doi.org/10.1103/PhysRevD.110.052004}{\doi{10.1103/PhysRevD.110.052004}.}}
\begin{document}\cmsNoteHeader{B2G-20-014}

\newlength\cmsTabSkip\setlength{\cmsTabSkip}{1ex}
\newcommand{\Hbb}{\ensuremath{\PH \to \PQb\PQb}\xspace}
\newcommand{\BbH}{\ensuremath{\PB \to \PQb\PH}\xspace}
\newcommand{\BbZ}{\ensuremath{\PB \to \PQb\PZ}\xspace}
\newcommand{\BtW}{\ensuremath{\PB \to \PQt\PW}\xspace}
\newcommand{\bH}{\ensuremath{\PQb\PH}\xspace}
\newcommand{\bZ}{\ensuremath{\PQb\PZ}\xspace}
\newcommand{\tW}{\ensuremath{\PQt\PW}\xspace}
\newcommand{\bHbH}{\ensuremath{\PQb\PH\PQb\PH}\xspace}
\newcommand{\bHbZ}{\ensuremath{\PQb\PH\PQb\PZ}\xspace}
\newcommand{\bZbZ}{\ensuremath{\PQb\PZ\PQb\PZ}\xspace}
\newcommand{\bHtW}{\ensuremath{\PQb\PH\PQt\PW}\xspace}
\newcommand{\bZtW}{\ensuremath{\PQb\PZ\PQt\PW}\xspace}
\newcommand{\tWtW}{\ensuremath{\PQt\PW\PQt\PW}\xspace}
\newcommand{\chimod}{\ensuremath{\chi^2_\text{mod}}\xspace}
\newcommand{\chimodndf}{\ensuremath{\chi^2_\text{mod}/\text{ndf}}\xspace}
\newcommand{\BrBbH}{\ensuremath{\mathcal{B}(\PB \to \bH)}\xspace}
\newcommand{\BrBbZ}{\ensuremath{\mathcal{B}(\PB \to \bZ)}\xspace}
\newcommand{\BrBtW}{\ensuremath{\mathcal{B}(\PB \to \tW)}\xspace}
\newcommand{\mvlq}{\ensuremath{m_{\text{VLQ}}}\xspace}
\newcommand{\mH}{\ensuremath{m_{\PH}}\xspace}
\newcommand{\muR}{\ensuremath{\mu_{\text{R}}}\xspace}
\newcommand{\muF}{\ensuremath{\mu_{\text{F}}}\xspace}

\DeclareRobustCommand{\VLQY}{{\HepParticle{Y}{}{}}\Xspace}
\DeclareRobustCommand{\VLQT}{{\HepParticle{T}{}{}}\Xspace}
\ifthenelse{\boolean{cms@external}}{\newcommand{\cmsTable}[1]{#1}}{\newcommand{\cmsTable}[1]{\resizebox{\textwidth}{!}{#1}}}
\ifthenelse{\boolean{cms@external}}{\newcommand{\cmsTableSmall}[1]{#1}}{\newcommand{\cmsTableSmall}[1]{\resizebox{0.75\textwidth}{!}{#1}}}
\newlength\cmsFigWidth\ifthenelse{\boolean{cms@external}}{\setlength\cmsFigWidth{\columnwidth}}{\setlength\cmsFigWidth{0.95\textwidth}}
\ifthenelse{\boolean{cms@external}}{\providecommand{\cmsLeft}{upper\xspace}}{\providecommand{\cmsLeft}{left\xspace}}
\ifthenelse{\boolean{cms@external}}{\providecommand{\cmsRight}{lower\xspace}}{\providecommand{\cmsRight}{right\xspace}} 
\setcounter{tocdepth}{3}

\cmsNoteHeader{B2G-20-014}

\title{Search for bottom-type vector-like quark pair production in dileptonic and fully hadronic final states in proton-proton collisions at \texorpdfstring{$\sqrt{s} = 13\TeV$}{sqrt(s) = 13 TeV}}

\date{\today}

\abstract{
     A search is described for the production of a pair of bottom-type vector-like quarks (\PB VLQs) with mass greater than 1000\GeV. Each \PB VLQ decays into a \PQb quark and a Higgs boson, a \PQb quark and a \PZ boson, or a \PQt quark and a \PW boson. This analysis considers both fully hadronic final states and those containing a charged lepton pair from a \PZ boson decay. The products of the \Hbb boson decay and of the hadronic \PZ or \PW boson decays can be resolved as two distinct jets or merged into a single jet, so the final states are classified by the number of reconstructed jets. The analysis uses data corresponding to an integrated luminosity of 138\fbinv collected in proton-proton collisions at $\sqrt{s} = 13\TeV$ with the CMS detector at the LHC from 2016 to 2018. No excess over the expected background is observed. Lower limits are set on the \PB VLQ mass at 95\% confidence level. These depend on the \PB VLQ branching fractions and are 1570 and 1540\GeV for 100\% \BbH and 100\% \BbZ, respectively. In most cases, the mass limits obtained exceed previous limits by at least 100 GeV.}

\hypersetup{pdfauthor={CMS Collaboration},pdftitle={A search for bottom-type vector-like quark pair production in dileptonic and fully hadronic final states in proton-proton collisions at sqrt(s) = 13 TeV},pdfsubject={CMS},pdfkeywords={CMS,  vector-like quarks}}

\maketitle

\section{Introduction}

A puzzle in elementary particle physics is the fact that the electroweak scale is so much lower than the Planck scale, which is connected to the stability of the Higgs boson (\PH) mass, \mH~\cite{tHooft:1979rat}. In the standard model (SM), the Higgs boson is a fundamental particle, first observed by the ATLAS and CMS Collaborations in 2012~\cite{Aad:2012tfa,Chatrchyan:2012xdj,Chatrchyan:2013lba} at the CERN LHC. In contrast to the leptons, quarks, and vector gauge bosons of the SM, the SM Higgs boson is a fundamental scalar particle with higher-order contributions to its mass due to vacuum fluctuations that diverge quadratically with the length scale. This results in \mH being driven to the cutoff value of these fluctuations. If there is no new physics below the Planck scale, this cutoff value would be about $10^{19}\GeV$, which predicts \mH to be many orders of magnitude greater than the observed value of 125\GeV~\cite{CMS:2020xrn,Georgi:1974yw}.

Several theories have been proposed to solve this problem, including supersymmetry~\cite{Wess:1973kz,Fayet:1976cr}, composite-Higgs models~\cite{Kaplan:1983sm,Agashe:2004rs} (where the Higgs boson is made up of smaller constituents bound by a new type of gauge interaction), and little-Higgs models~\cite{ArkaniHamed:2001nc,ArkaniHamed:2002qy} (where the Higgs boson is a pseudo-Nambu--Goldstone boson arising from spontaneous breaking of a global symmetry at the \TeVns energy scale). In the composite-Higgs case, the mass divergence does not arise, since the UV cutoff is the scale set by the composite size, while, in the case of little-Higgs models, the Higgs boson is ``naturally'' light because of the collective breaking of the global symmetry via two couplings~\cite{Schmaltz:2002wx}.

In both the composite- and little-Higgs models, a new type of particle arises~\cite{delAguila:1982fs}: a vector-like fermion that couples to the weak gauge bosons through purely vector current couplings. The strongly interacting vector-like fermions are referred to as vector-like quarks (VLQs). Unlike heavy fourth-generation sequential chiral quarks, these are not excluded by the measured \PH production cross section~\cite{Aad:2019mbh,Sirunyan:2018sgc}  because the nonchirality of the VLQs allows for gauge-invariant mass terms in the Lagrangian and thus does not require Yukawa couplings. 

There are four types of VLQs with renormalizable couplings to SM quarks. These are defined by their electrical charge: $-1/3$ (\PB), $+2/3$ (\VLQT), $+5/3$ (\PX), or $-4/3$ (\VLQY)~\cite{Aguilar-Saavedra:2013qpa}.  The \PB VLQ mixes with the chiral down-type quarks, and similarly, the \VLQT VLQ mixes with the chiral up-type quarks. Precision measurements that constrain additional couplings of the first- and second-generation SM quarks~\cite{Atre:2008iu,Atre:2011ae,delAguila:2000rc} indicate that the primary couplings of the \VLQT and \PB VLQs are to third-generation SM quarks. The values of the \VLQT and \PB VLQ branching fractions are determined by unknown parameters in the theory: the VLQ multiplet configuration, the VLQ mass, and the coupling of the VLQ to chiral quarks~\cite{Aguilar-Saavedra:2013wba}. The possible decays of the \PB VLQ are \BbH, \BbZ, and \BtW, so the branching fractions of these three modes sum to 100\%. In $\VLQT\PB$ doublet models, the branching fractions \BrBbH and \BrBbZ are approximately equal with values determined by the relative mixing of the \VLQT and \PB VLQs with SM quarks~\cite{Aguilar-Saavedra:2013qpa}. They range from 50\% each for the case of no  $\VLQT\PQt$ mixing to 0\% each for the case of no $\PB\PQb$ mixing. In $\PX\VLQT\PB$ triplet models, \BrBbZ and \BrBbH are each approximately 50\% for \PB VLQ masses above 2\TeV, while in $\VLQT\PB\VLQY$ triplet models and \PB singlet models \BrBbZ and \BrBbH, they are each approximately 25\% for \PB VLQ masses above 2\TeV. In this analysis, lower limits on the \PB VLQ mass are presented over the full range of branching fractions.

Previously, the best \PB mass limits from the CMS Collaboration for pair-produced \PB VLQs~\cite{CMS:2020ttz} were obtained by analyzing events with fully hadronic final states using data from proton-proton ($\Pp\Pp$) collisions at $\sqrt{s} = 13\TeV$, corresponding to an integrated luminosity of 137\fbinv. In that analysis, \PB masses were excluded at 95\% confidence level (\CL) up to 1570, 1390, and 1450 \GeV in the 100\% \BbH, 100\% \BbZ, and $\VLQT\PB$ doublet with no $\VLQT\PQt$ mixing (50\% \BbH and 50\% \BbZ) models, respectively. The best \PB mass limits from the ATLAS Collaboration in pair-produced \PB events  using $\Pp\Pp$ collisions at $\sqrt{s} = 13\TeV$ at 95\% \CL are 1030\GeV (based on 36.1\fbinv)~for the 100\% \BbH case~\cite{ATLAS:2018ziw} and 1420 and 1320\GeV (based on 136\fbinv)~\cite{ATLAS:2022hnn} for the 100\% \BbZ and $\VLQT\PB$ doublet with no $\VLQT\PQt$ mixing models, respectively. 

The analysis presented here combines the fully hadronic and leptonic categories and improves upon the previous CMS analysis in two ways. First, the fully hadronic category is extended by including two additional modes in which one of the VLQs in an event decays to \PQt and \PW so that, in addition to the \bHbH, \bHbZ, and \bZbZ decay modes considered in \cite{CMS:2020ttz},  the \bHtW and \bZtW decay modes are also included. Second, dileptonic events in which a \PZ boson in either the \bHbZ or \bZbZ mode decays to a pair of opposite-sign electrons or muons are included. Because the dileptonic branching fraction of the \PZ boson is an order of magnitude smaller than its hadronic branching fraction, the number of events in these channels is significantly less than in the fully hadronic ones. However, this is offset by the substantially lower background in these channels and results in significantly improved sensitivity to events in which a VLQ decays to \bZ, leading to improved sensitivity for large \BrBbZ.

\section{Analysis overview}
\label{sec:analysis_overview}
\begin{figure*}[hbtp]\centering
    \includegraphics[width=0.36\textwidth]{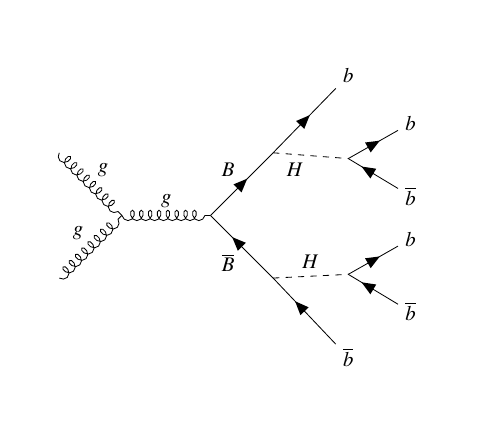}\hspace{0.05\textwidth}  
    \includegraphics[width=0.36\textwidth]{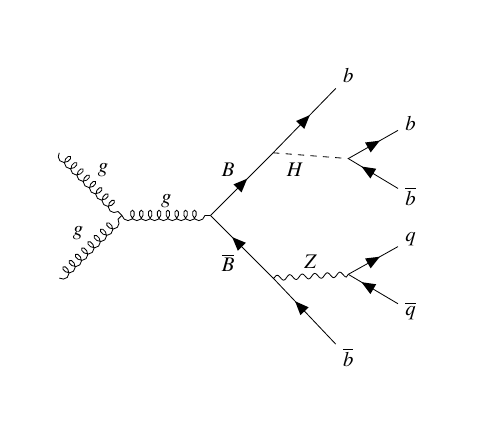}
    \includegraphics[width=0.36\textwidth]{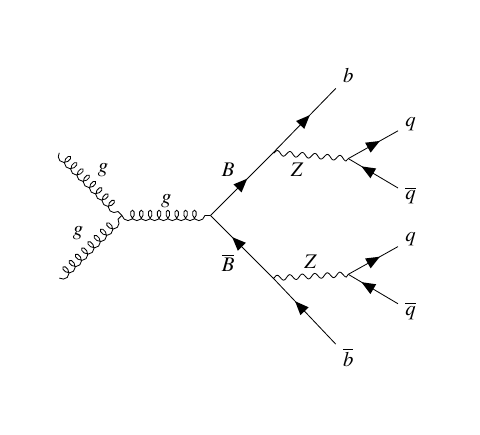}\hspace{0.05\textwidth}
    \includegraphics[width=0.36\textwidth]{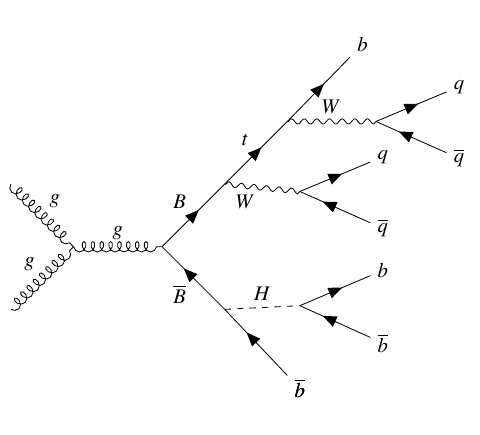}
    \includegraphics[width=0.36\textwidth]{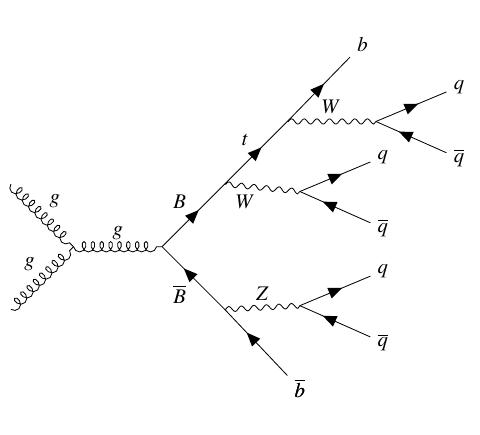}
    \caption{Feynman diagrams for the pair production of bottom-type VLQs that decay into a \PQb or \PQt quark or antiquark and either a Higgs, \PZ, or \PW boson with fully hadronic final states. Upper row: \bHbH and \bHbZ; middle row: \bZbZ and \bHtW; lower row: \bZtW. Modes with decays to \tWtW (not shown) are also possible but not considered in this analysis. The \PB and \PAB can be exchanged in the decays.}
    \label{fig:feynman_hadronic}
\end{figure*}

\begin{figure}[hbtp]\centering
  \includegraphics[width=0.33\textwidth]{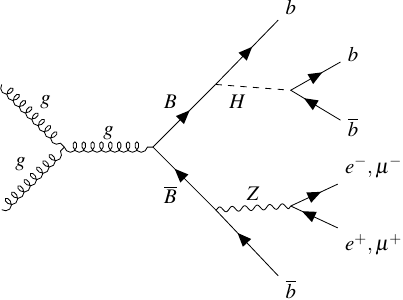}\hspace{0.05\textwidth}
  \includegraphics[width=0.33\textwidth]{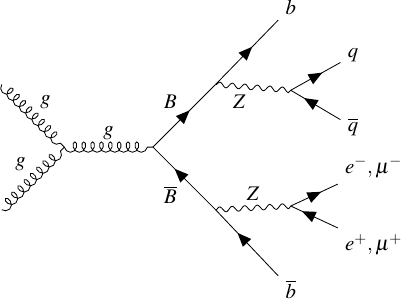}
  \caption{Feynman diagrams of the pair production of bottom-type VLQs that decay into a \PQb quark or antiquark and either a Higgs or \PZ boson with a dilepton final state: \bHbZ mode (\cmsLeft) and \bZbZ mode (\cmsRight). The \PB and \PAB can be exchanged in the decays.}
  \label{fig:feynman_dilepton}
\end{figure}

In this analysis, we search for events featuring the production of a pair of bottom-type VLQs, \PB, with mass greater than 1000\GeV, in data collected with the CMS detector from $\Pp\Pp$ collisions at $\sqrt{s} = 13\TeV$ at the LHC in 2016--2018, corresponding to an integrated luminosity of 138\fbinv~\cite{CMS-LUM-17-003,CMS-PAS-LUM-17-004,CMS-PAS-LUM-18-002}. We consider three possible decays of the \PB: a \PQb quark and a Higgs boson (\BbH), a \PQb quark and a \PZ boson (\BbZ), and a \PQt quark  and a \PW boson (\BtW). Events are classified into different channels depending on three properties. The first is whether the event falls into the ``fully hadronic'' category, in which all of the bosons in an event decay into hadronic jets, or the ``leptonic'' category, where an event contains a \BbZ decay in which the \PZ boson decays into a pair of leptons, either electrons ($\Pe\Pe$) or muons ($\PGm\PGm$). Since an essential element of the analysis is the categorization of events based on their kinematical reconstruction, events with leptonic \PW boson decays are not considered. Figures~\ref{fig:feynman_hadronic} and~\ref{fig:feynman_dilepton} show the representative Feynman diagrams for the different decay modes in the fully hadronic and leptonic categories, respectively. 

The second property determining the event classification is the reconstructed decay mode, which is determined by the decays of the two \PB VLQs. The leptonic category includes only \bHbZ and \bZbZ decay modes (\ie, events with one \BbH and one \BbZ decay, and events with two \BbZ decays, respectively), while the fully hadronic category includes, in addition, \bHbH, \bHtW, and \bZtW decay modes. The \tWtW mode is not considered since its high jet multiplicity results in a large number of possible jet combinations.

The third property determining the event classification is the number of reconstructed jets. Because the bosons produced in the \PB VLQ decay often have a significant Lorentz boost, the two jets produced in the hadronic decay of a \PW, \PZ, or Higgs boson can merge and be reconstructed as a single merged jet. In the \tW case, the \PQt quark decay into $\PQb\PW$ can produce three jets (one from the \PQb quark and two from the \PW boson), two jets (one from the \PQb quark and a merged jet from the \PW boson), or one fully merged jet. As a result, the final state can have a different jet multiplicity, depending on the number of merged jets. In the leptonic category, we also include events where an additional jet is produced from initial- or final-state radiation (ISR or FSR). Events containing ISR and FSR are not included in the hadronic category due to the large number of possible jet combinations resulting from the higher jet multiplicities.

The reconstruction of events and assignment of jets to parent particles is performed using a modified $\chi^2$ metric, \chimod, which uses as input the mass differences between the two reconstructed bosons and the Higgs, \PZ, or \PW boson, normalized by their resolutions, and the fractional mass difference of the two reconstructed VLQs. The value of \chimodndf, where ndf is the number of degrees of freedom, is calculated for each possible reconstructed decay mode and jet assignment, and the mode with the lowest value is selected as the reconstructed mode. Thus, each event is assigned to exactly one reconstructed mode. In the leptonic case, the \chimodndf value is also used to identify extra jets that are likely to be from ISR and FSR.

The signature in the leptonic category consists of three or four final-state jets and a pair of oppositely charged, same-flavor electrons or muons with an invariant mass consistent with the \PZ boson mass. In the fully hadronic category, the signature consists of four, five, or six final-state jets. Table~\ref{table:eventmodes} summarizes the final event classification, showing the set of ``channels'' defined by the event category, decay mode, and jet multiplicity.

\begin{table*}[htbp]
\centering
\topcaption{Summary of channels considered for each category and jet multiplicity. Although events with a jet from ISR or FSR are included in the leptonic category, for these events the extra jet is not included in the categorization of the jet multiplicity of the event.}
\label{table:eventmodes}
\begin{scotch}{ccc}
Jet          & Leptonic & Fully hadronic \\
multiplicity & category & category \\
\hline
3 & \bHbZ, \bZbZ & \NA \\
4 & \bHbZ, \bZbZ & \bHbH, \bHbZ, \bZbZ \\
5 & \NA & \bHbH, \bHbZ, \bZbZ, \bHtW, \bZtW \\
6 & \NA & \bHbH, \bHbZ, \bZbZ, \bHtW, \bZtW \\
\end{scotch}
\end{table*}

The main background in the leptonic category consists of Drell--Yan (DY) dilepton production in association with jets, while in the fully hadronic category the background is predominantly from SM events composed uniquely of jets produced through the strong interaction, referred to as quantum chromodynamics (QCD) multijet events.

The \chimod requirement also aids in separating potential signal events from background by ensuring that the jets are kinematically consistent with production from a Higgs, \PZ, or \PW boson and that the reconstructed VLQs have masses consistent with each other. Further signal separation is achieved by requiring that some of the jets are tagged as originating from \PQb quarks, either using tagging of individual jets or tagging of $\bbbar$ pairs in merged jets from $\PH \to \PQb\PAQb$ decays. In addition, in the dileptonic case, the two leptons are also required to be kinematically consistent with the decay of a \PZ boson. 

\section{The CMS detector}

 The central feature of the CMS apparatus is a superconducting solenoid of 6\unit{m} internal diameter, providing a magnetic field of 3.8\unit{T}. Within the solenoid volume are a silicon pixel and strip tracker, a lead tungstate crystal electromagnetic calorimeter (ECAL), and a brass and scintillator hadron calorimeter (HCAL), each composed of a barrel and two endcap sections. Forward calorimeters extend the pseudorapidity ($\eta$) coverage provided by the barrel and endcap detectors. Muons are measured in gaseous detectors embedded in the steel flux-return yoke outside the solenoid.

The ECAL consists of 75\,848 lead tungstate crystals, which provide coverage in $\abs{\eta} < 1.48$ in a barrel region and $1.48 < \abs{\eta} < 3.0$ in two endcap regions. In the region $\abs{\eta} < 1.74$, the HCAL cells have widths of 0.087 in $\eta$ and 0.087 in azimuth ($\phi$). For $\abs{\eta} < 1.48$, the HCAL cells map in the $\eta$-$\phi$ plane onto $5{\times}5$ arrays of ECAL crystals to form calorimeter towers projecting radially outwards from close to the nominal interaction point. For $\abs{\eta} > 1.74$, the coverage of the towers increases progressively to a maximum of 0.174 in $\Delta \eta$ and $\Delta \phi$. Within each tower, the energy deposits in ECAL and HCAL cells are summed to define the calorimeter tower energies, which are subsequently used to provide the energies and directions of hadronic jets. When combining information from the entire detector, the jet energy resolution typically amounts to 15--20\% at 30\GeV, 10\% at 100\GeV, and 5\% at 1\TeV. The electron momentum is estimated by combining the energy measurement in the ECAL with the momentum measurement in the tracker. The momentum resolution for electrons with $\pt \approx 45\GeV$ from $\PZ \to \Pe \Pe$ decays ranges from 1.6 to 5\%. It is generally better in the barrel region than in the endcaps, and also depends on the bremsstrahlung energy emitted by the electron as it traverses the material in front of the ECAL~\cite{CMS:2020uim,CMS-DP-2020-021}. 

Muons are measured in the pseudorapidity range $\abs{\eta} < 2.4$, with detection planes made using three technologies: drift tubes, cathode strip chambers, and resistive-plate chambers. The single muon trigger efficiency exceeds 90\% over the full $\eta$ range, and the efficiency to reconstruct and identify muons is greater than 96\%. Matching muons to tracks measured in the silicon tracker results in a relative transverse momentum (\pt) resolution, for muons with \pt up to 100\GeV, of 1\% in the barrel and 3\% in the endcaps. The \pt resolution in the barrel is better than 7\% for muons with \pt up to 1\TeV~\cite{CMS:2018rym}. 

Events of interest are selected using a two-tiered trigger system. The first level, composed of custom hardware processors, uses information from the calorimeters and muon detectors to select events at a rate of around 100\unit{kHz} within a fixed latency of about 4\mus~\cite{Sirunyan:2020zal}. The second level, known as the high-level trigger, consists of a farm of processors running a version of the full event reconstruction software optimized for fast processing, and reduces the event rate to around 1\unit{kHz} before data storage~\cite{Khachatryan:2016bia}.

A more detailed description of the CMS detector, together with a definition of the coordinate system used and the relevant kinematic variables, can be found in Ref.~\cite{Chatrchyan:2008zzk}.

\section{Data and simulated events}

Signal events with pair production of VLQs were simulated using the Monte Carlo generator \MGvATNLO~\cite{Alwall:2014hca}. For samples corresponding to 2016 data, version v2.3.3 was used with NNPDF3.0 next-to-leading order (NLO) parton distribution functions (PDFs)~\cite{Ball:2014uwa}; for samples corresponding to 2017--2018 data, v2.4.2 was used with NNPDF3.1 next-to-NLO PDFs~\cite{NNPDF:2017mvq}. The generated VLQ masses $m_{\PB}$ cover the range 1000--1800\GeV in steps of 100\GeV. Hadronization of the underlying partons was simulated using \PYTHIA v8.212~\cite{Sjostrand:2014zea} with the CUETP8M1 tune~\cite{Khachatryan:2015pea} for samples corresponding to 2016 data, and with the CP5 tune~\cite{Sirunyan:2019dfx} for samples corresponding to 2017 and 2018 data. Corrections to the cross sections to next-to-NLO and next-to-next-to-leading logarithmic soft-gluon resummation were obtained using \textsc{Top++} 2.0~\cite{Czakon:2013goa,Czakon:2011xx,Cacciari:2011hy} with the MSTW2008NNLO68CL PDF set from the \textsc{lhapdf} 5.9.0 library~\cite{Whalley:2005nh,Bourilkov:2006cj,Sirunyan:2018omb}.

Although the background estimate is derived purely from data, simulated background samples are used for cross-checks of the data distributions. The background processes considered include $\text{DY} + \text{jets}$, QCD multijet, $\PW + \text{jets}$, $\PZ + \text{jets}$, and $\ttbar + \text{jets}$. The $\text{DY} + \text{jets}$ process is simulated with \MGvATNLO v2.6.0 at NLO, using the FxFx prescription~\cite{Frederix:2012ps} for jet merging, while the other four backgrounds are simulated using \MGvATNLO v2.4.2 at leading order with the MLM prescription~\cite{Alwall:2007fs} for jet merging. The hadronization is simulated in the same way as for the signal samples. 

In order to simulate the effect of additional $\Pp\Pp$ interactions within the same or nearby bunch crossings (``pileup''), \PYTHIA v8.226 with a total inelastic $\Pp\Pp$ cross section of 69.2\unit{mb}~\cite{CMS:2018mlc} was used to simulate minimum bias events to overlay on the hard scattering process. Following event generation, the \GEANTfour package~\cite{Agostinelli:2002hh,Allison:2006ve} was used to simulate the CMS detector response. Scale factors corresponding to jet energy corrections, jet energy resolutions~\cite{Khachatryan:2016kdb}, QCD renormalization and factorization scales, pileup, and jet tagging~\cite{CMS-DP-2018-058,Sirunyan:2017ezt} were applied to the simulated signal events so that the corresponding distributions would agree with those in the data.

\section{Lepton and jet reconstruction and jet tagging}\label{sec:lepton_jet_reconstruction}

The particle-flow algorithm~\cite{Sirunyan:2017ulk} aims to reconstruct and identify each individual particle in an event, with an optimized combination of information from the various elements of the CMS detector. The primary vertex is taken to be the vertex corresponding to the hardest scattering in the event, evaluated using tracking information alone, as described in Section 9.4.1 of Ref.~\cite{CMS-TDR-15-02}. The energy of photons is obtained from the ECAL measurement. The energy of electrons is determined from a combination of the electron momentum from the track, the energy of the corresponding ECAL cluster, and the energy sum of all bremsstrahlung photons spatially compatible with originating from the electron track. The energy of muons is obtained from the curvature of the corresponding track. The energy of charged hadrons is determined from a combination of their momentum measured in the tracker and the matching ECAL and HCAL energy deposits, corrected for the response function of the calorimeters to hadronic showers. Finally, the energy of neutral hadrons is obtained from the corresponding corrected ECAL and HCAL energies.

For each event, hadronic jets are clustered from these reconstructed particles using the infrared and collinear safe anti-\kt algorithm via \textsc{FastJet}~\cite{Cacciari:2008gp, Cacciari:2011ma}. To account for the difference between merged and resolved jets, two separate clusterings are performed on the input particles, one using a distance parameter of 0.4 (``AK4 jets'') and the other using a distance parameter of 0.8 (``AK8 jets''). Jet momentum is determined as the vectorial sum of all particle momenta in the jet, and from simulation it is found to be, on average, within 5 to 10\% of the true momentum over the entire \pt spectrum and detector acceptance. Pileup can contribute additional tracks and calorimetric energy depositions to the jet momentum. The pileup-per-particle identification algorithm (PUPPI)~\cite{Sirunyan:2020foa,Bertolini:2014bba} is used to mitigate the effect of pileup at the reconstructed-particle level, making use of local shape information, event pileup properties, and tracking information. A local shape variable is defined, which distinguishes between collinear and soft diffuse distributions of other particles surrounding the particle under consideration. The former is attributed to particles originating from the hard scatter and the latter to particles originating from pileup interactions. Charged particles identified to be originating from pileup vertices are discarded. For each neutral particle, a local shape variable is computed using the surrounding charged particles compatible with the primary vertex within the tracker acceptance ($\abs{\eta} < 2.5$), and using both charged and neutral particles in the region outside of the tracker coverage. The momenta of the neutral particles are then rescaled according to their probability to originate from the primary interaction vertex deduced from the local shape variable, superseding the need for jet-based pileup corrections~\cite{Sirunyan:2020foa}.

Jet energy corrections are derived from simulation studies so that the average measured energy of jets becomes identical to that of particle-level jets. In situ measurements of the momentum balance in dijet, $\text{photon} + \text{jet}$, $\PZ + \text{jet}$, and multijet events are used to account for any residual differences in the jet energy scale between data and simulation~\cite{Khachatryan:2016kdb}. Additional selection criteria are applied to each jet to remove jets potentially dominated by anomalous contributions from various subdetector components or reconstruction failures.

We require AK4 jets to have $\pt > 50\GeV$ and AK8 jets to have $\pt > 200\GeV$, with $\abs{\eta} < 2.4$ in both cases. The event jet multiplicity is determined by the number of AK4 jets in the event passing these requirements. For the AK8 jets, a grooming algorithm is applied. In this algorithm, the constituents of the AK8 jets are reclustered using the Cambridge--Aachen algorithm~\cite{Dokshitzer:1997in,Wobisch:1998wt}. The ``modified mass drop tagger'' algorithm~\cite{Dasgupta:2013ihk,Butterworth:2008iy}, also known as the ``soft-drop'' (SD) algorithm, with angular exponent $\beta = 0$, soft cutoff threshold $z_{\mathrm{cut}} < 0.1$, and characteristic radius $R_{0} = 0.8$~\cite{Larkoski:2014wba}, is applied to remove soft, wide-angle radiation from the jet. The resulting ``soft-drop mass'' provides a more accurate estimate of the mass of the parent Higgs, \PZ, or \PW boson in the case of merged jets, and so is used in these cases.

The final state is expected to contain two \PQb jets from the decays of the VLQs, and may contain additional \bbbar pairs from $\PH \to \bbbar$ or $\PZ \to \bbbar$ decays. Consequently, identification of these \PQb jets provides an effective way to discriminate signal events from SM background. For individual jets, the \textsc{DeepJet} \PQb discriminant~\cite{Bols:2020bkb,CMS-DP-2018-058,Sirunyan:2017ezt} is applied to AK4 jets to obtain single \PQb tags, while for merged jets containing $\bbbar$ pairs, an algorithm developed for $\PH \to \bbbar$ tagging~\cite{Sirunyan:2017ezt} is applied to AK8 jets to obtain double-\PQb tags.

In the leptonic category, electrons or muons are selected with $\pt > 50\GeV$ and $\abs{\eta} < 2.4$. Electrons and muons are required to pass a loose set of identification requirements, and muons are additionally required to pass a loose isolation requirement to reject muons within jets from, \eg, semileptonic decays of \PQb quarks~\cite{CMS:2020uim,CMS:2018rym}. Studies of the dileptonic mass distributions in simulation showed that this additional loose isolation requirement on muons improved the \PZ boson purity, while a similar loose isolation requirement on electrons was not needed.

\section{Event selection, reconstruction, and categorization}
\label{sec:eventreco}

In this section, the selection and categorization of events is described. This consists of the online selection of events by selected triggers; the offline requirements on jets, leptons and \PQb-tagged jets; and the reconstruction of the VLQ candidates and selection of the event category. 

\subsection{Online event selection}
Events are selected online with the following high-level trigger requirements:
\begin{itemize}
\item For the fully hadronic category, the trigger requires the scalar sum of the measured jet \pt values to be greater than 900 (1050)\GeV in the 2016 (2017--2018) data set.
\item For leptonic events containing muons, the trigger requires a muon with $\pt > 50\GeV$. For 2017--2018 data, a trigger that accepts muons with $\pt > 100\GeV$ identified from their signature in the tracker is also used as a backup to the primary trigger for consistency with other analyses~\cite{CMS:2021yvr}.
\item For leptonic events containing electrons, a set of triggers is used to increase the efficiency, selecting any of the following: an electron passing a tight set of identification criteria with $\pt > 27$, 35, or 32\GeV in 2016, 2017, or 2018, respectively; an electron with $\pt > 115\GeV$; a photon with $\pt > 200\GeV$ (175\GeV in 2016); a pair of isolated electrons with $\pt > 23\: (12)\GeV$ for the leading (subleading) electron; a pair of  electrons with $\pt > 33\GeV$; or a pair of photons with $\pt > 70\GeV$ (60\GeV in 2016).
\end{itemize}

The trigger efficiencies are measured using data sets collected by triggers independent of those used in the analysis. For the fully hadronic category, the trigger efficiencies are parameterized by\HT, defined as the scalar sum of the \pt of all offline AK4 jets satisfying the above \pt and $\eta$ requirements. The efficiency exceeds 99\% for values of \HT greater than 1350 GeV. For the leptonic category, the trigger efficiencies are parameterized by the highest \pt lepton. Both the electron and muon trigger efficiencies exceed 98\% for values of the highest lepton \pt greater than 50 GeV.

\subsection{Offline requirements on jets, leptons, and \texorpdfstring{\PQb}{b}-tagged jets}
Offline selection requirements are then applied. For the leptonic category, the event is required to have at least three and no more than five AK4 jets with $\pt > 50\GeV$ and $\abs{\eta} < 2.4$, and to have at least one pair of opposite-sign, same-flavor leptons with an invariant mass in the range $80 < m_{\Pell\Pell} < 102\GeV$, consistent with the \PZ boson mass. If there are two such pairs each with invariant mass within the \PZ boson mass window, the pair with the mass closest to that of the \PZ boson is used. For the fully hadronic category, there must be at least four and no more than six AK4 jets with $\pt > 50\GeV$ and $\abs{\eta} < 2.4$, and the event must have $\HT > 1350\GeV$. In addition, to ensure orthogonality with both the leptonic analysis and with other \PB VLQ searches using a single lepton, any event with an isolated electron or muon with $\pt > 50\GeV$, or at least one lepton pair meeting the criteria for the leptonic category, is rejected.

A requirement is also placed on the minimum number of jets tagged as originating from a \PQb quark. For the fully hadronic category, the number of tagged \PQb jets and the working points (WPs) for the taggers are optimized for discovery of a 1400\GeV VLQ signal. The WPs used for the single-\PQb tagger, ``loose'', ``medium'', and ``tight'', correspond to an identification efficiency of approximately 93, 82, and 65\%, respectively, for \PQb jets with $\pt > 30\GeV$ in simulated \ttbar events, with a probability of 10\%, 1\%, and 0.1\%, respectively, for light-quark (\PQu, \PQd, \PQs) and gluon jets to be misidentified as \PQb jets~\cite{CMS-DP-2018-058}. For the $\bbbar$ tagger, two WPs, ``loose'' and ``medium 2'', are used, corresponding to an identification efficiency of approximately 75 and 45\%, respectively, with a misidentification probability of 11 and 3\%, respectively~\cite{Sirunyan:2017ezt}. For the leptonic category, the medium WP is used for the single \PQb tagger in all channels. Table~\ref{table:btag_requirements} summarizes the final requirements.

\begin{table*}[hbtp]
    \centering
    \topcaption{Required minimum number of single ($N_{\PQb}$) and double ($N_{\bbbar}$) \PQb tags, and working points (WPs) used for each category, decay mode, jet multiplicity ($N_{\text{j}}$), and number of merged jets ($N_{\text{AK8}}$). The working points are described in the text. For a given event mode, there can be several jet multiplicities depending on the number of merged jets.}
    \label{table:btag_requirements}
    \begin{scotch}{lcccccc}
        Decay mode & $N_{\text{j}}$ & $N_{\text{AK8}}$ & Min $N_{\PQb}$   & \PQb tagger WP & Min $N_{\bbbar}$   & $\bbbar$ tagger WP \\
        \hline
        Fully hadronic \bHbH & 4 & 2 & 2 & Medium & 1 & Loose \\
        Fully hadronic \bHbH & 5 & 1 & 3 & Medium & 1 & Loose \\
        Fully hadronic \bHbH & 6 & 0 & 4 & Medium & \NA & \NA \\ [\cmsTabSkip]
        Fully hadronic \bHbZ & 4 & 2 & 2 & Loose & 1 & Medium 2 \\
        Fully hadronic \bHbZ & 5 & 1 & 3 & Medium & 0 & Loose \\
        Fully hadronic \bHbZ & 6 & 0 & 4 & Medium & \NA & \NA \\ [\cmsTabSkip]
        Fully hadronic \bZbZ & 4 & 2 & 2 & Medium & 0 & Loose \\
        Fully hadronic \bZbZ & 5 & 1 & 2 & Medium & 0 & Loose \\
        Fully hadronic \bZbZ & 6 & 0 & 2 & Tight & \NA & \NA \\ [\cmsTabSkip]
        Fully hadronic \bHtW & 5 & 3 & 2 & Medium & 0 & Loose \\
        Fully hadronic \bHtW & 6 & 2 & 3 & Medium & 0 & Loose \\ [\cmsTabSkip]
        Fully hadronic \bZtW & 5 & 3 & 2 & Medium & 0 & Loose \\
        Fully hadronic \bZtW & 6 & 2 & 2 & Tight & 0 & Loose \\ [\cmsTabSkip]
        Leptonic \bHbZ & 3 & 1 & 1 & Medium & \NA & \NA \\
        Leptonic \bHbZ & 4 & 0 & 1 & Medium & \NA & \NA \\ [\cmsTabSkip]
        Leptonic \bZbZ & 3 & 1 & 1 & Medium & \NA & \NA \\
        Leptonic \bZbZ & 4 & 0 & 1 & Medium & \NA & \NA \\
    \end{scotch}
\end{table*}

\subsection{Event reconstruction and categorization}\label{sec:eventcat}We  define a ``preselection'' sample, which consists of events that pass all of the selection requirements except for the \PQb tagging. Before the \PQb tagging requirements are applied, the event sample is dominated by SM backgrounds, primarily QCD multijet and $\ttbar$ processes, and any potential contribution from the signal is negligible. Using the ``preselection'' events, candidate bosons (Higgs, \PZ, or \PW) and VLQs are reconstructed. A candidate boson is formed from either two AK4 jets or a single AK8 jet, with the mass taken as the invariant mass of the two jets in the former case and the soft-drop mass of the jet in the latter case. We only use AK8 jets that have a soft-drop mass greater than 50\GeV and match to an AK4 jet within $\Delta R = \sqrt{\smash[b]{(\Delta\eta)^2 + (\Delta\phi)^2}} < 0.3$. In order to avoid overlap with another AK4 jet which could affect the AK8 jet mass, the AK8 jet is discarded if there is another AK4 jet that matches to it within $\Delta R < 0.6$.

A VLQ candidate decaying to \bH or \bZ is then formed by combining an additional jet with a boson candidate, with its reconstructed mass \mvlq taken to be the invariant mass of this combination. In the \tW case, a \PQt quark candidate can be reconstructed either from a single AK8 jet, with the mass taken to be the soft-drop mass of the jet, or from two or three AK4 jets, with the mass taken to be the invariant mass of these jets.

The decay mode, as well as the assignment of reconstructed jets to their parent particles, is determined using the modified $\chi^2$ parameter, \chimod, evaluated for all possible decay modes and jet assignments. Because the inputs to \chimod are not necessarily distributed as Gaussian variables, the resulting expression does not follow a true $\chi^2$ distribution, and thus \chimod is used to denote it. However, since \chimod is used for the event selection to choose the channel and jet configuration as well as provide a discriminant, this difference does not affect the outcome of the analysis.

The general form of \chimod is as follows in the fully hadronic category:
\begin{linenomath}
\begin{equation}
\label{eq:chi2_equation_hadronic}
\chimod = \frac{(\Delta\mvlq - \overline{\Delta m}_{\text{VLQ}})^2}{\sigma^2_{\Delta\mvlq}} + \frac{(m_1 - \overline{m_1})^2}{\sigma^2_{m_1}} + \frac{(m_2 - \overline{m_2})^2}{\sigma^2_{m_2}},
\end{equation}
\end{linenomath}
where $\Delta\mvlq$ is the fractional mass difference of the two reconstructed VLQ candidates in the event, given by $\Delta\mvlq = 2({\mvlq}_1 - {\mvlq}_2)/({\mvlq}_1 + {\mvlq}_2)$; $m_{1,2}$ are the masses of the two reconstructed bosons; $\overline{m}_{1,2}$ and $\overline{\Delta m}_{\text{VLQ}}$ are the average masses of the bosons and $\Delta\mvlq$, respectively; and $\sigma$ are the standard deviations. The values of $\overline{m}_i$, $\sigma_{m_i}$, $\overline{\Delta m}_{\text{VLQ}}$, and  $\sigma_{\Delta\mvlq}$ are computed by using simulated signal samples with $m_{\PB} = 1400\GeV$ and selecting events where the jets are closely matched to the generated partons in simulation.

In the case where the decay mode contains a \BtW decay, another term is added to the \chimod expression for each reconstructed \PQt quark: $(m_{\PQt} - \overline{m}_{\PQt})/\sigma^2_{m_{\PQt}}$, where $m_{\PQt}$ is the reconstructed \PQt quark mass, and $\overline{m}_{\PQt}$ and $\sigma_{m_{\PQt}}$ are the average and standard deviation of the \PQt quark mass in simulation. This term is added to aid in distinguishing such events from events where one of the \PB VLQs decays to \bZ.  Because of the missing momentum in events with a leptonic $\PW$ boson decay, the $m_{\PW}$ term and, subsequently, the $\Delta m_{\text{VLQ}}$ term used in the $\chimod$ calculation would not be accurate enough to allow adequate categorization of  the event. As a result, events with leptonic \PW boson decays are not considered in this analysis.

In the leptonic category, only the mass of the hadronically decaying boson is included in \chimod, so the expression is simply:
\begin{linenomath}
\begin{equation}
\label{eq:chi2_equation_leptonic}
\chimod = \frac{(\Delta\mvlq - \overline{\Delta m}_{\text{VLQ}})^2}{\sigma^2_{\Delta\mvlq}} + \frac{(m_1 - \overline{m_1})^2}{\sigma^2_{m_1}},
\end{equation}
\end{linenomath}
where $m_1$ is the mass of the hadronically decaying Higgs or \PZ boson.

Mass distributions for the parent \PZ, \PW, \PH, or \PQt are obtained and then fitted with a bifurcated Gaussian (\ie, a Gaussian that has a different $\sigma$ parameter above and below the mean). Consequently, when the \chimod value is computed, the appropriate $\sigma$ value is used depending on whether the reconstructed mass is above or below $\overline{m}$. These distributions are computed separately depending on whether the final state is reconstructed as two resolved jets or one merged jet, and also separately for each jet multiplicity.
The values for $\sigma_{\Delta\mvlq}$ are typically around 0.1, while the values for $\sigma_{m_i}$ are roughly 10--20\GeV.

Once the \chimod value has been computed for all channels within an event category, the channel with the smallest \chimodndf is selected. The value of ndf is 2 for the leptonic channels; 3 for the fully hadronic modes \bHbH, \bHbZ, and \bZbZ; and 4 for the fully hadronic modes \bHtW and \bZtW. This method was found to have good accuracy, as determined by tests with the signal simulations. Although the mode of some events is not correctly identified, these events are not rejected but are included as part of the signal under a different mode, while the selection based on \chimod still provides good background rejection. 

\begin{figure*}[!htbp]
\centering 
\includegraphics[width=0.4\textwidth]{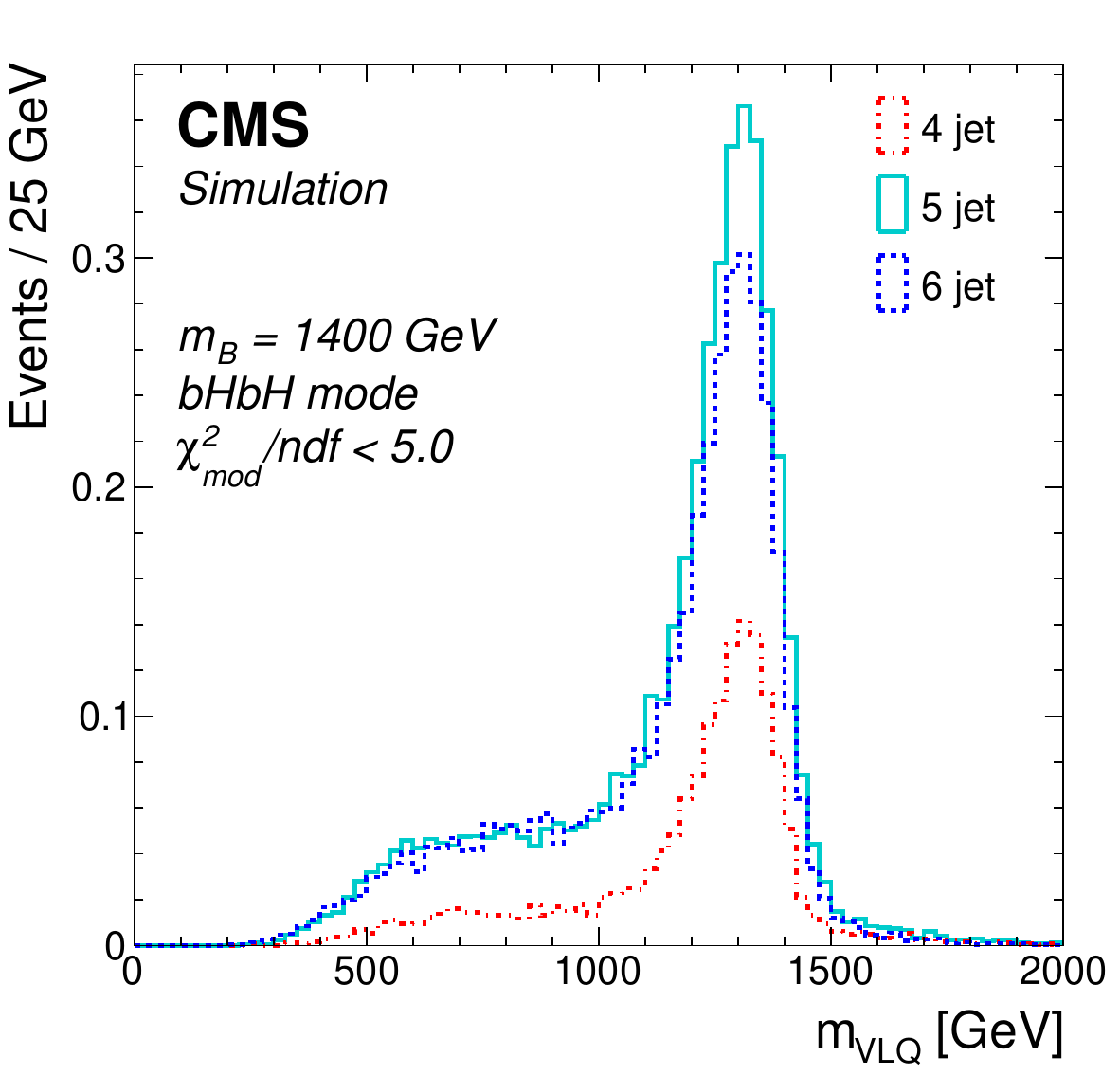}
\includegraphics[width=0.4\textwidth]{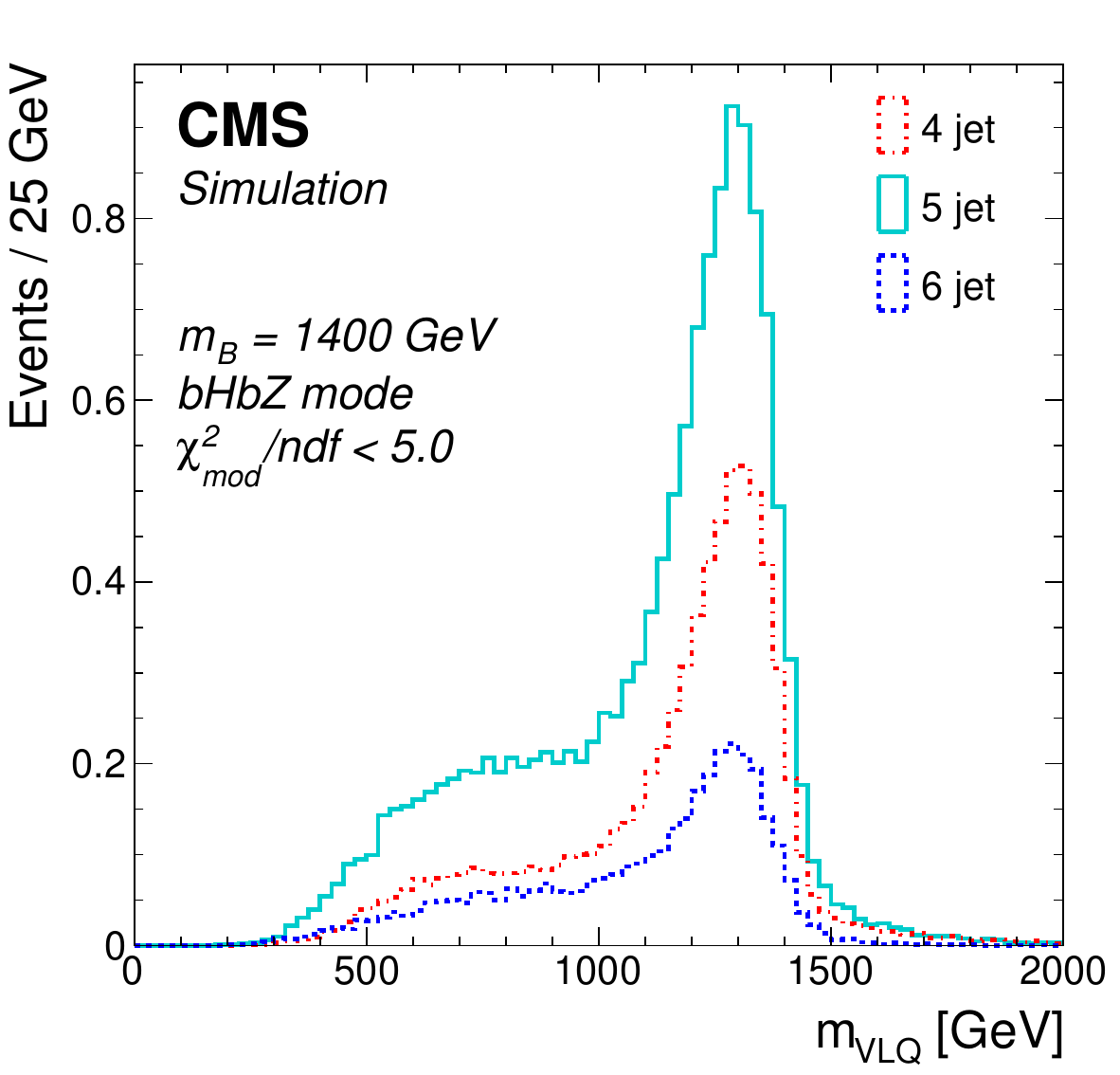}
\includegraphics[width=0.4\textwidth]{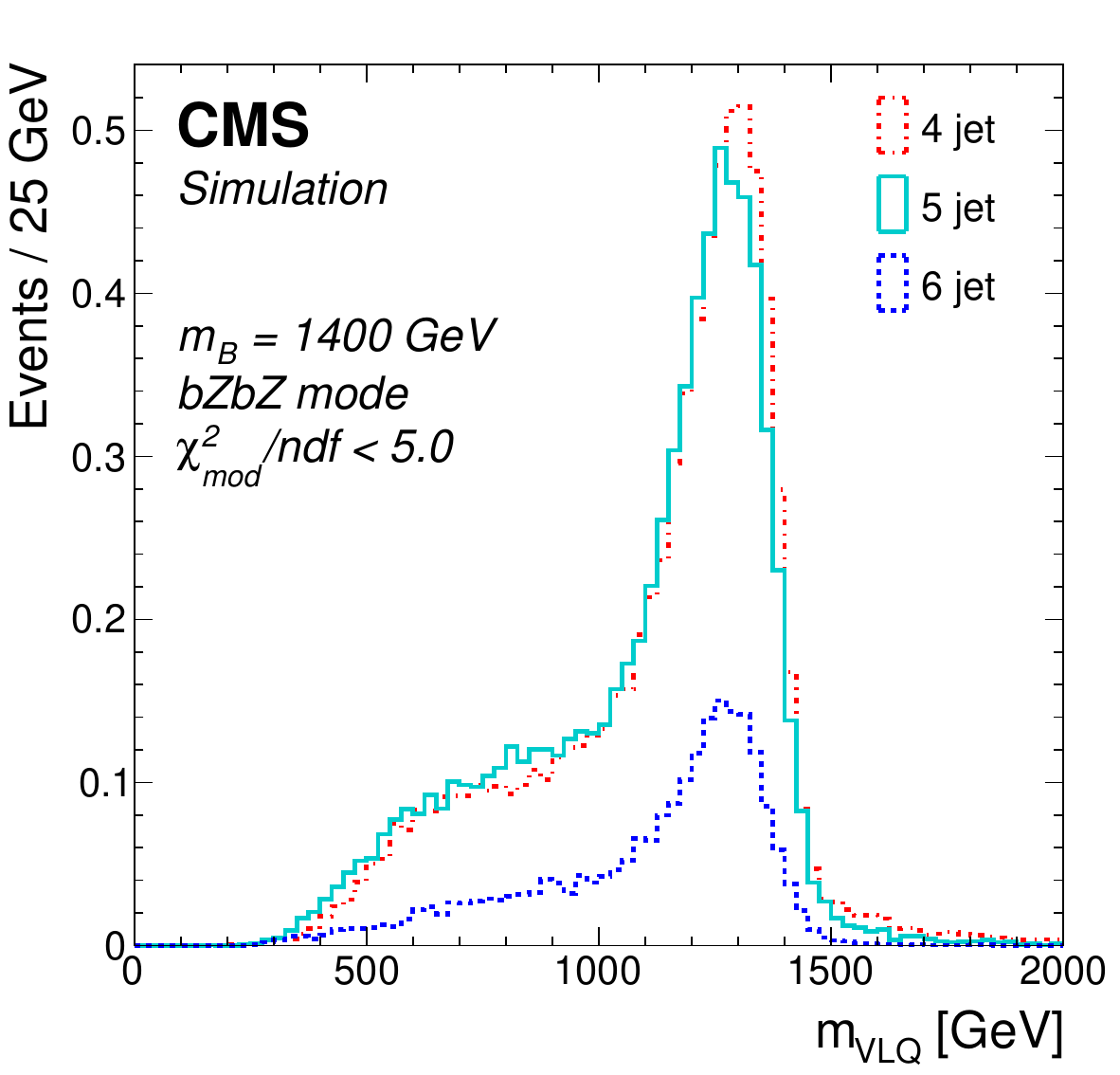}
\includegraphics[width=0.4\textwidth]{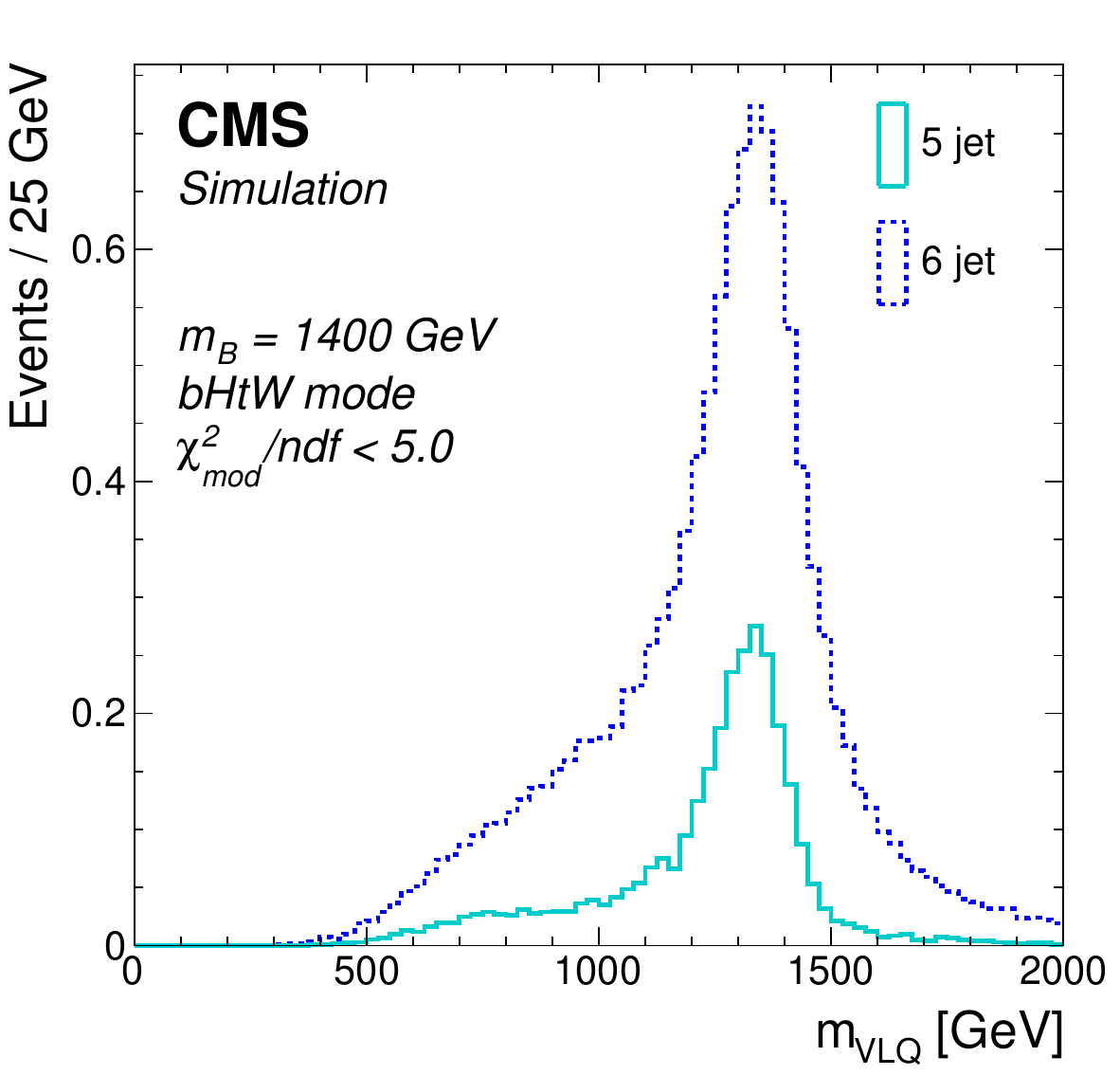}
\includegraphics[width=0.4\textwidth]{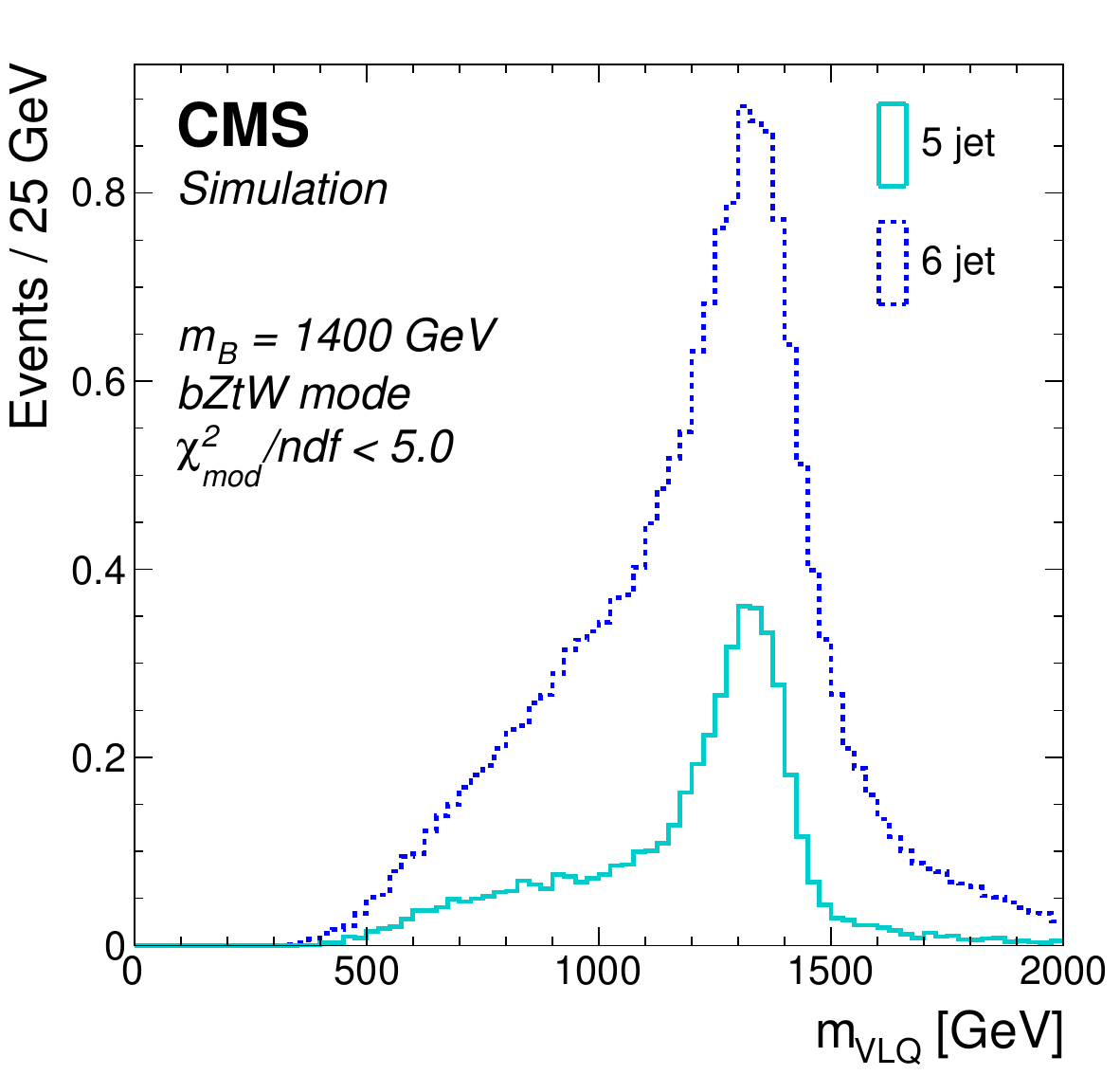}
\caption{Reconstructed VLQ mass distributions for simulated events for the channels in the fully hadronic category with $m_{\PB} = 1400\GeV$. Upper row: Channels in the \bHbH (left) and \bHbZ (right) decay modes. Middle row: Channels in the \bZbZ (left) and \bHtW (right) decay modes. Lower row: Channels in the \bZtW decay mode. The different colors indicate the different jet multiplicities. A selection of $\chimodndf < 5$ has been applied. The values represent the expected number of events over the background in the 2016--2018 data sample. }
\label{fig:signal_mass_had}
\end{figure*}

\begin{figure}[!htbp]
\centering 
\includegraphics[width=0.49\textwidth]{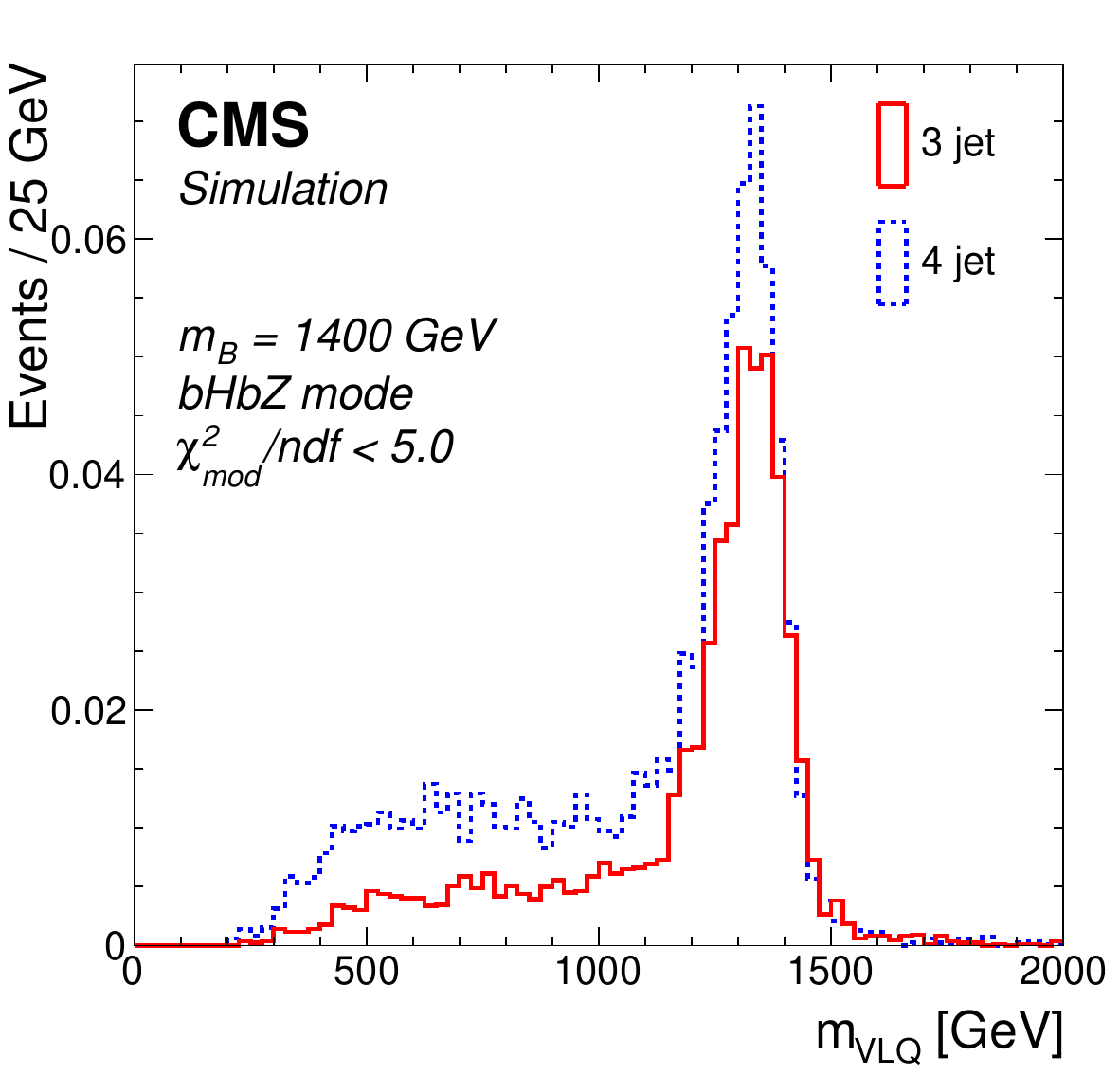}
\includegraphics[width=0.49\textwidth]{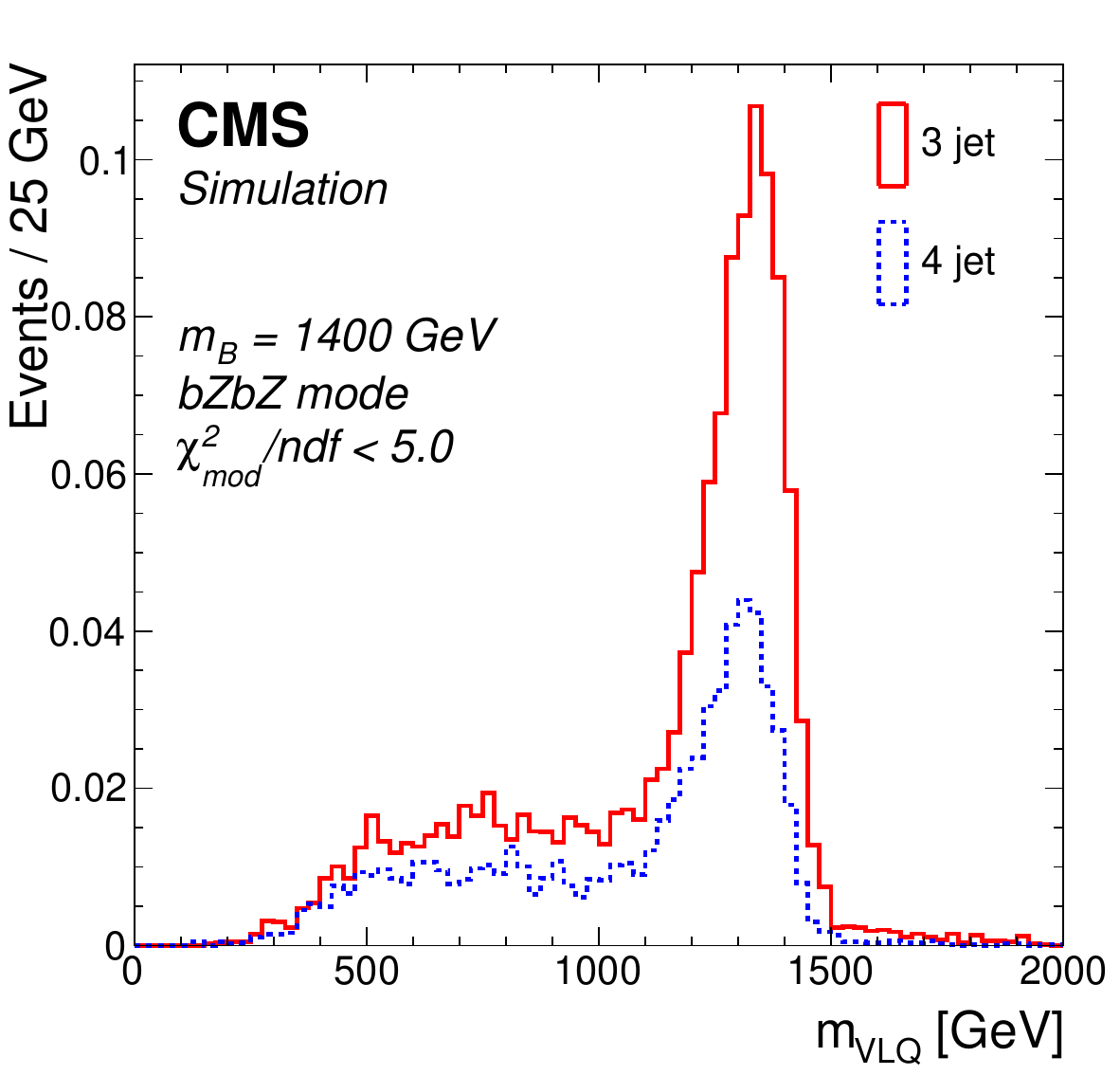}
\caption{Reconstructed VLQ mass distributions for simulated events passing the \PQb tag requirement for the channels in the dileptonic category with $m_{\PB} = 1400\GeV$ in the \bHbZ (\cmsLeft) and \bZbZ (\cmsRight) event modes. A selection of $\chimodndf < 5$ has been applied. The values represent the expected number of events over the background in the 2016--2018 data sample. }
\label{fig:signal_mass_dilep}
\end{figure}

The performance of the \chimod-based reconstruction is indicated by the average of the two reconstructed VLQ masses, referred to as \mvlq, which is shown for simulated signal samples with $m_{\PB} = 1400\GeV$ in Fig.~\ref{fig:signal_mass_had} (channels in the fully hadronic category) and Fig.~\ref{fig:signal_mass_dilep} (channels in the dileptonic category). The mass is well reconstructed for events near the peak; the low-mass tail corresponds to events in which the selected jet permutation is incorrect. Tests on simulated signal samples with different \PB VLQ masses showed similar performance.

In the leptonic category, in order to account for a possible jet due to ISR or FSR, an extra jet reconstructed by the method described in Section~\ref{sec:lepton_jet_reconstruction} but not originating from VLQ pair production is allowed. Specifically, for events that contain five reconstructed jets, the \chimodndf is evaluated for all possible assignments of four jets with one jet discarded, and the event is then treated as a four-jet event using the four jets that resulted in the best \chimodndf. For events that contain four reconstructed jets, the \chimodndf is calculated for these four jets, and then also for each possible jet assignment for a three-jet event. If the four-jet \chimodndf is less than each of the three-jet \chimodndf values, then it is treated as a four-jet event. Otherwise, it is treated as a three-jet event using those three jets.

\begin{figure*}[hbtp]
    \centering
    \includegraphics[width=0.42\textwidth]{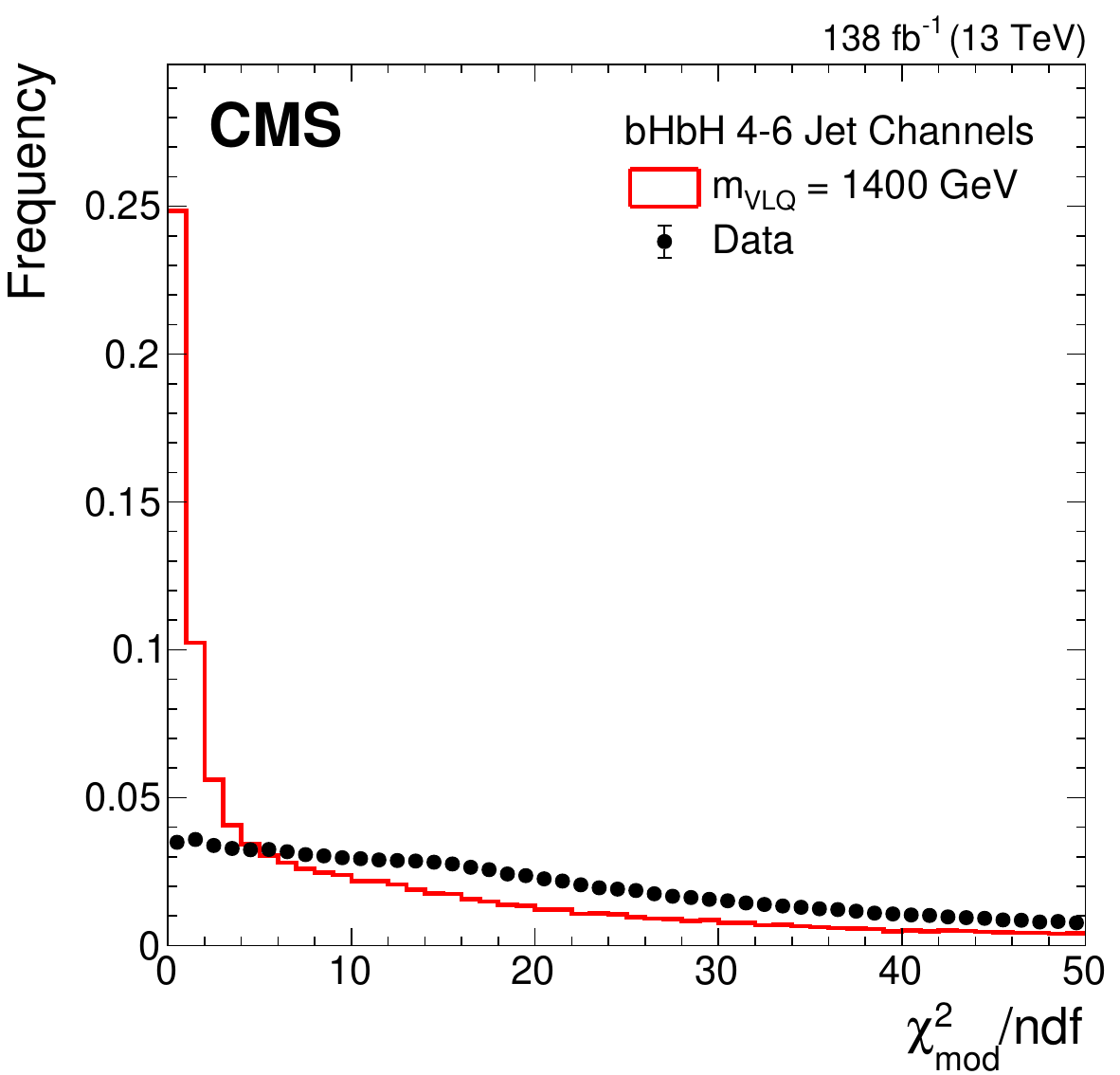}
    \includegraphics[width=0.42\textwidth]{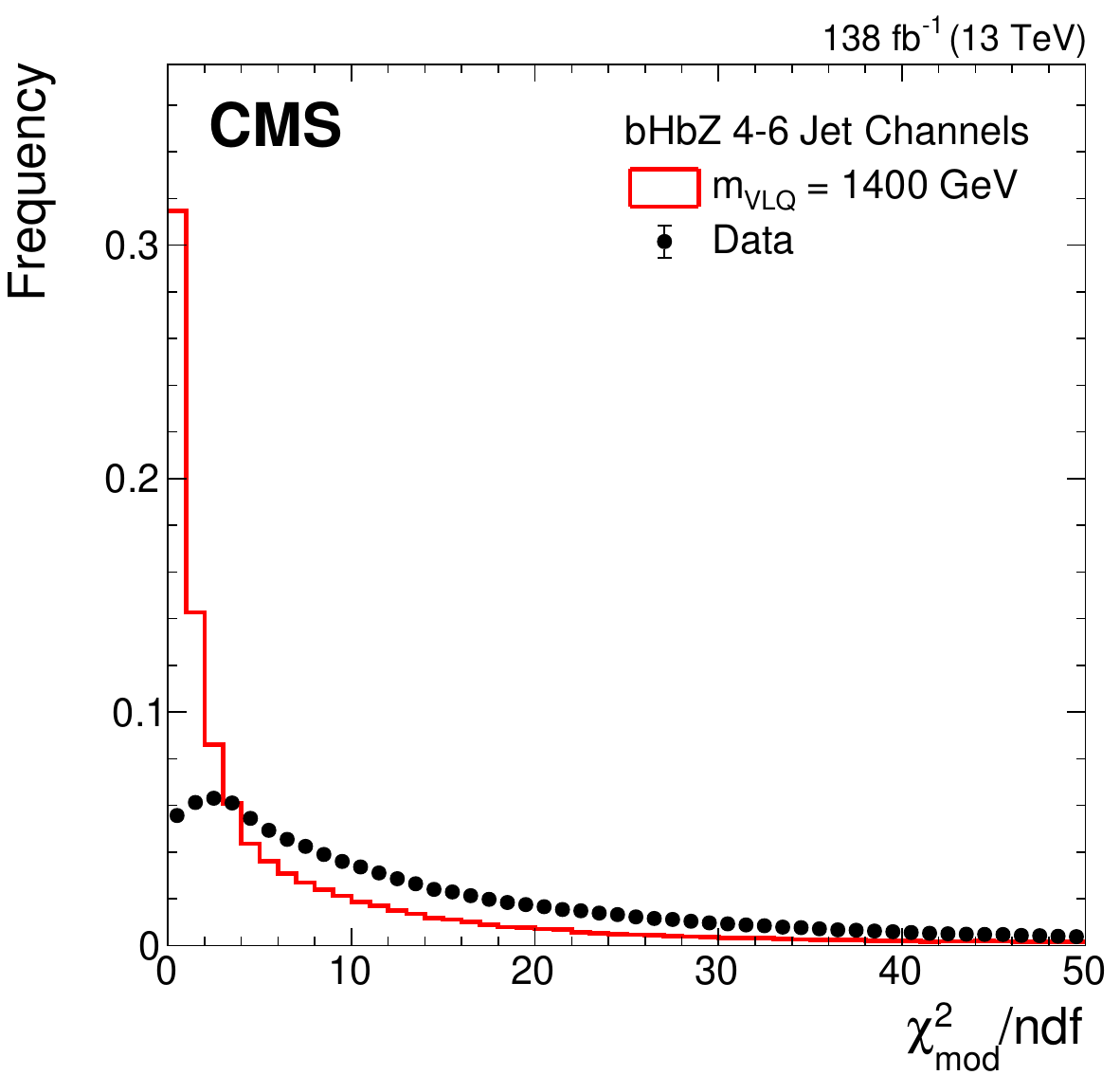}
    \includegraphics[width=0.42\textwidth]{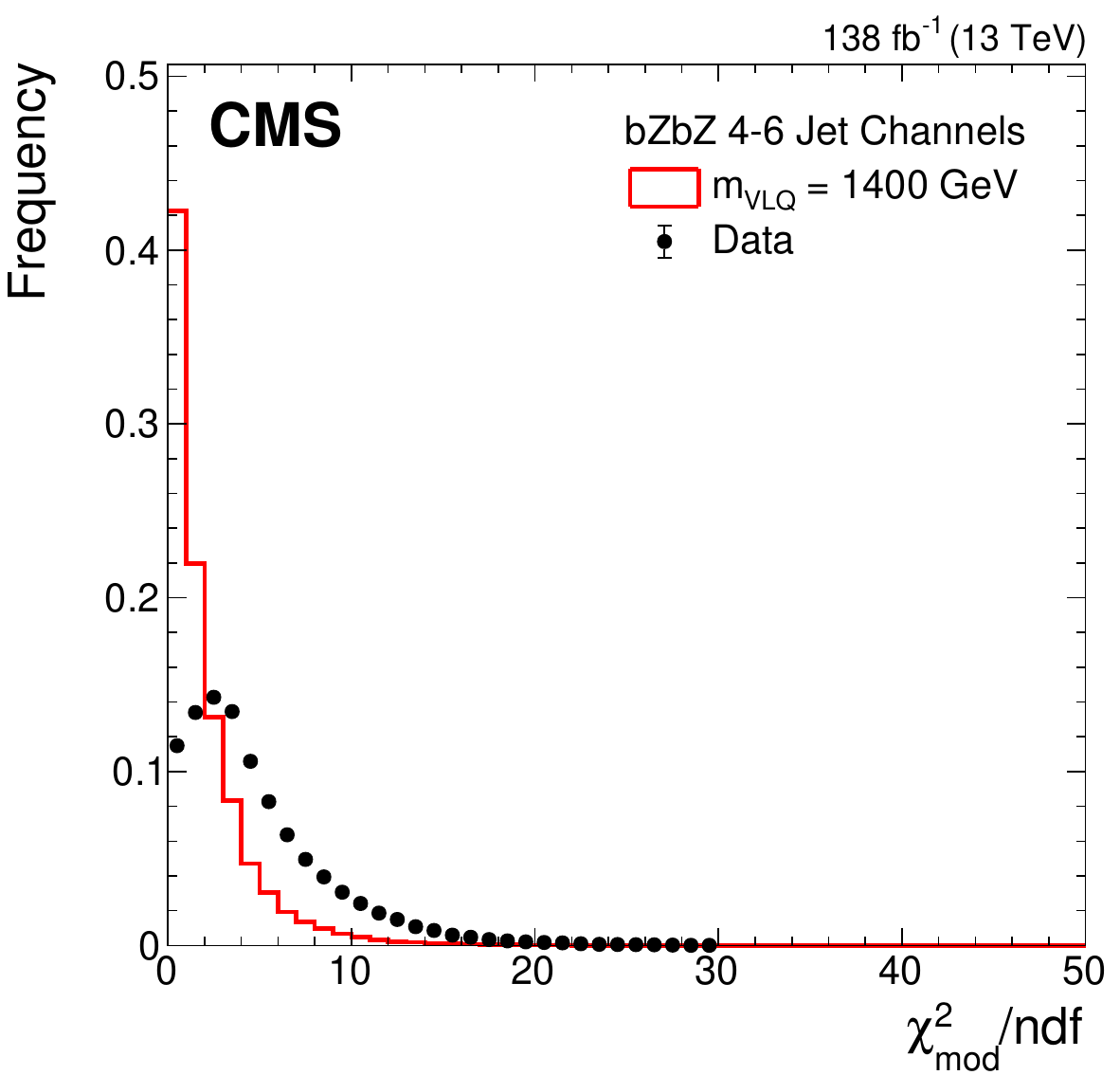}
    \includegraphics[width=0.42\textwidth]{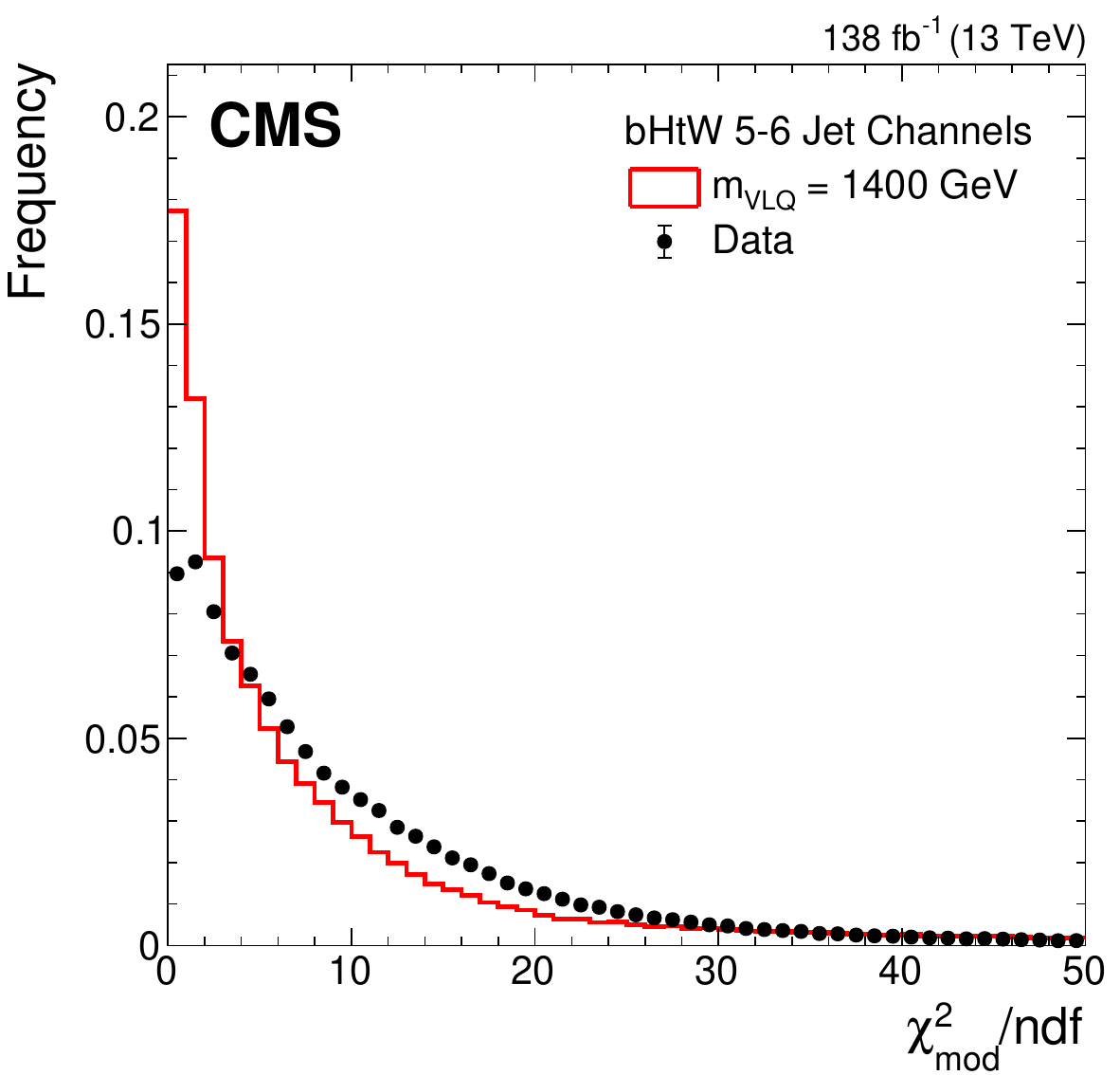}
    \includegraphics[width=0.42\textwidth]{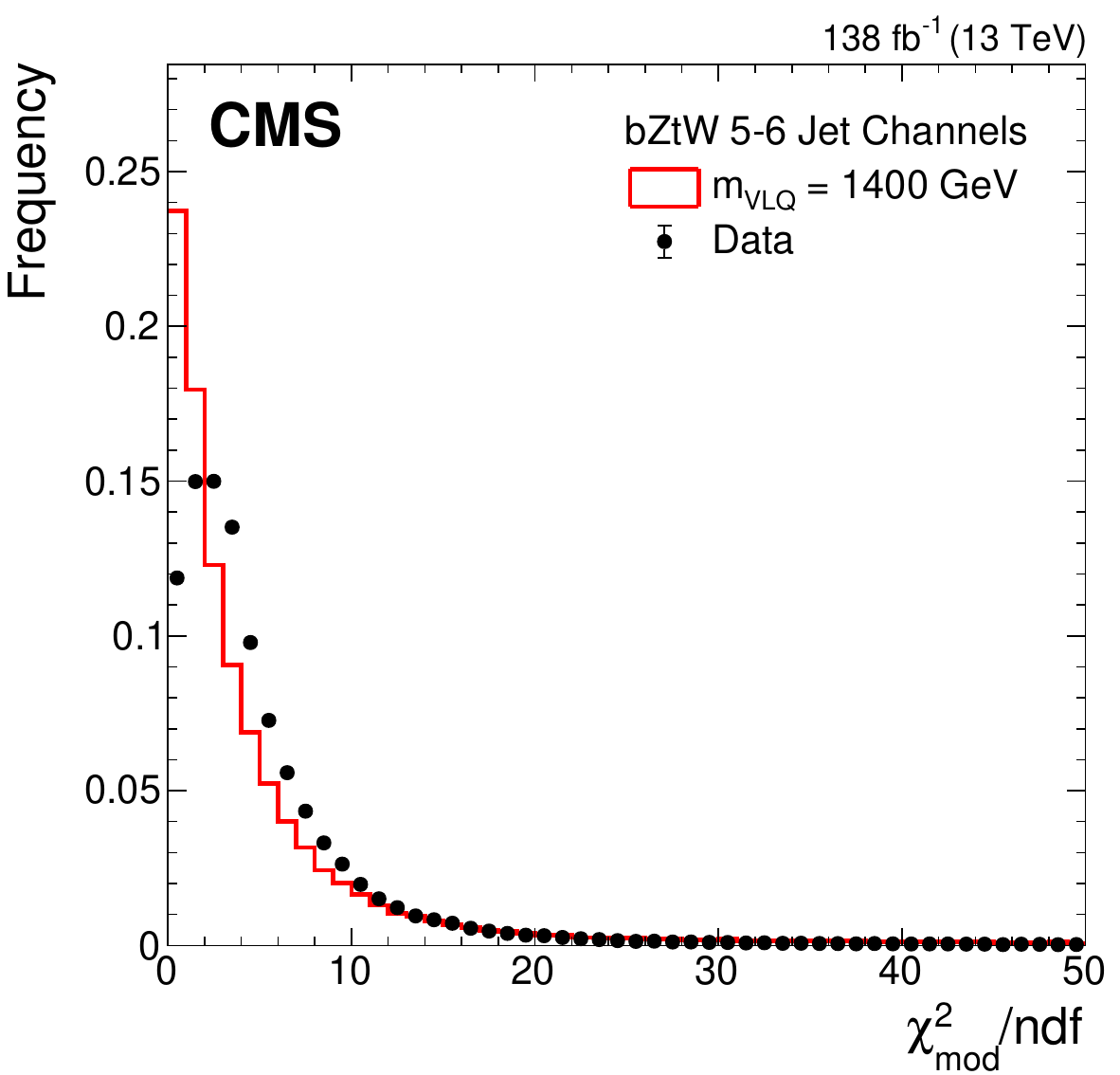}
    \caption{Normalized distributions of the value of the least \chimodndf for simulated signal events and data events in the fully hadronic category before any \PQb tagging requirements are applied in the fully hadronic category. Upper row: \bHbH (left), \bHbZ (center), and \bZbZ (right) decay modes. Lower row: \bHtW (left) and \bZtW (right) decay modes. A signal mass of $m_{\PB} = 1400\GeV$ is used and compared against all three years of data. All jet multiplicities have been combined together.}
    \label{fig:chimod_dist_h}
\end{figure*}

\begin{figure*}[hbtp]
    \centering
    \includegraphics[width=0.42\textwidth]{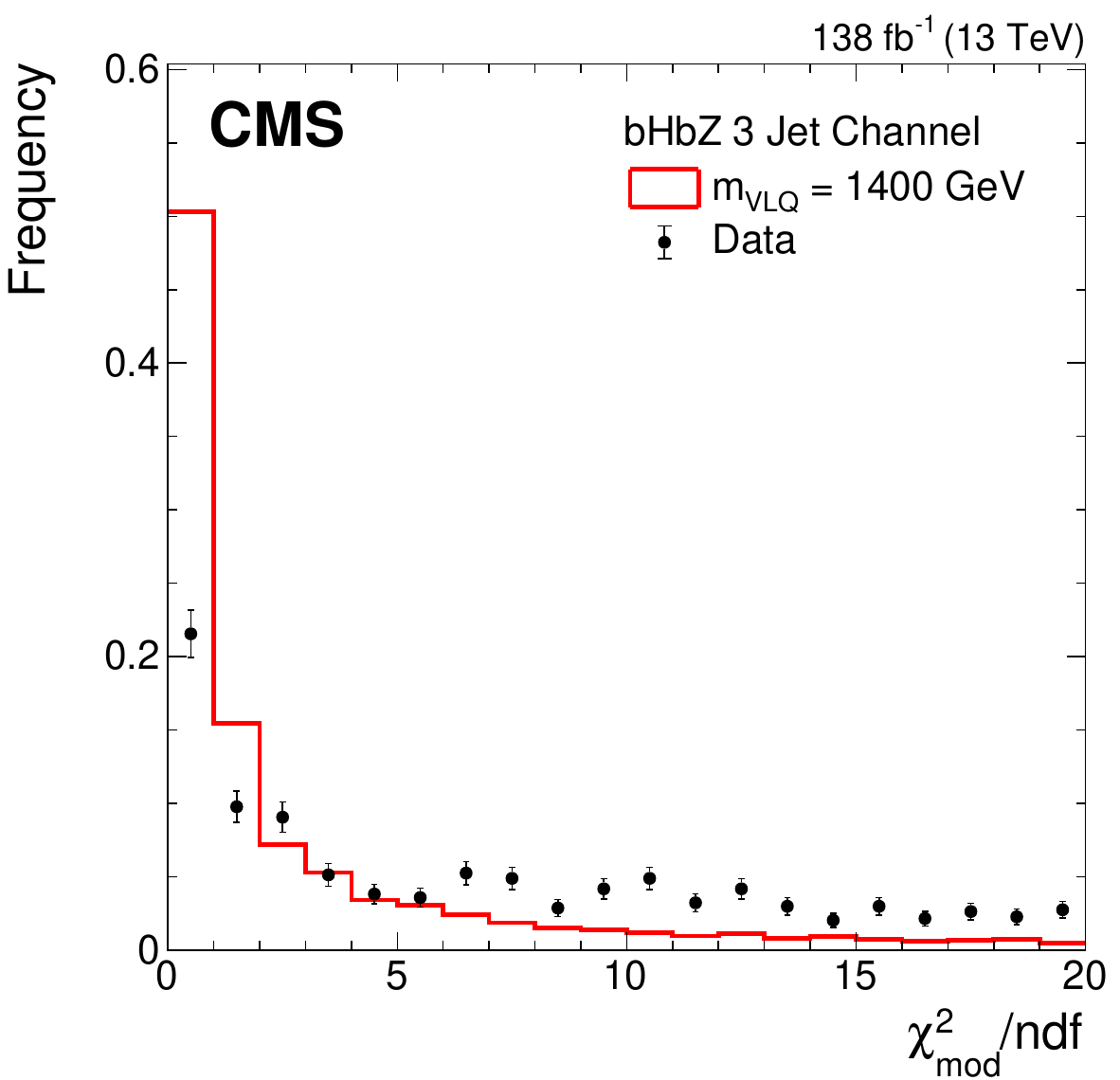}
    \includegraphics[width=0.42\textwidth]{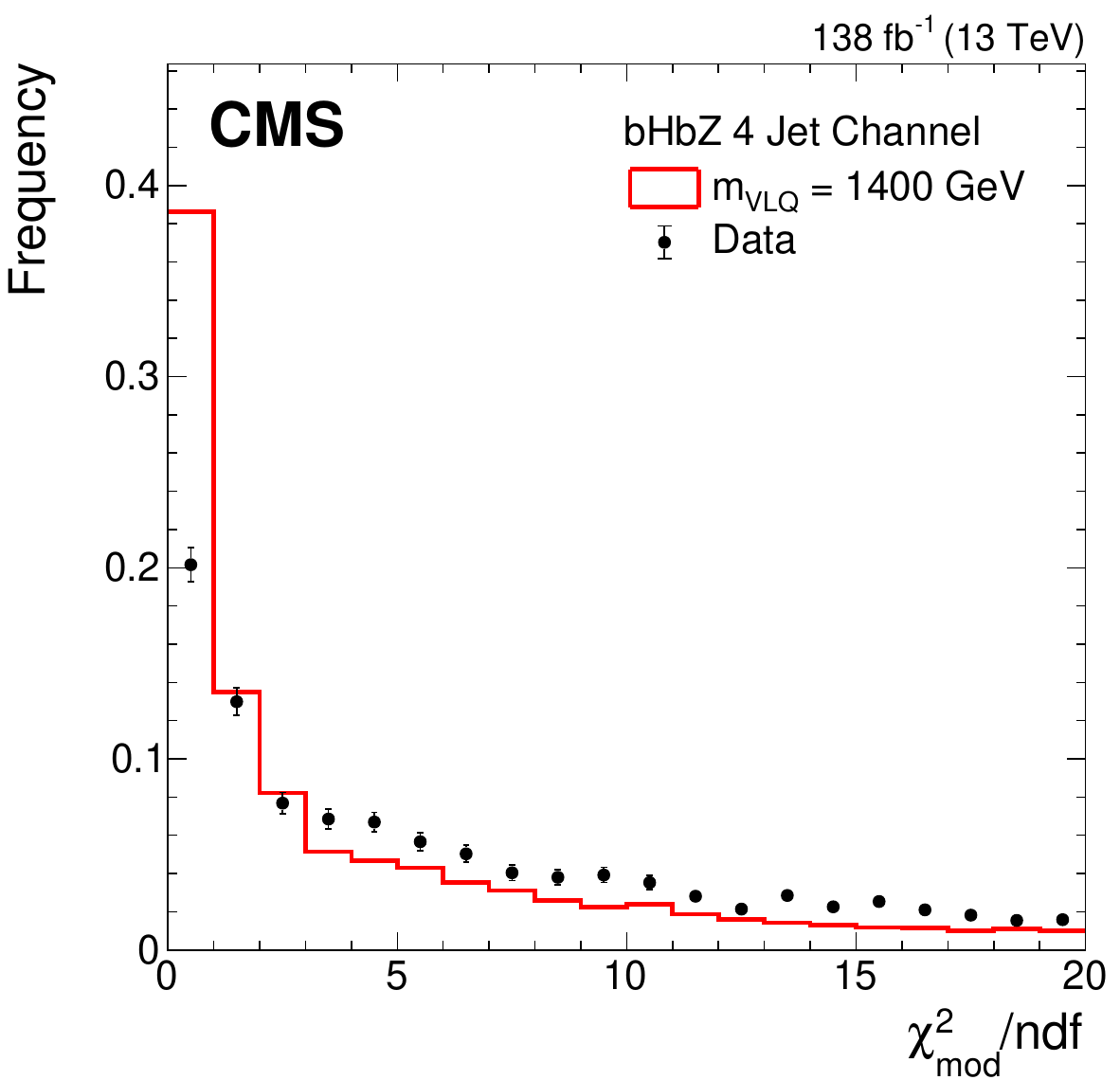}
    \includegraphics[width=0.42\textwidth]{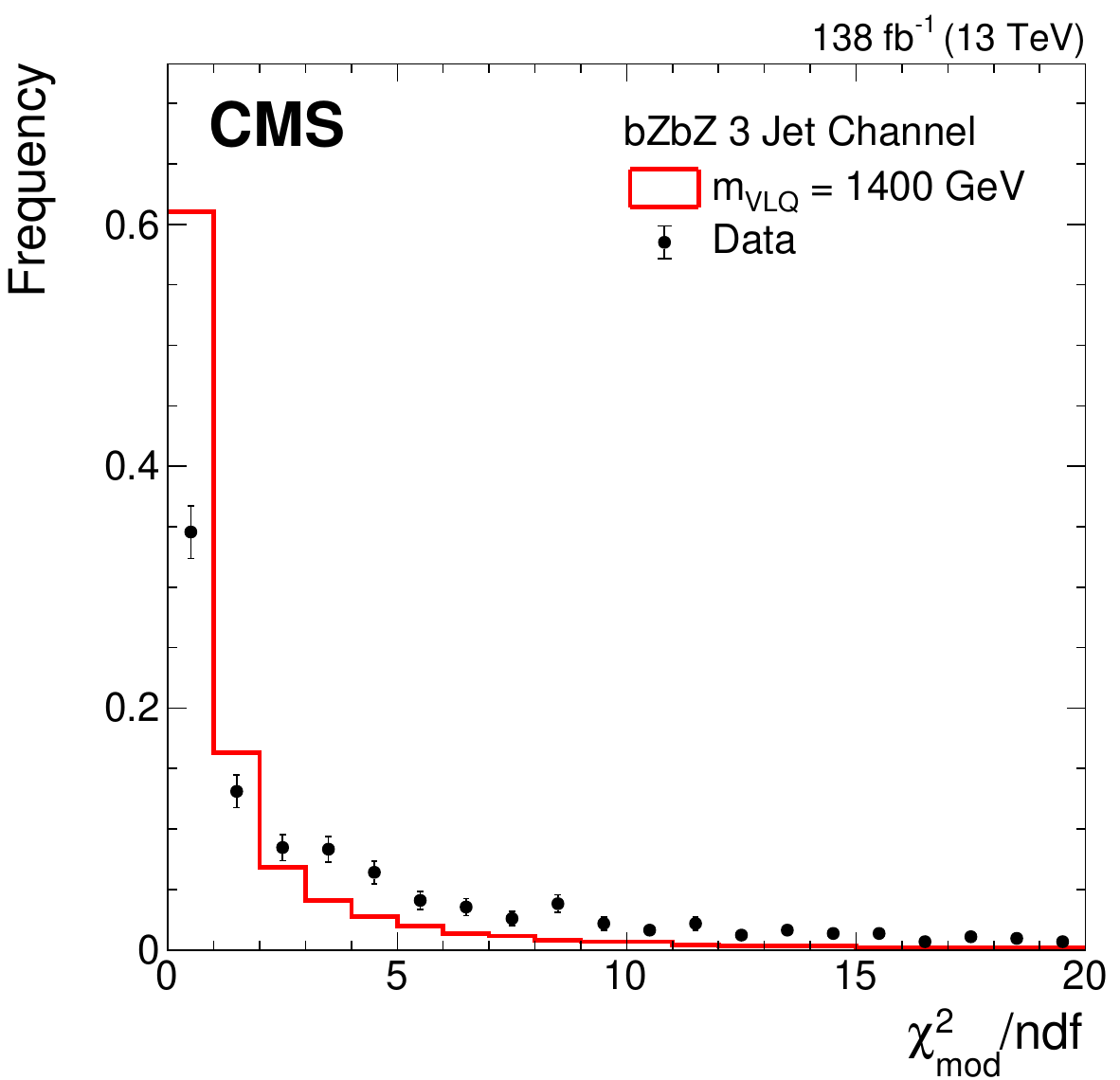}
    \includegraphics[width=0.42\textwidth]{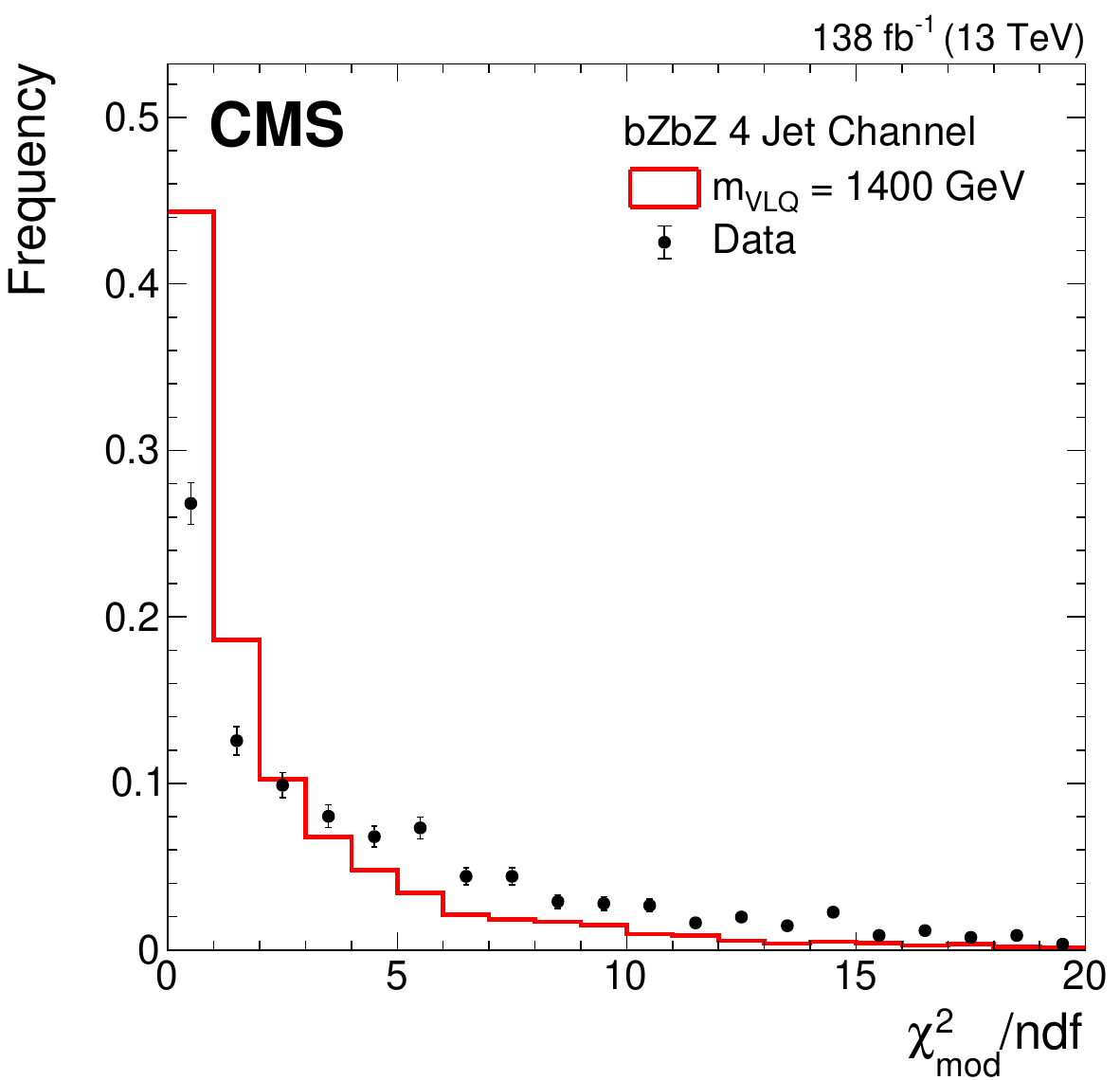}
    \caption{Normalized distributions of the value of the least \chimodndf for simulated signal events and data events in the leptonic category before any \PQb tagging requirements are applied. Upper row: \bHbZ decay mode, 3-jet (left) and 4-jet (right) events. Lower row: \bZbZ decay mode, 3-jet (left) and 4-jet (right) events. A signal mass of $m_{\PB} = 1400\GeV$ is used and compared against all three years of data.}
    \label{fig:chimod_dist_l}
\end{figure*}

Figures~\ref{fig:chimod_dist_h} and~\ref{fig:chimod_dist_l} show the resulting distributions in the fully hadronic and leptonic categories, respectively, of the value of the least \chimodndf for simulated signal events compared to the distributions for data events before any \PQb tagging requirements are applied. In these plots, the two distributions are normalized to the same area.

Since the \chimodndf values tend to be lower for signal events than for background events, an upper bound is set on the least \chimodndf in order to provide background rejection. This limit is optimized for signal sensitivity separately for each category and channel, using a \PB mass of 1400\GeV. For the fully hadronic category, equal branching fractions for \BbH, \BbZ, and \BtW are assumed, and for the leptonic category, branching fractions of $\BrBbH = 30\%$ and $\BrBbZ = 70\%$ are used. Table~\ref{table:chi2_optimization} shows the optimized  \chimodndf limits obtained for each category. These values do not depend significantly on the \PB mass assumed in the optimization since the \chimodndf measure in Eqs.~(\ref{eq:chi2_equation_hadronic}) and~(\ref{eq:chi2_equation_leptonic}) depends on the mass difference between the two reconstructed VLQ candidates, and not the VLQ mass itself. As a cross check, the optimization was performed with other \PB masses, resulting in only small changes in the optimal \chimodndf limit values. These optimized \chimodndf limits are then used in conjunction with \PQb tagging, to provide signal discrimination in the reconstructed \mvlq spectrum, which is used to obtain the background estimate described in the next section.

\begin{table}[hbtp]
    \centering
    \topcaption{Optimized upper bound values of the \chimodndf selection as a function of jet multiplicity and decay mode.}
    \label{table:chi2_optimization}
    \begin{scotch}{lcccc}
        \multirow{2}{*}{Decay mode} & \multicolumn{4}{c}{Jet multiplicity} \\
         & 3 & 4 & 5 & 6 \\ \hline
         Fully hadronic \bHbH & \NA & 1.5 & 2.75 & 1.0 \\
         Fully hadronic \bHbZ & \NA & 2.0 & 1.25 & 1.25 \\
         Fully hadronic \bZbZ & \NA & 0.75 & 1.25 & 1.75 \\
         Fully hadronic \bHtW & \NA & \NA & 2.5 & 5.0 \\
         Fully hadronic \bZtW & \NA & \NA & 1.5 & 6.0 \\ [\cmsTabSkip]
         Leptonic \bHbZ & 2.9 & 2.5 & \NA & \NA \\
         Leptonic \bZbZ & 2.0 & 2.6 & \NA & \NA \\ 
    \end{scotch}
\end{table}

\section{Background estimation}
\label{sec:background}

As the QCD multijet background is difficult to precisely model in simulation, the background estimation is based exclusively on control samples in data. Since the composition and behavior of the background are different between the fully hadronic and dileptonic categories, separate methods are used for each category.

\subsection{Background estimation in the fully hadronic category}
\label{sec:background_h}

\begin{figure*}[hbtp]
    \centering
    \includegraphics[width=0.44\textwidth]{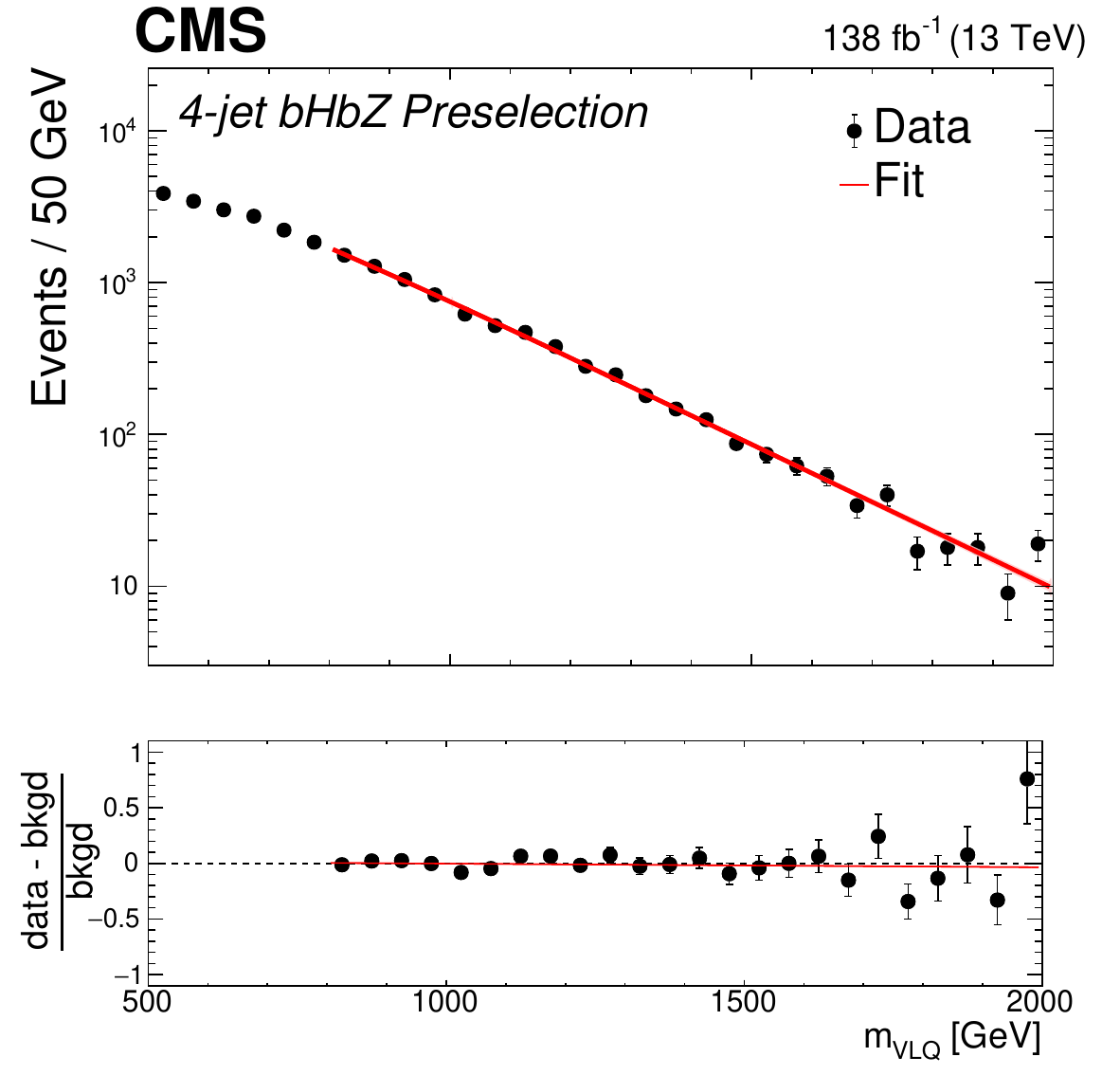}
    \includegraphics[width=0.44\textwidth]{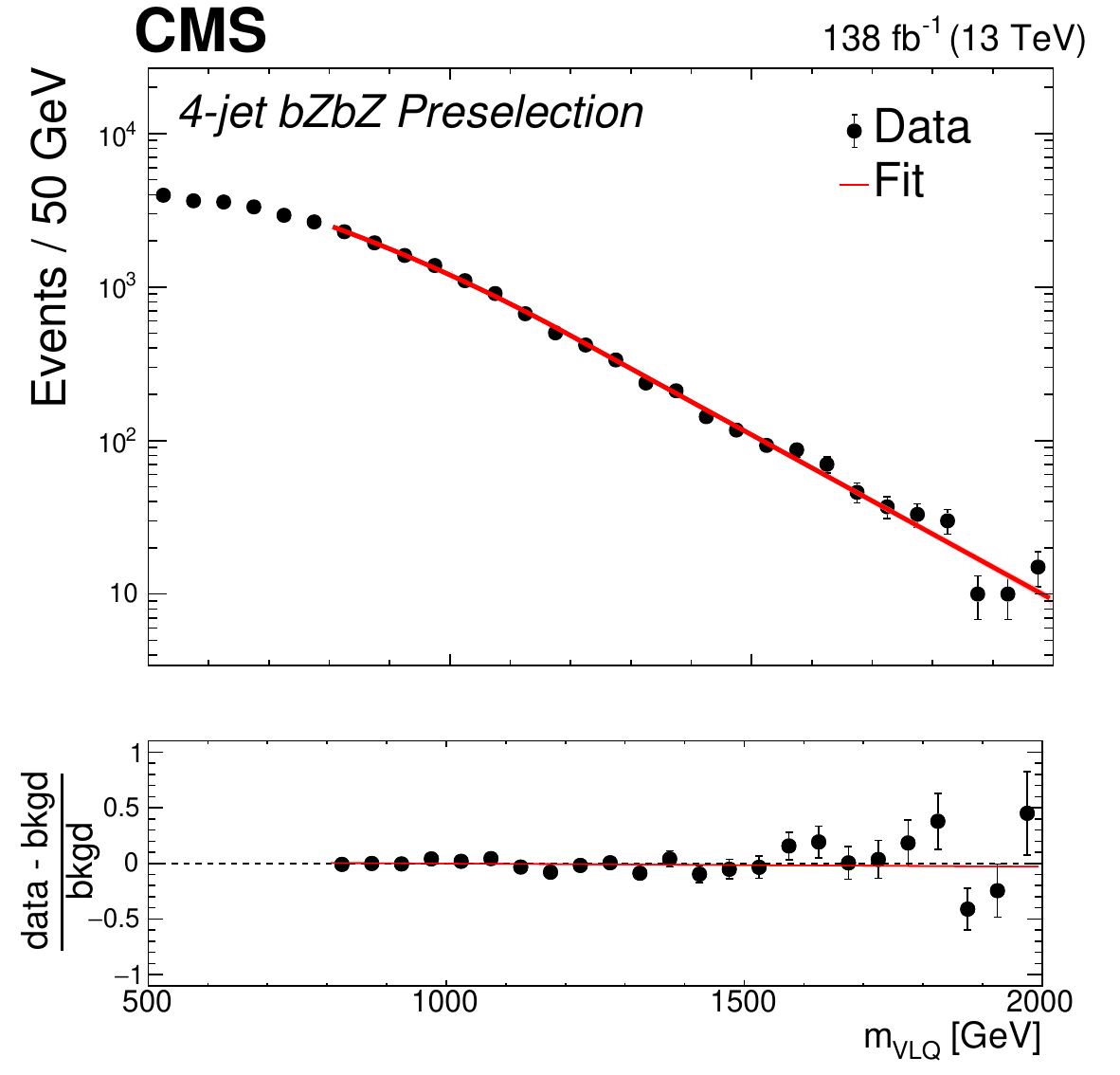}
    \includegraphics[width=0.44\textwidth]{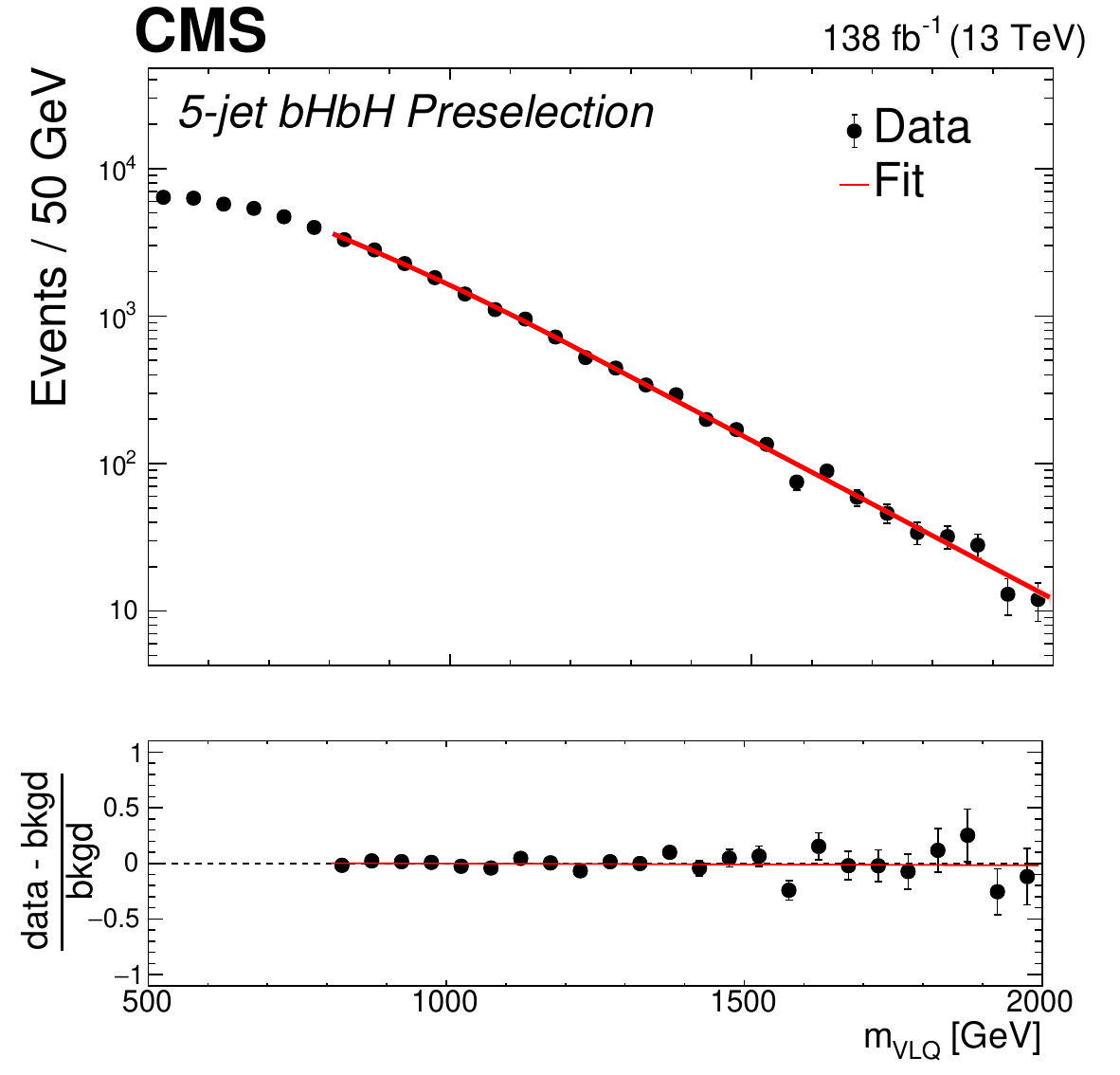}
    \includegraphics[width=0.44\textwidth]{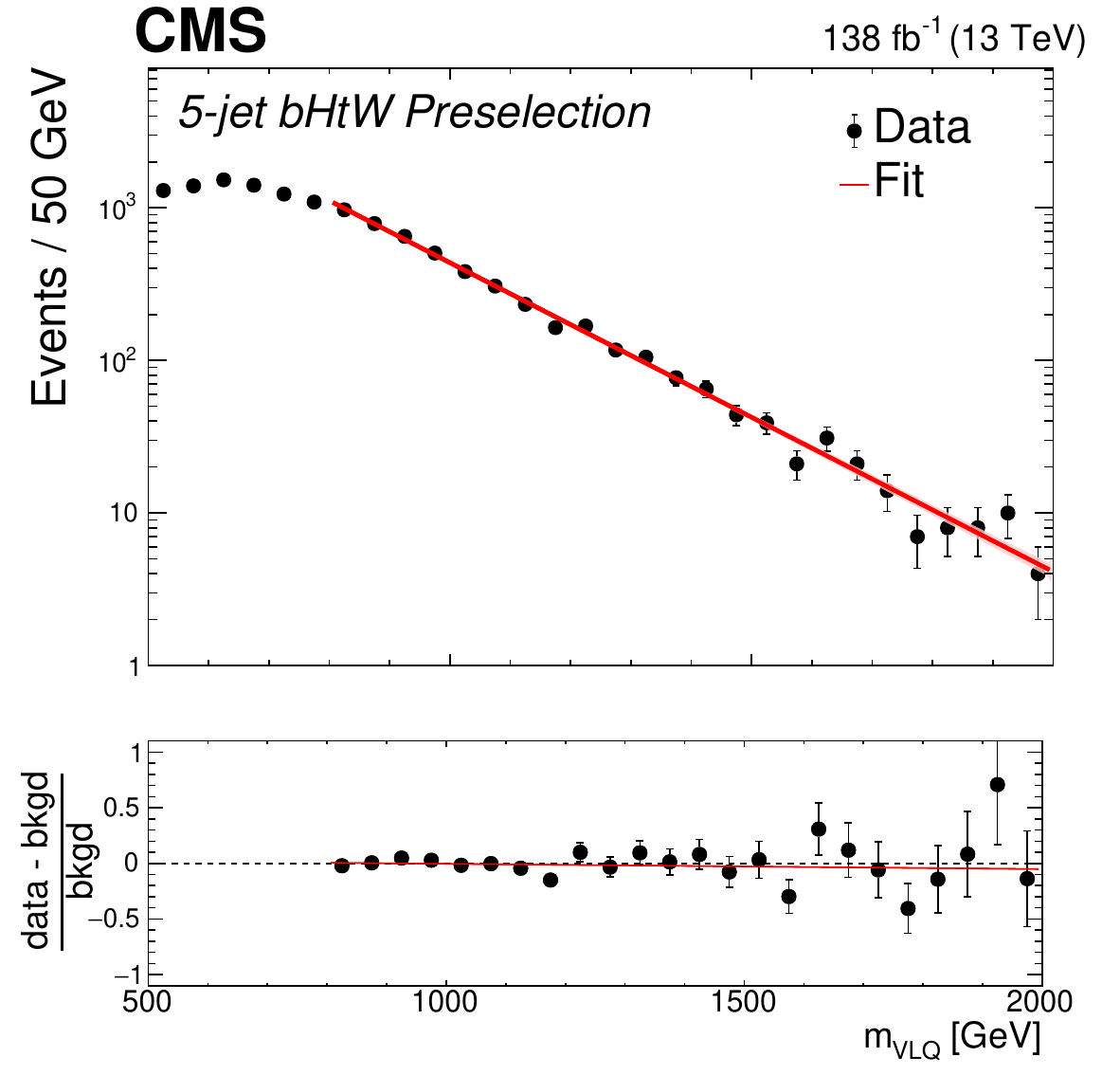}
    \includegraphics[width=0.44\textwidth]{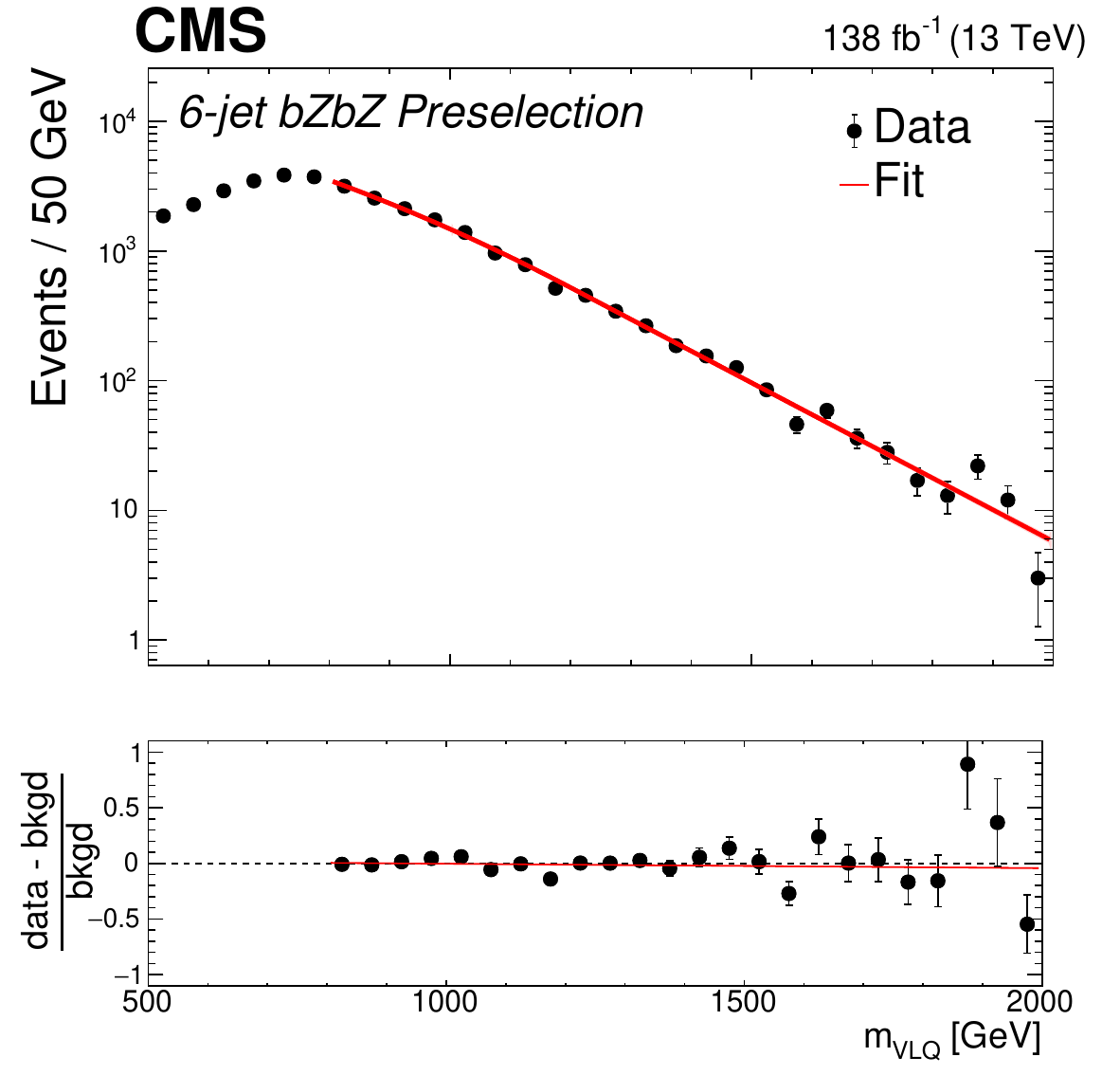}
    \includegraphics[width=0.44\textwidth]{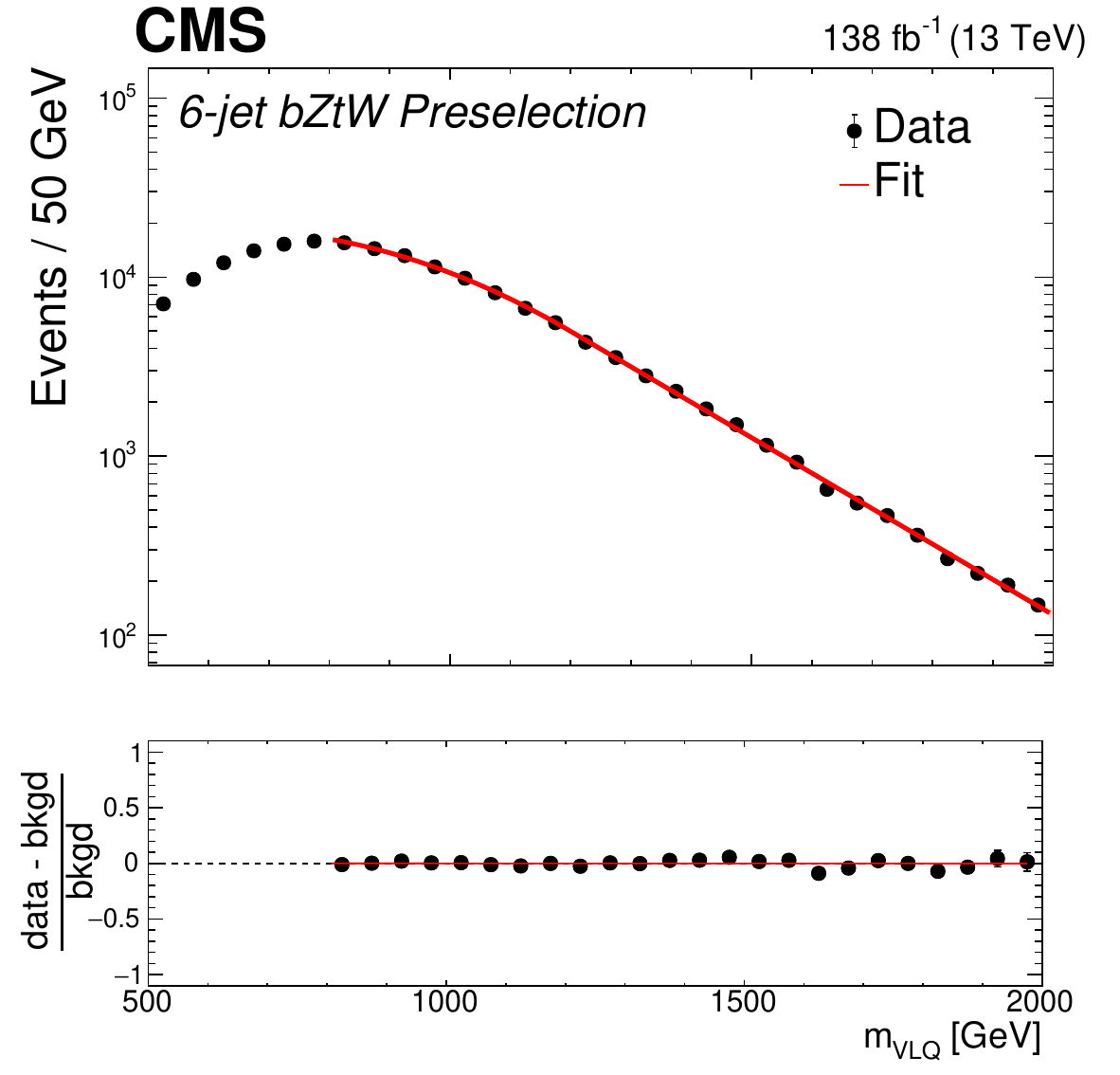}
    \caption{Distributions of \mvlq for the preselected data sample in the fully hadronic category for some selected channels. Upper row: 4-jet \bHbZ (left) and 4-jet \bZbZ (right) modes. Middle row: 5-jet \bHbH (left) and 5-jet \bHtW (right) modes. Lower row: 6-jet \bZbZ (left) and 6-jet \bZtW (right) modes. The fit to the data (shown by the black points) is given by the red line, and the bottom panel displays the fractional difference between the data and fit, (data-fit)/fit.}
    \label{fig:mvlq_distributions_h}
\end{figure*}

The background estimation for the fully hadronic category generally follows the procedure described in Ref.~\cite{CMS:2020ttz}. Using the preselection sample defined in Section~\ref{sec:eventcat}, the expected distribution of the number of events as a function of the VLQ candidate mass is obtained and then fit with the function
\begin{linenomath}
\begin{equation}
n(x) = p_0 \exp{[-( p_1 x+\theta(-x) p_2 x^2)]}
\end{equation}
\end{linenomath}
in the range $\mvlq > 800\GeV$, where $x \equiv m/\GeVns - 1200$ and $\theta(x)$ is the Heaviside step function. The second-order term is applied only for mass values less than 1200\GeV to account for the shape change of the distribution in the low-mass region compared to a traditional exponential function caused by the trigger turn on. This is also the reason the fit only includes $\mvlq > 800\GeV$. Smoothness at the $\mvlq = 1200\GeV$ interface is enforced by employing the same parameter $p_1$ for both $\mvlq < 1200\GeV$ and $\mvlq > 1200\GeV$. Figure~\ref{fig:mvlq_distributions_h} shows the resulting fit results for six representative fully hadronic channels.

The second step is to estimate the reduction in background that results from applying the jet \PQb tagging requirements. This factor is called the background jet-tagged fraction (BJTF). It is obtained by using events in a sideband region defined by the mass range $500 < \mvlq < 800\GeV$. This mass range is chosen to ensure that the sideband region is free of any potential signal, as this range is below the current lower exclusion limits on the VLQ mass~\cite{Sirunyan:2019sza,Aaboud:2018wxv,CMS:2020ttz}. Table~\ref{table:bjtf_h} summarizes the values of the BJTF derived from this sideband region, denoted $\varepsilon_0$ in Eq.~(\ref{eq:background_h}), for the different channels considered.

\begin{table*}[hbtp]
\centering
\topcaption{Values of the BJTF for data events in the control region with $500 < \mvlq < 800\GeV$ for each of the fully hadronic channels considered. The uncertainties shown are statistical.}
\label{table:bjtf_h}
\begin{scotch}{cccccc}
& \bHbH & \bHbZ & \bZbZ & \bHtW & \bZtW \\
\hline
4 jets & $0.0010 \pm 0.0007$ & $0.0032 \pm 0.0005$ & $0.0075 \pm 0.0016$ & \NA & \NA \\
5 jets & $0.0009 \pm 0.0002$ & $0.0040 \pm 0.0004$ & $0.0492 \pm 0.0020$ & $0.0243 \pm 0.0019$ & $0.0223 \pm 0.0019$\\
6 jets & $0.0020 \pm 0.0003$ & $0.0018 \pm 0.0003$ & $0.0368 \pm 0.0018$ & $0.0089 \pm 0.0003$ & $0.0311 \pm 0.0005$\\
\end{scotch}
\end{table*}

\begin{figure*}[hbtp]
    \centering
    \includegraphics[width=0.32\textwidth]{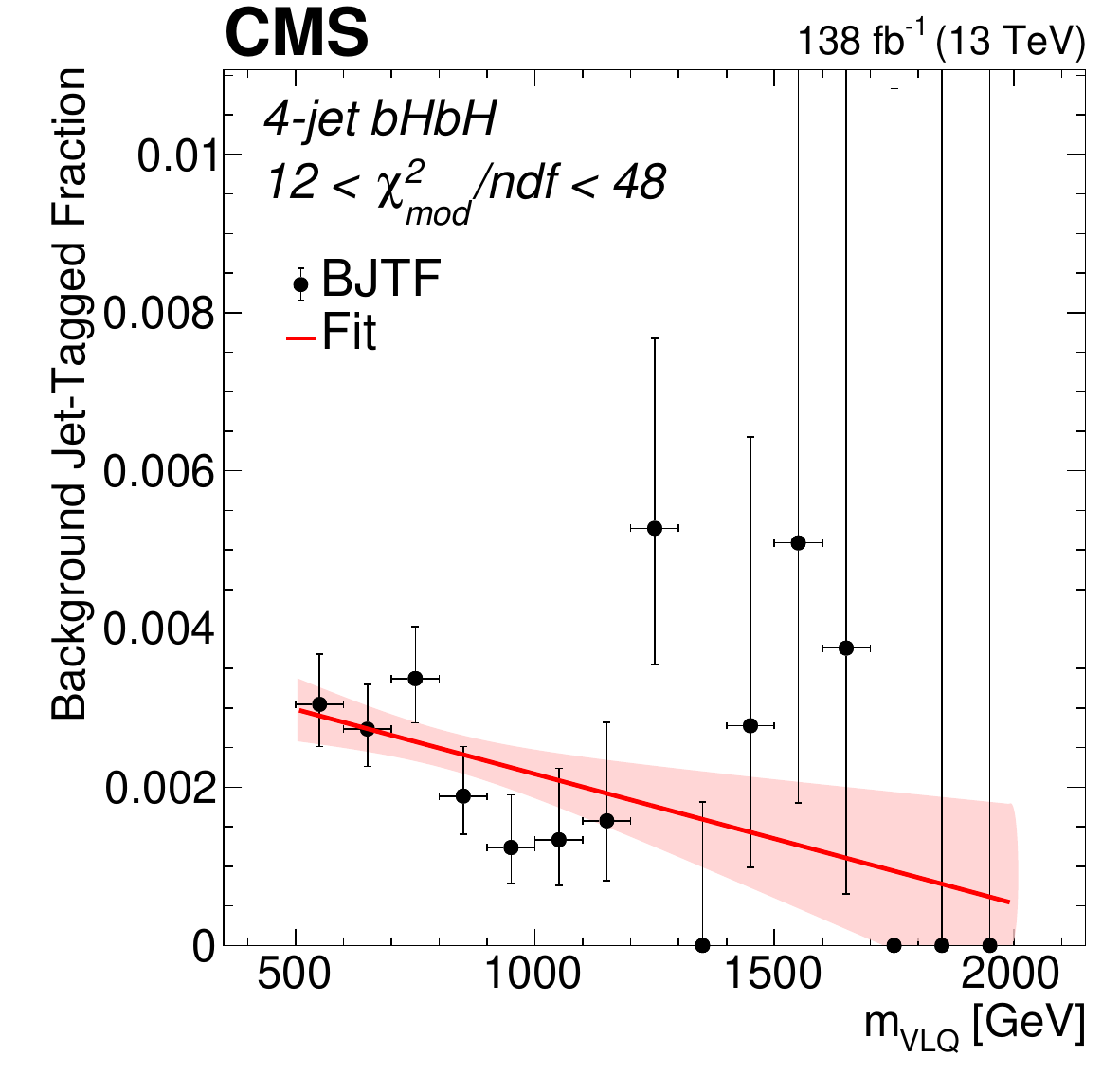}  
    \includegraphics[width=0.32\textwidth]{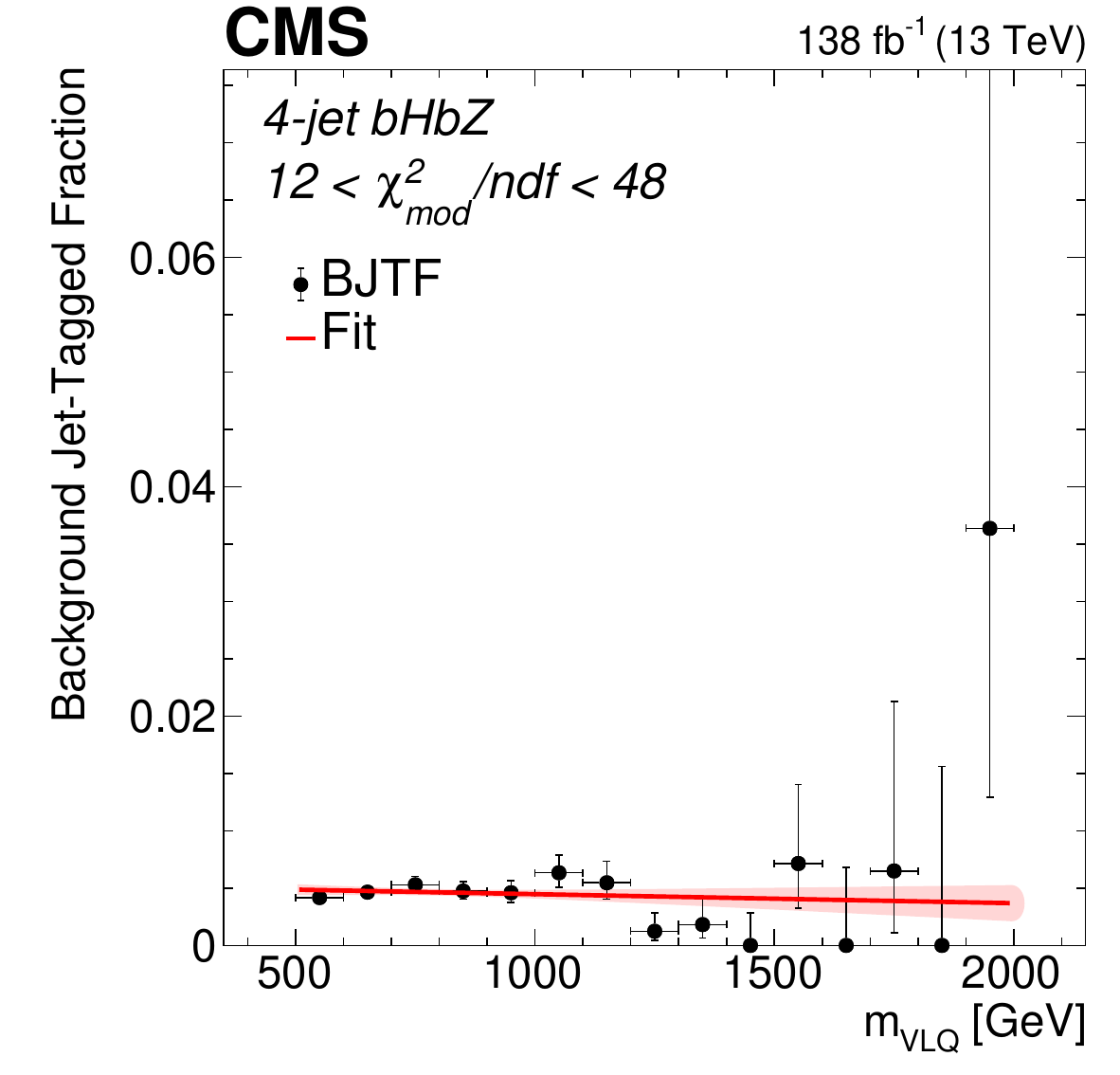}  
    \includegraphics[width=0.32\textwidth]{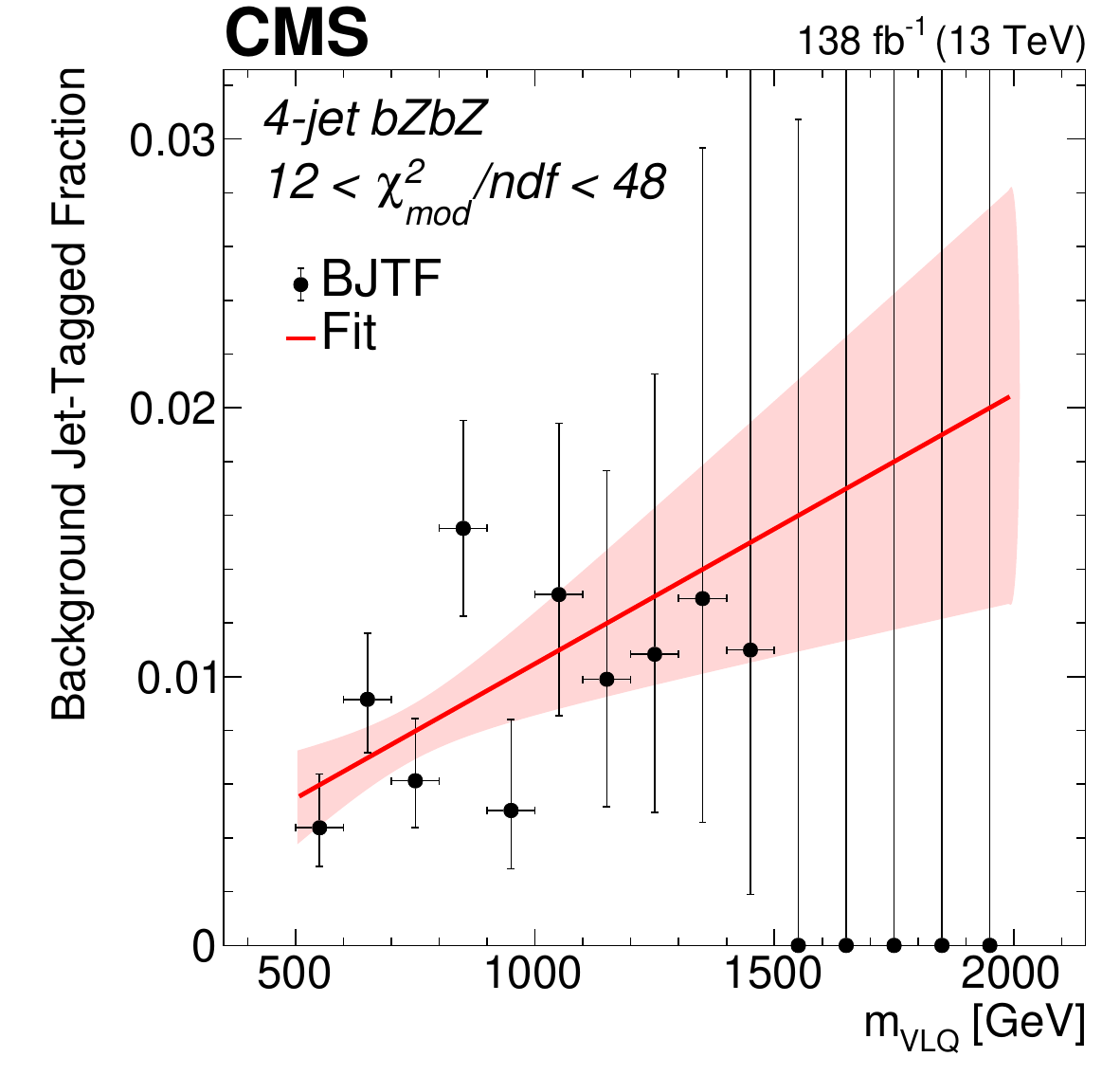}  
    \includegraphics[width=0.32\textwidth]{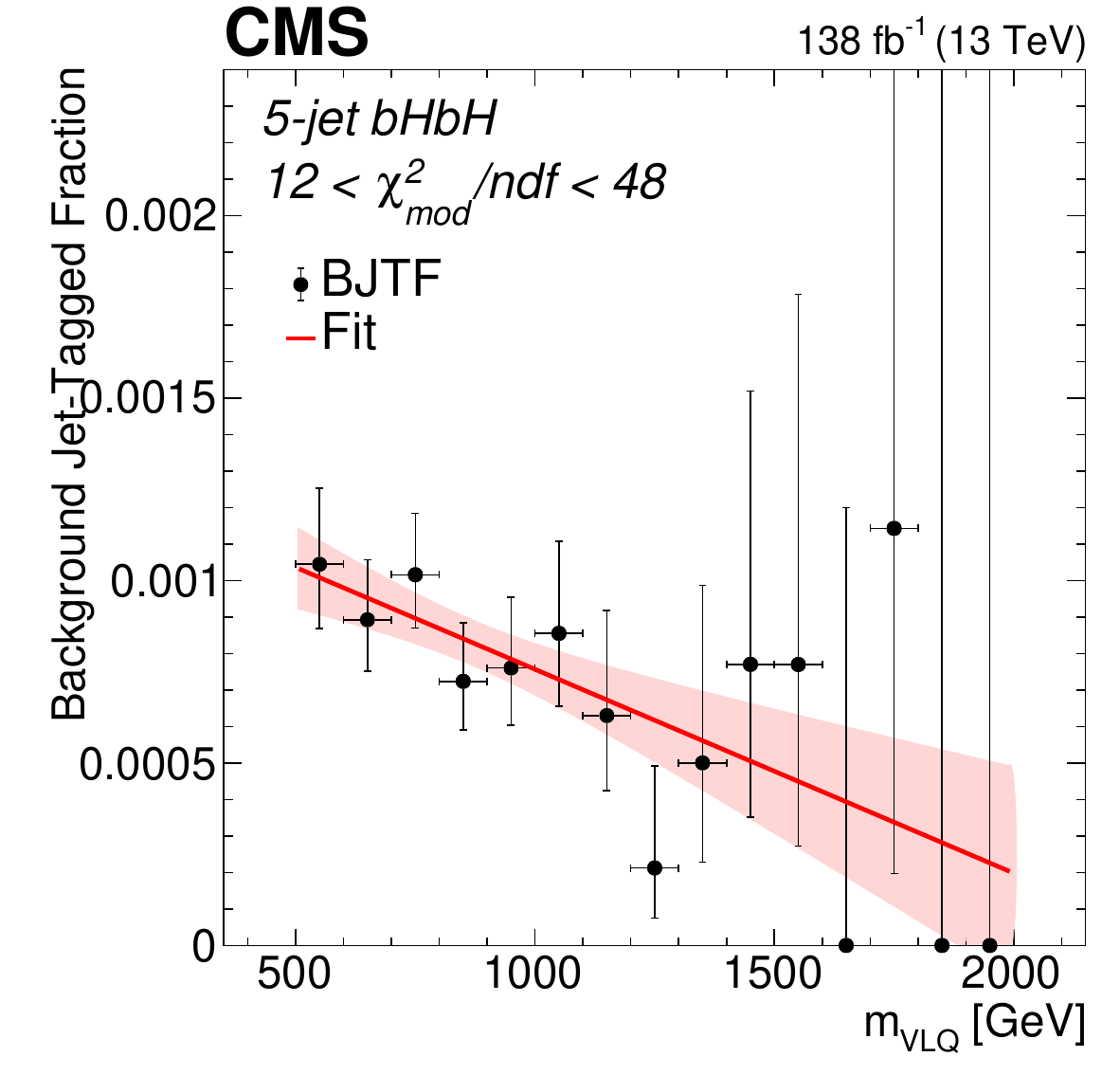}  
    \includegraphics[width=0.32\textwidth]{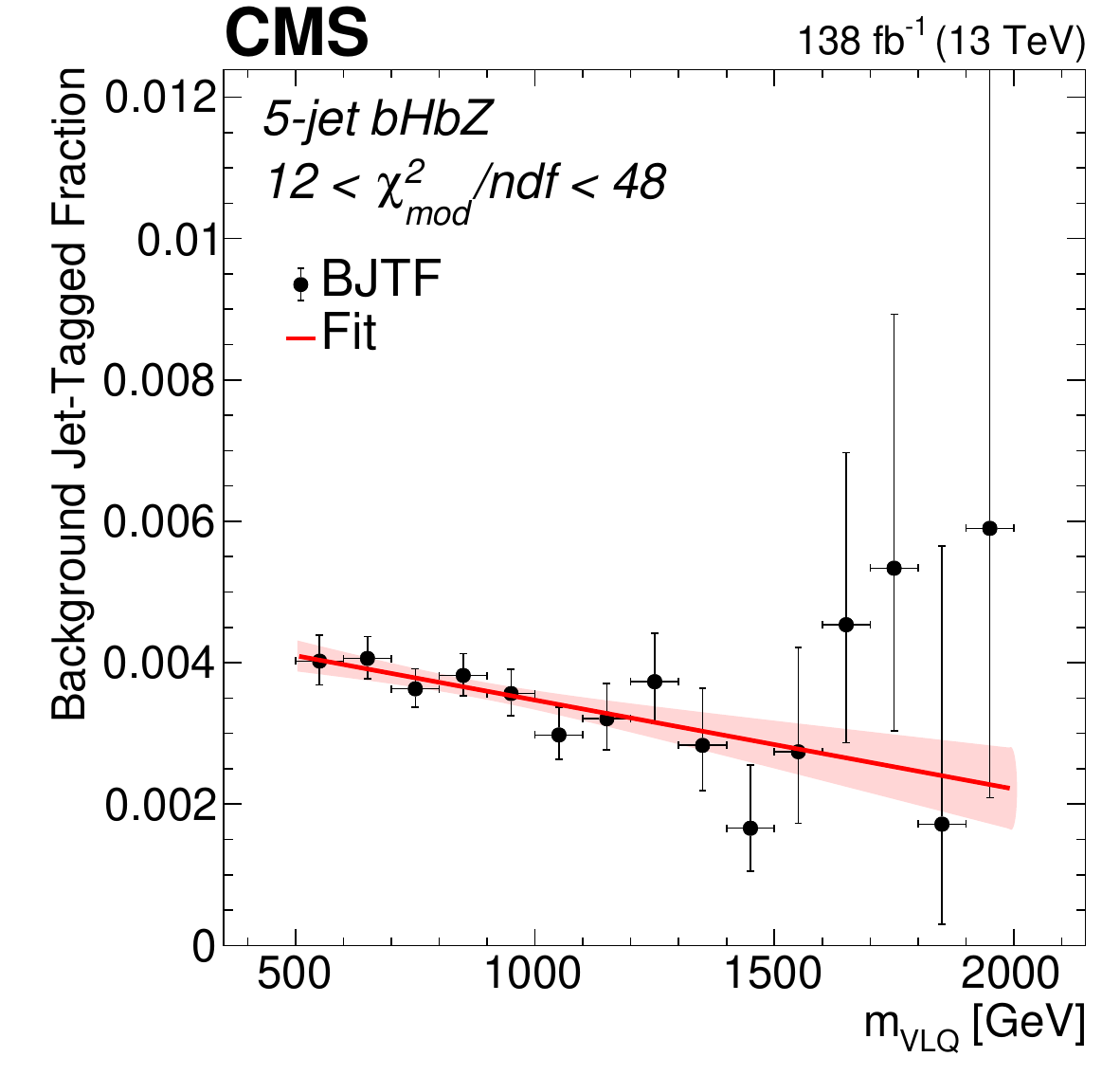}  
    \includegraphics[width=0.32\textwidth]{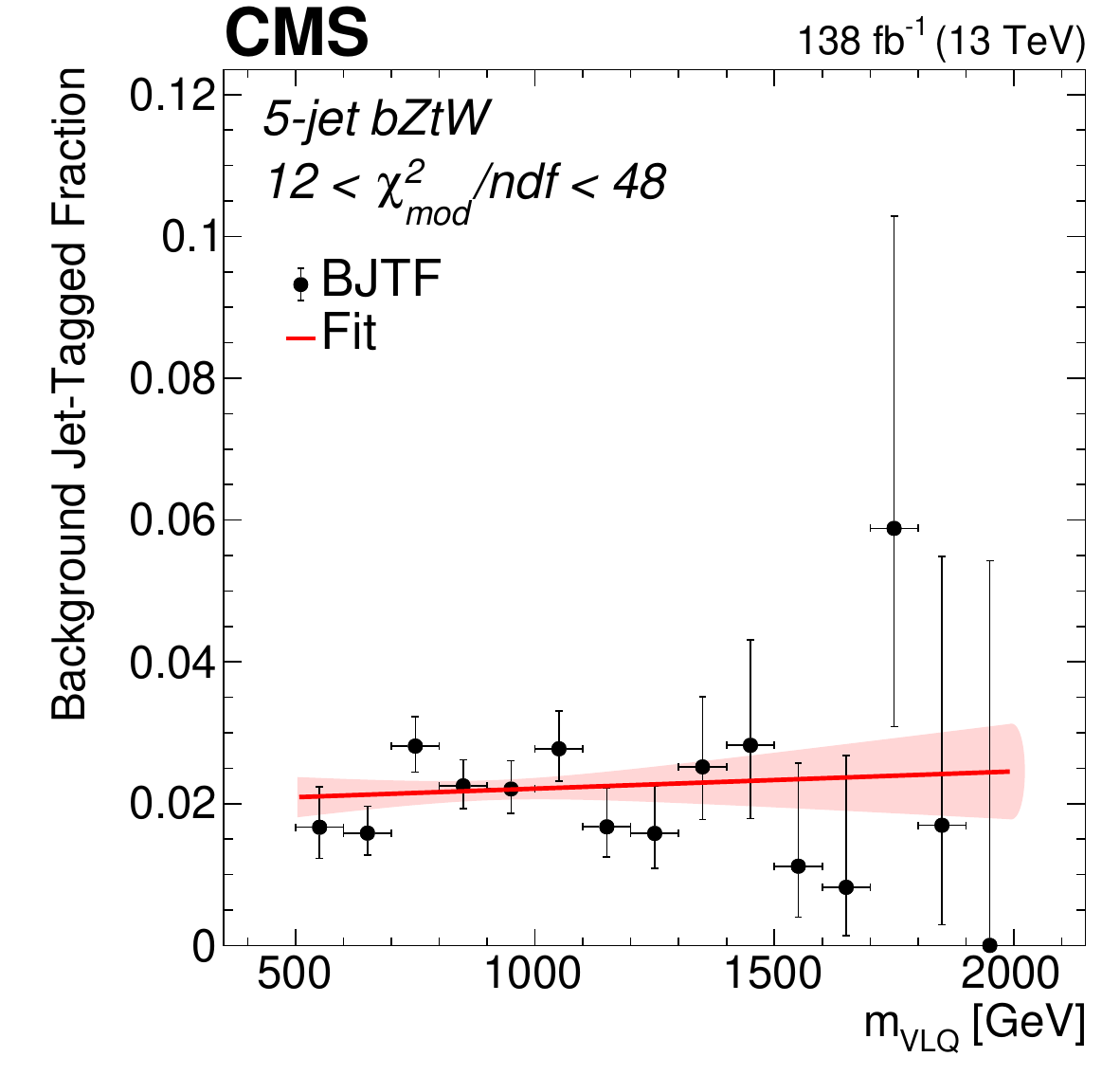}  
    \includegraphics[width=0.32\textwidth]{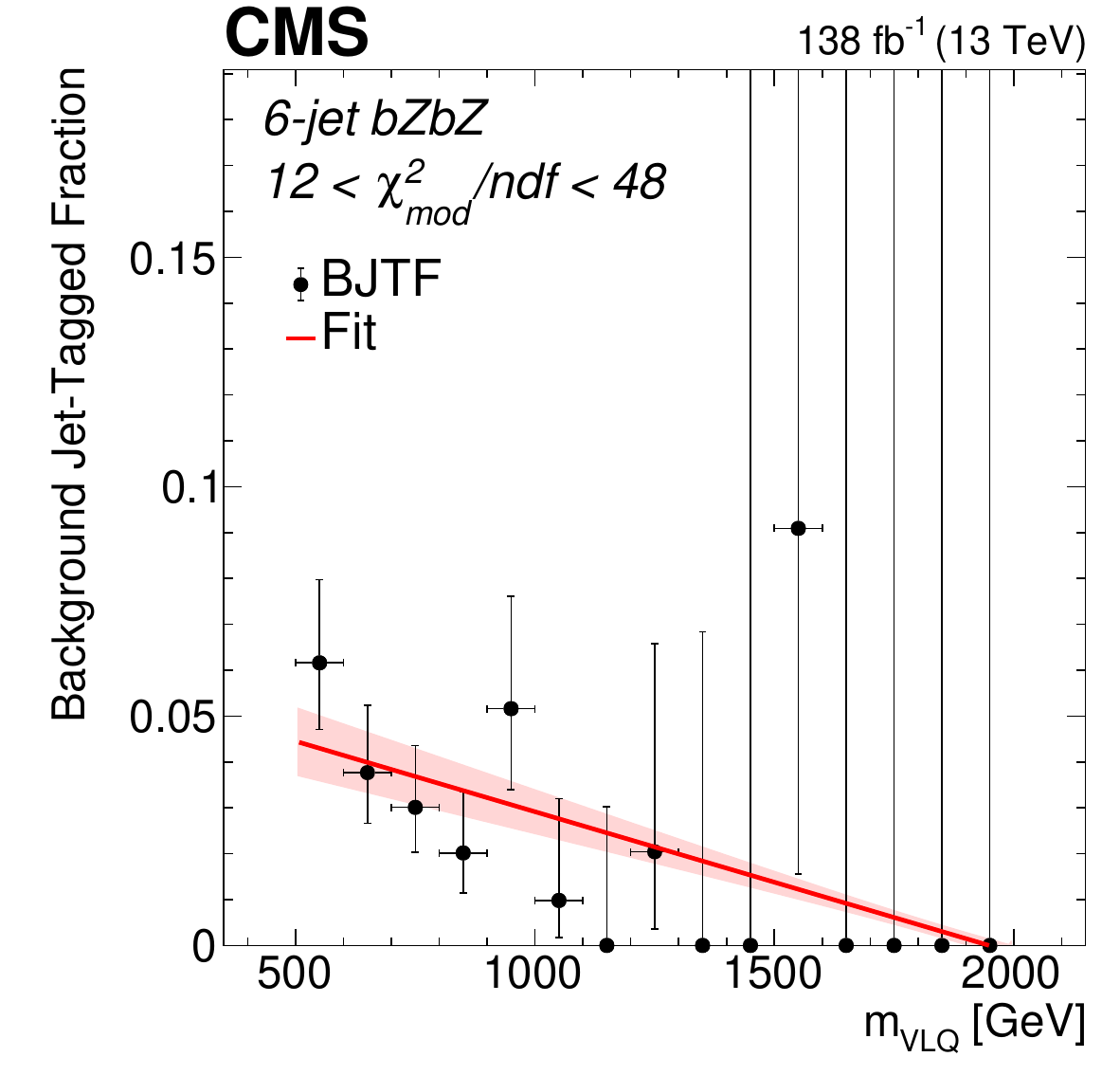}  
    \includegraphics[width=0.32\textwidth]{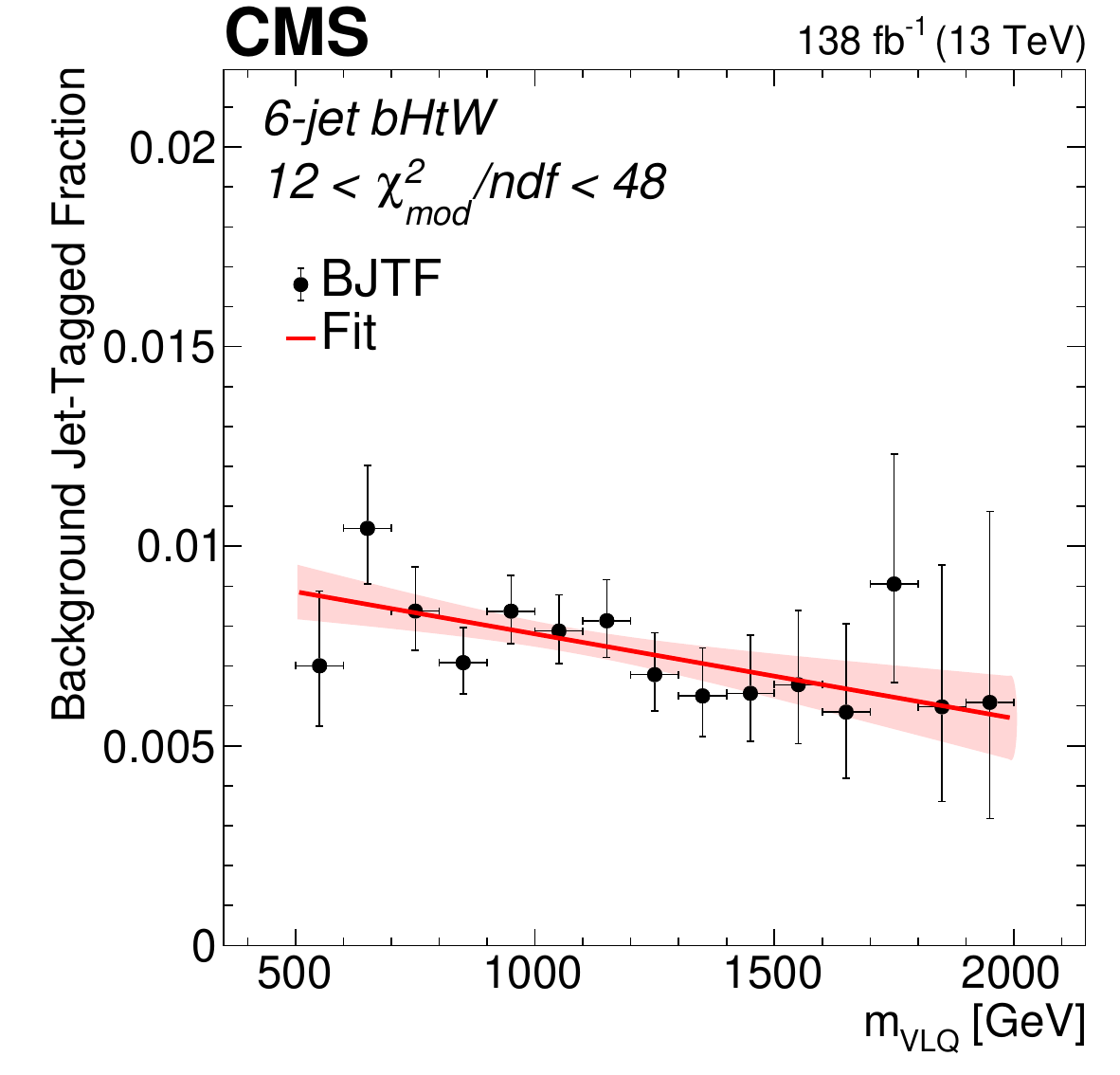}  
    \includegraphics[width=0.32\textwidth]{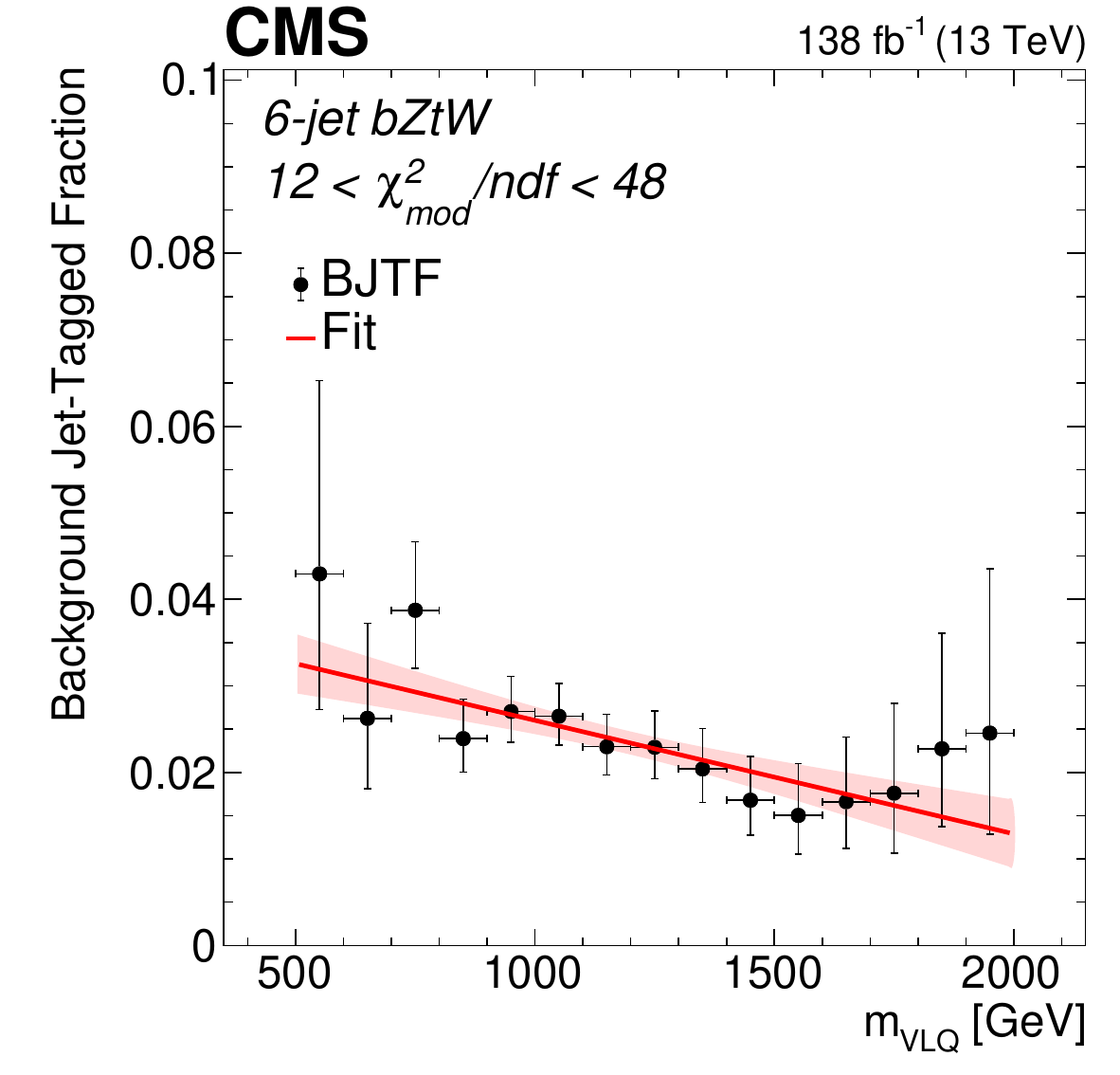}  
    \caption{Value of BJTF as a function of \mvlq in the control region with $12 < \chimodndf < 48$ for some selected fully hadronic channels. Upper row: 4-jet events in the \bHbH (left), \bHbZ (center), and \bZbZ (right) modes. Middle row: 5-jet events in the \bHbH (left), \bHbZ (center), and \bZtW (right) modes. Lower row: 6-jet events in the \bZbZ (left), \bHtW (center), and \bZtW (right) modes. The linear fit is shown by the red line, and the associated uncertainty in the fit is shown by the shaded band.}
    \label{fig:bjtf_control}
\end{figure*}

We allow for a dependence of the BJTF factor on \mvlq that could occur, for example, because the jet tagging efficiency depends on the \pt of the jet, and VLQs with higher mass will generally produce jets with higher \pt. For this, we define a control region as events with \chimodndf in the range $12 < \chimodndf < 48$. The dependence of the BJTF on \mvlq in this control region is then fit with a first-order polynomial, denoted $\varepsilon(m) = p_0+p_1 m$. Checks with $F$-tests indicate that the first-order polynomial fit is preferred over higher-order fits. Figure~\ref{fig:bjtf_control} shows the BJTF distribution in the control regions along with the first-order polynomial fits for a few representative fully hadronic channels.

The estimate of the number of background events in the signal region, $n_{\mathrm{bkg}}$, as a function of \mvlq is then given by the following expression:
\begin{linenomath}
\begin{equation}
\label{eq:background_h}
n_{\mathrm{bkg}}(\mvlq) = n(\mvlq) \varepsilon_0 \frac{\varepsilon(\mvlq)}{\big(\int_{500\GeV}^{800\GeV}\varepsilon(m')\,\rd m'\big)/(300\GeV)}\,,
\end{equation}
\end{linenomath}
where $n(\mvlq)$ is the number of candidates as a function of \mvlq in the preselected sample, as shown in Fig.~\ref{fig:mvlq_distributions_h}, $\varepsilon_0$ is the BJTF determined from the sideband with low VLQ mass, as shown in Table~\ref{table:bjtf_h}, and the last factor is the correction for the dependence of the BJTF on \mvlq, as shown in Fig.~\ref{fig:bjtf_control}.

\subsection{Background estimation in the leptonic category}
\label{sec:background_l}

The background estimation for the leptonic category is based on a procedure similar to that for the fully hadronic category, but with different definitions of the signal and control regions. In particular, the use of \PQb tagging is different for the two categories. For the leptonic category, the BJTF factor described above is not used. Instead, for the leptonic category, the signal region is defined to be events with $\chimodndf < 5$  for which  at least one of the jets directly originating from the \PB VLQ candidate (\ie, not associated with an \PH or a \PZ boson decay) is \PQb tagged (called \PQb-tag events). The control region is defined to be events with the same $\chimodndf < 5$ requirement but where neither of the jets directly originating from a \PB VLQ candidate is \PQb tagged (called \PQb-veto events). Notably, these requirements are both independent of the hadronically decaying \PH or \PZ boson.

\begin{figure*}[hbtp]
    \centering
    \includegraphics[width=0.49\textwidth]{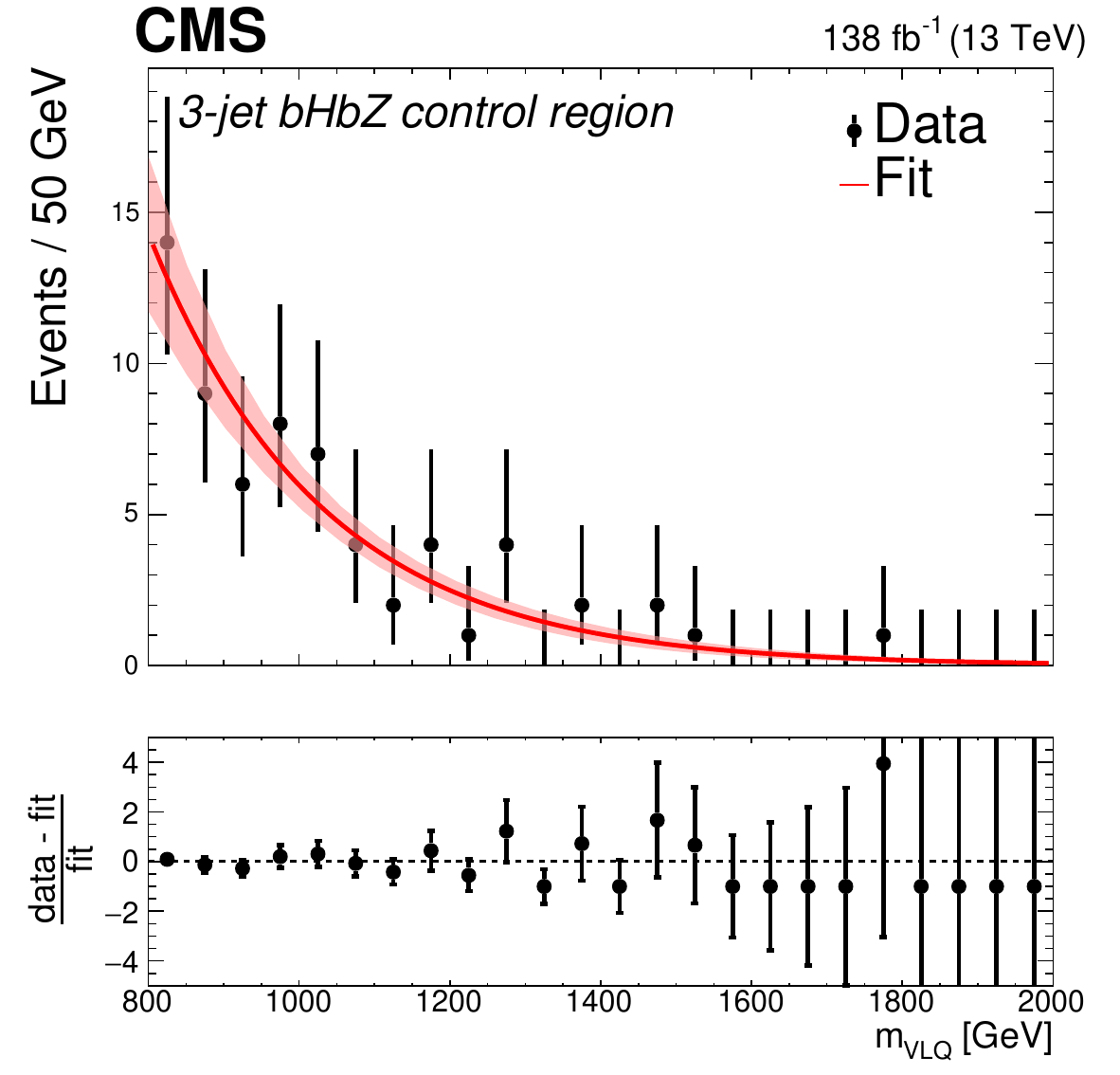}
    \includegraphics[width=0.49\textwidth]{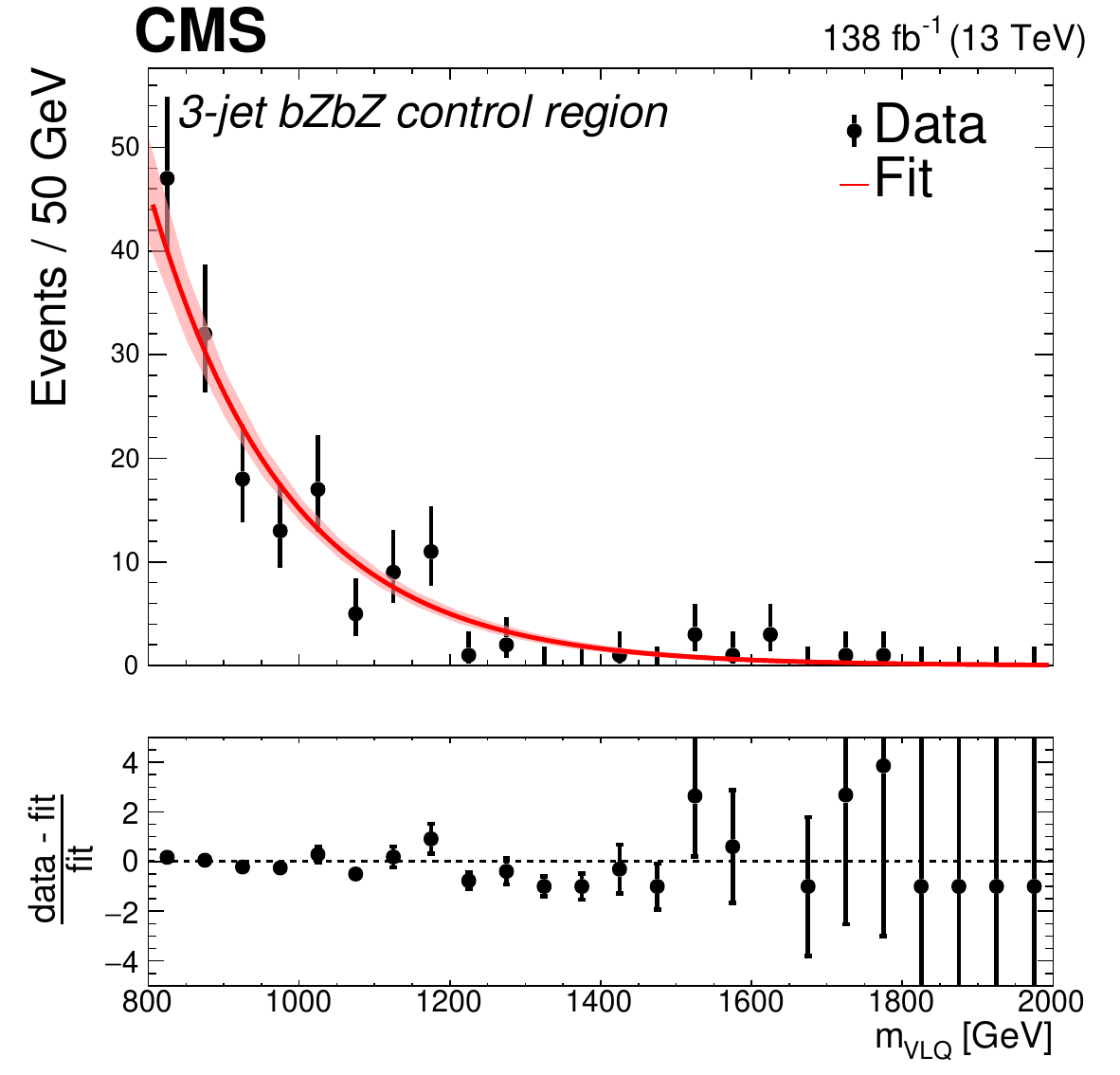}
    \includegraphics[width=0.49\textwidth]{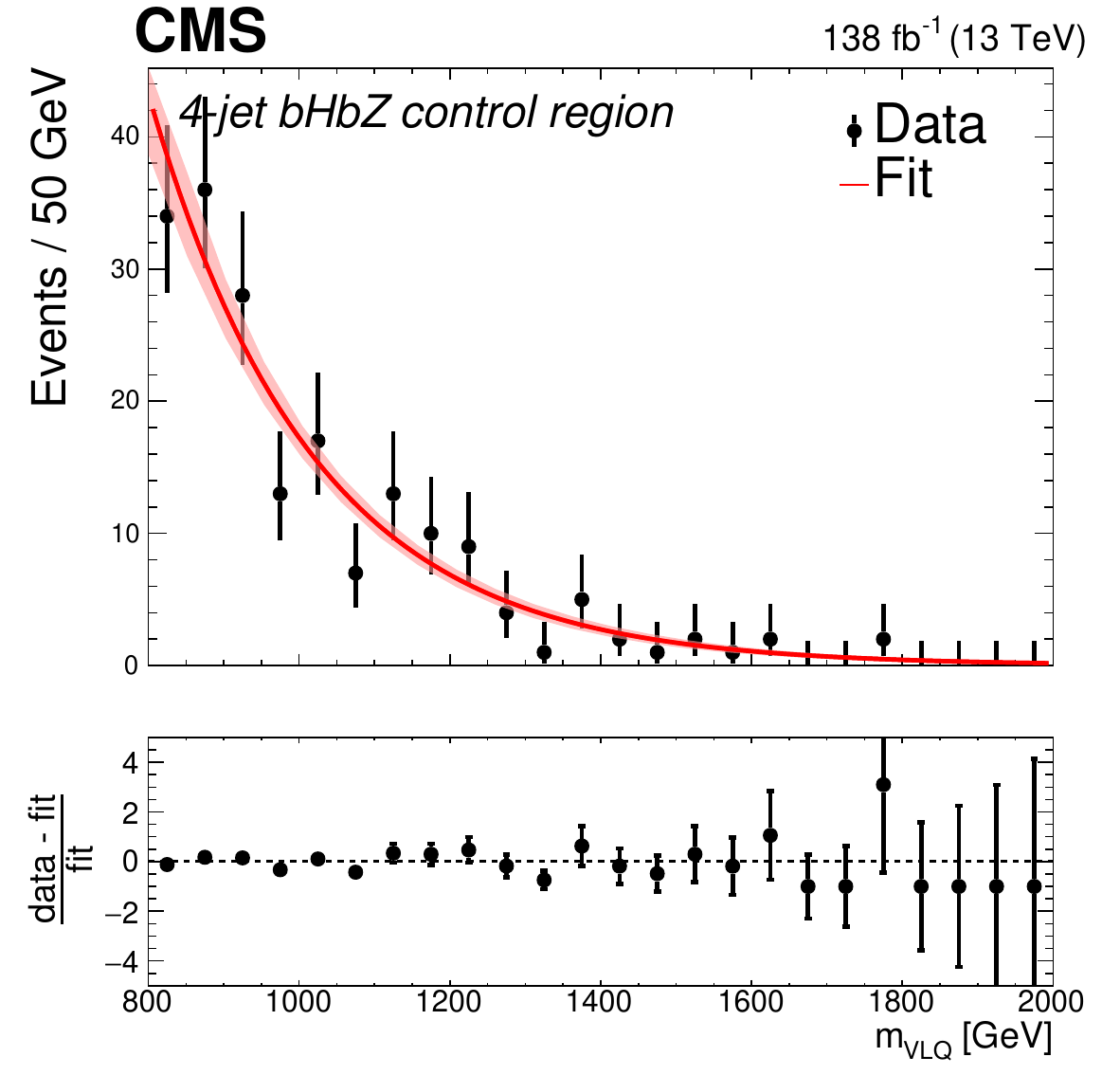} 
    \includegraphics[width=0.49\textwidth]{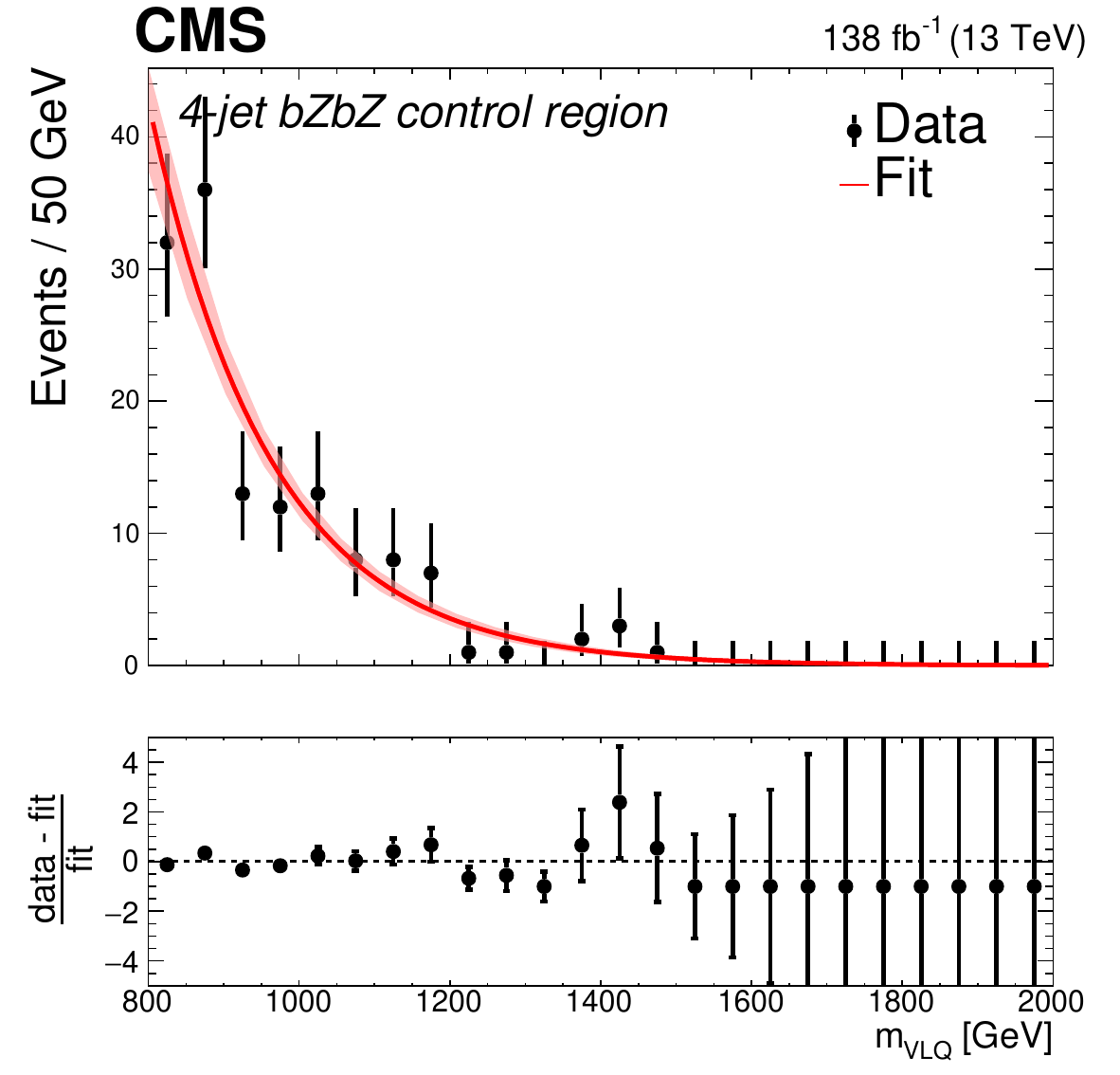}
    \caption{Distributions of \mvlq for events in the control region for the channels in the leptonic category. Upper row: 3-jet events in the \bHbZ (left) and \bZbZ (right) modes. Lower row: 4-jet events in the \bHbZ (left) and \bZbZ (right) modes. The exponential fit and its uncertainty are shown by the red line and the light red shaded band, respectively. The bottom panel shows the fractional difference between the data and fit, (data-fit)/fit.}
    \label{fig:mvlq_distributions_l}
\end{figure*}

First, the VLQ mass distribution in the control region is fit with an exponential function over the mass range $800 < \mvlq < 2000\GeV$. Figure~\ref{fig:mvlq_distributions_l} shows the resulting distributions and the exponential fits for each of the four leptonic channels. Then, this distribution is normalized to the expected distribution in the signal region by multiplying by a normalization factor, the ratio of \PQb-tag events to \PQb-veto events in the low-mass range of $450 < \mvlq < 900\GeV$. Table \ref{table:bjtf_d} shows the resulting normalization factors for the four channels.

\begin{table}[hbtp]
\centering
\topcaption{Values of the \PQb-tag to \PQb-veto ratio for events in the mass range $450 < \mvlq < 900\GeV$ with $\chimodndf < 5$,  for each of the dileptonic channels. The uncertainties shown are statistical only.}
\label{table:bjtf_d}
\begin{scotch}{ccc}
& \bHbZ & \bZbZ\\
\hline
3 jets & $0.101 \pm 0.010$ & $0.130 \pm 0.006$\\
4 jets & $0.127 \pm 0.009$ & $0.125 \pm 0.074$\\
\end{scotch}
\end{table}

\begin{figure*}[hbtp]
    \centering
    \includegraphics[width=0.49\textwidth]{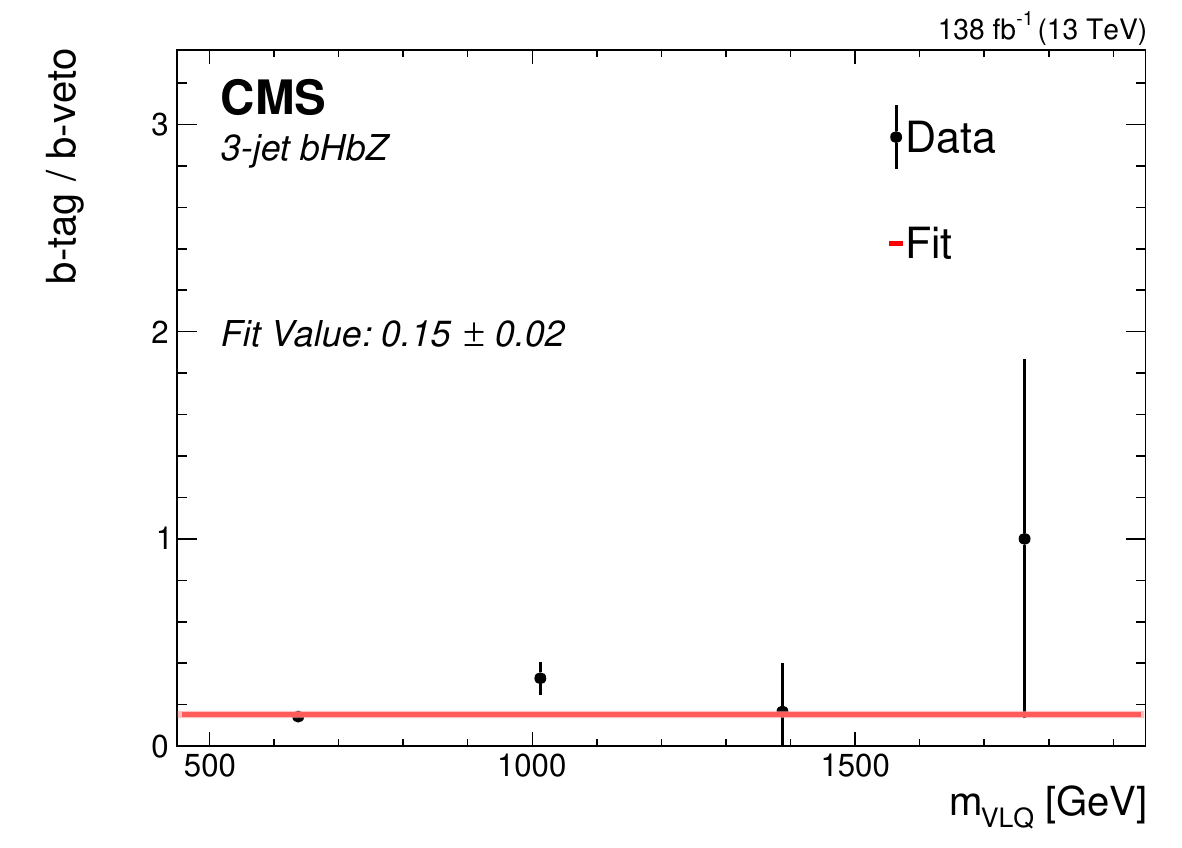}
    \includegraphics[width=0.49\textwidth]{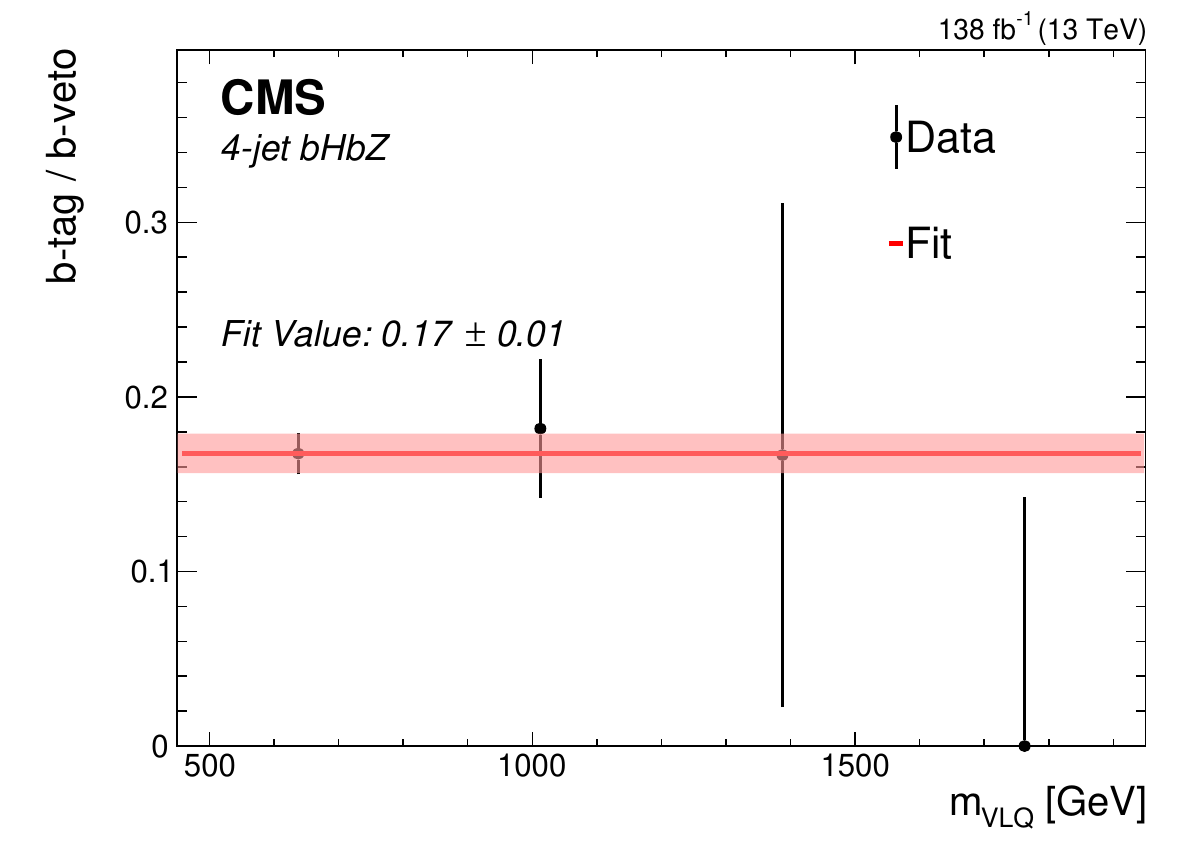}
    \includegraphics[width=0.49\textwidth]{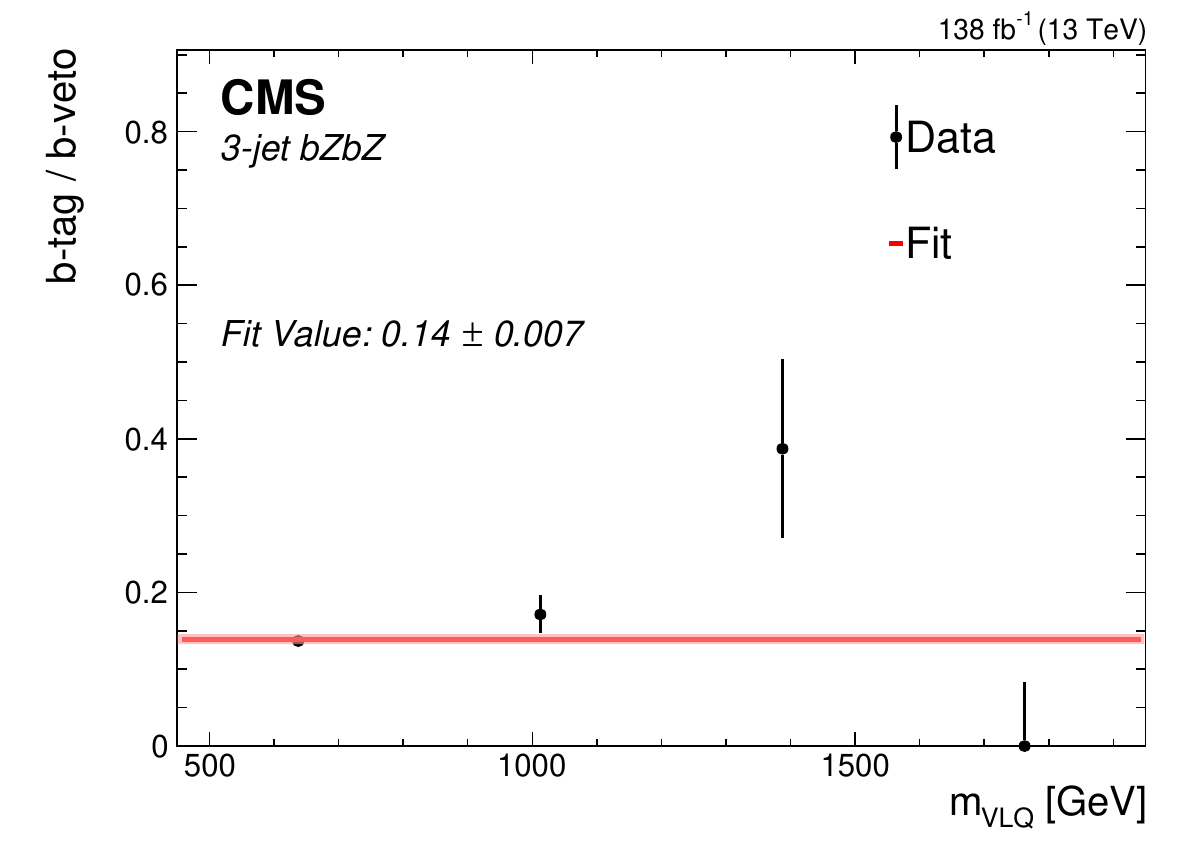}
    \includegraphics[width=0.49\textwidth]{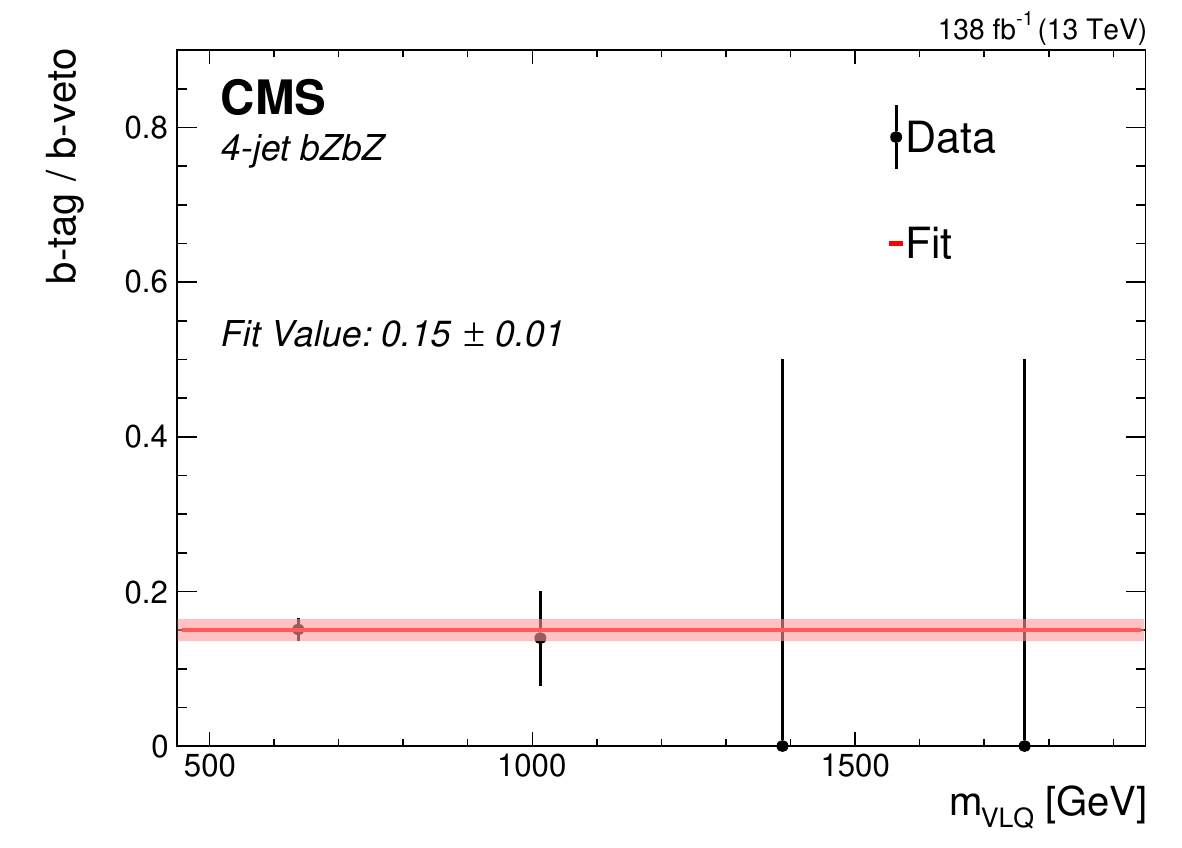}
    \caption{Normalization factor as a function of \mvlq for leptonic events in the $5 < \chimodndf < 20$ region. Upper row: \bHbZ events in the 3-jet (left) and 4-jet (right) channels. Lower row: \bZbZ events in the 3-jet (left) and 4-jet (right) channels. The fit to a constant value and its uncertainty are shown by the red line and the light red shaded band, respectively.}
    \label{fig:normalization}
\end{figure*}

\begin{figure*}[hbtp]
    \centering
    \includegraphics[width=0.49\textwidth]{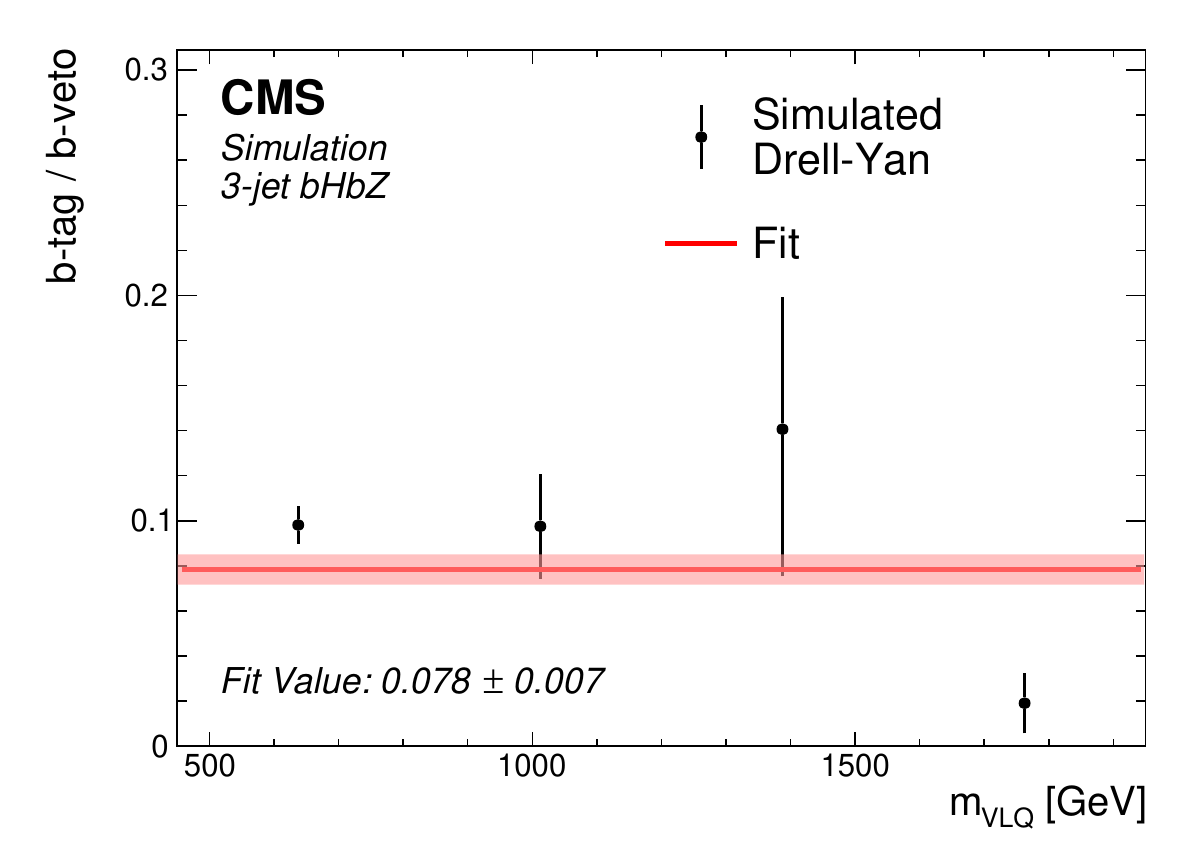}
    \includegraphics[width=0.49\textwidth]{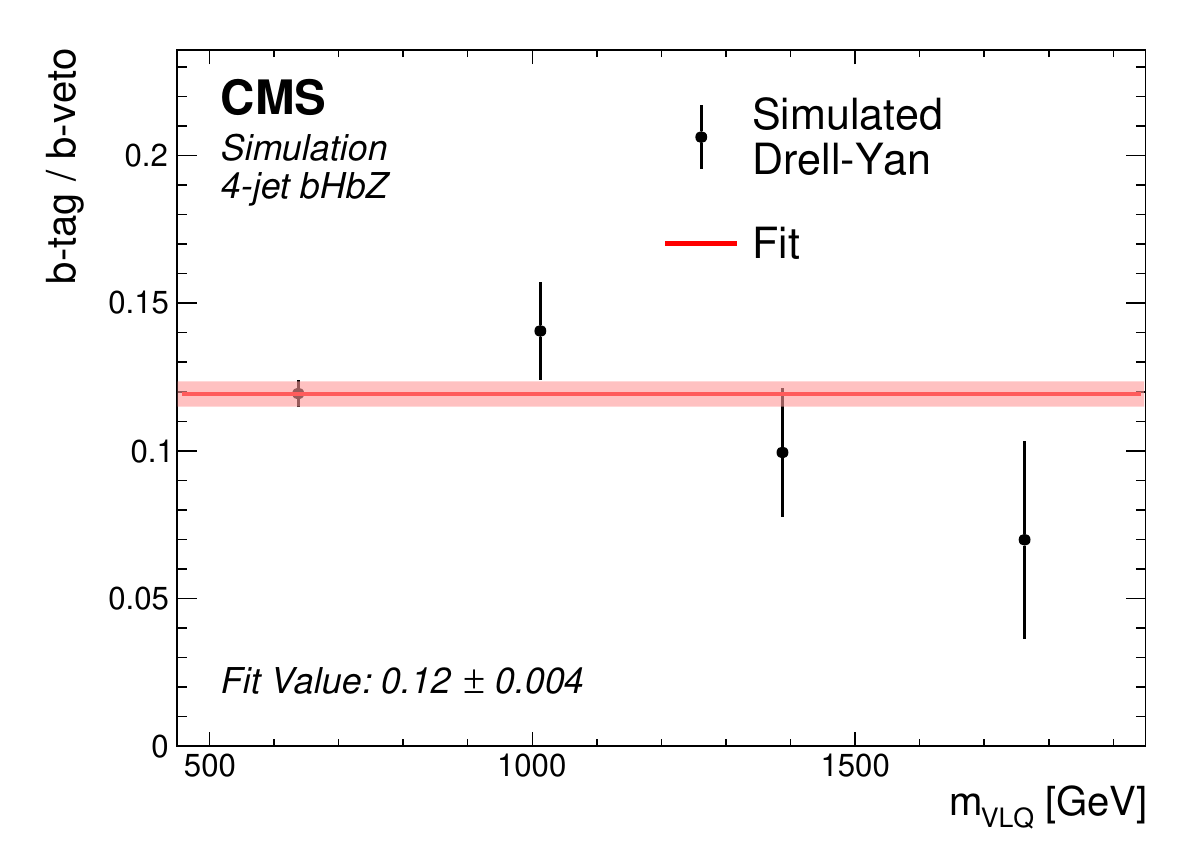}
    \includegraphics[width=0.49\textwidth]{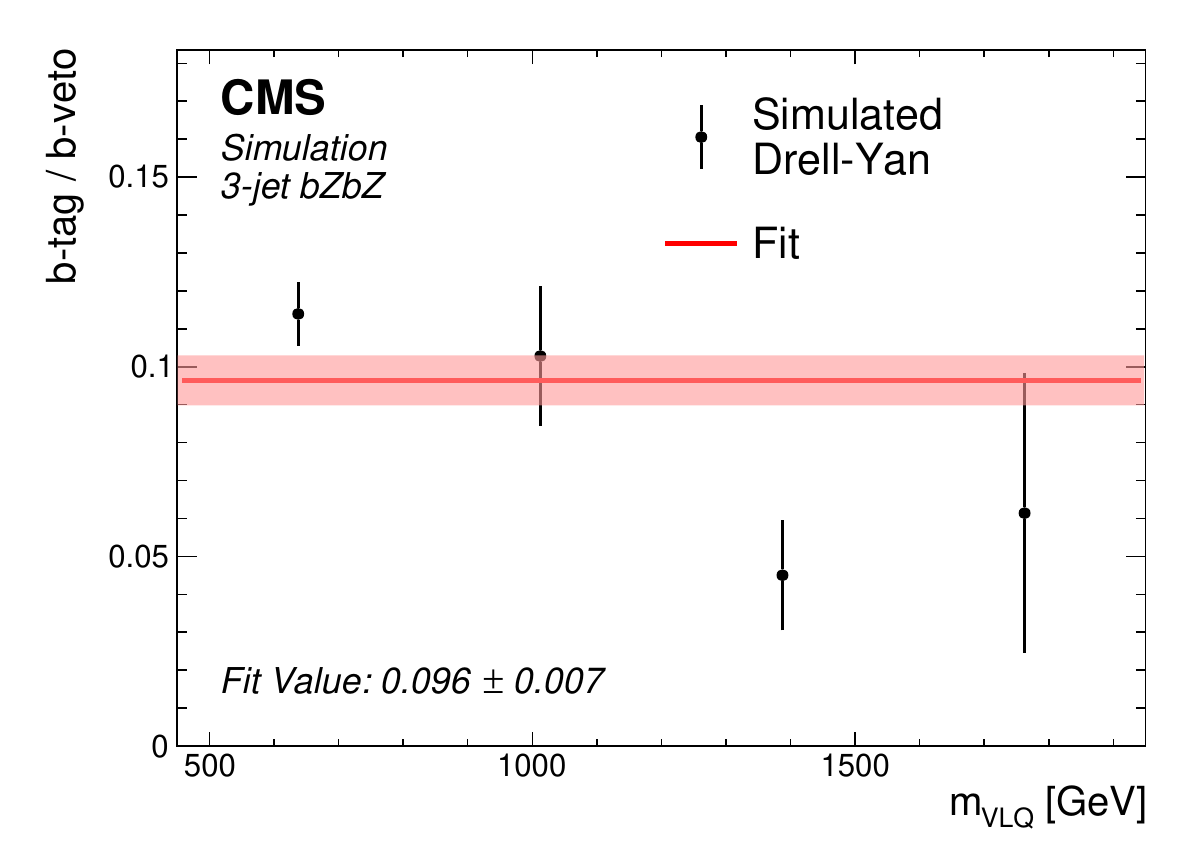}
    \includegraphics[width=0.49\textwidth]{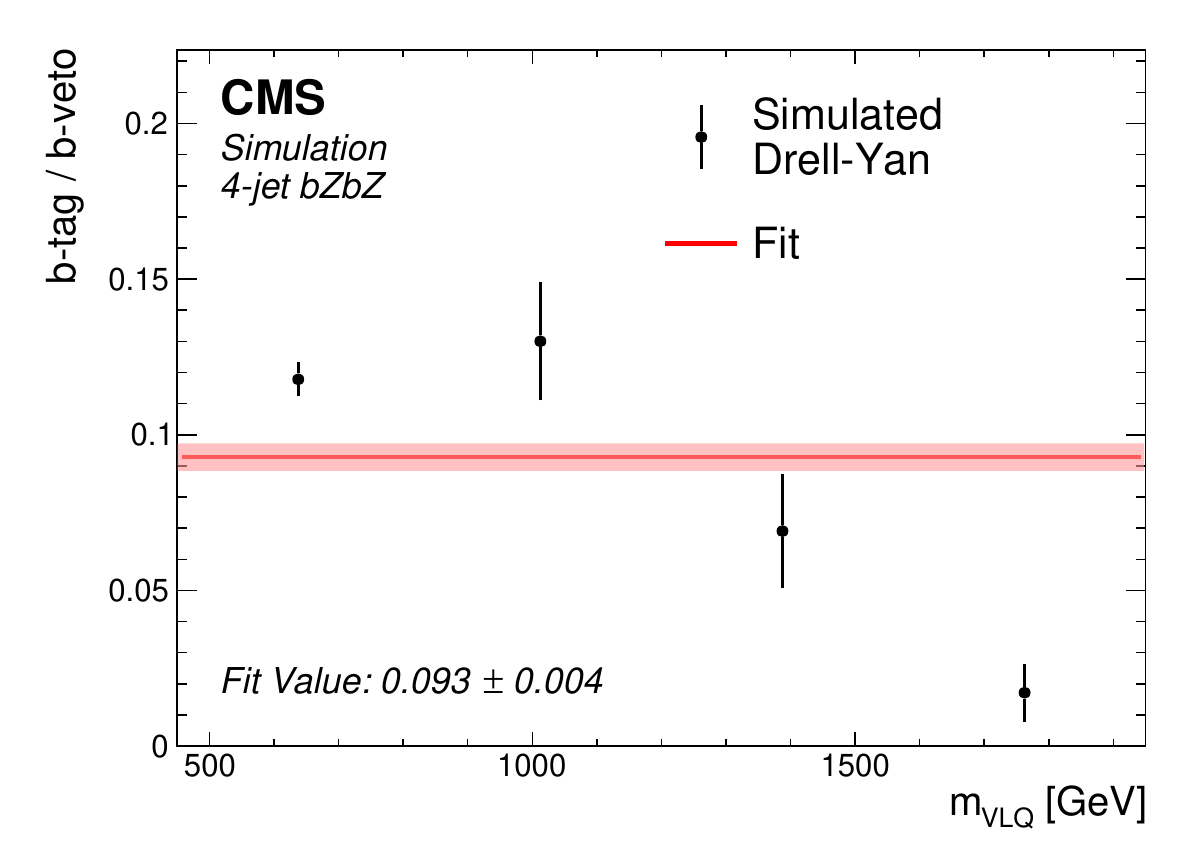}
    \caption{Normalization factor in the leptonic category as a function of \mvlq for simulated Drell--Yan events with $\chimodndf < 5$. Upper row: \bHbZ events in the 3-jet (left) and 4-jet (right) channels. Lower row: \bZbZ events in the 3-jet (left) and 4-jet (right) channels. The fit to a constant value and its uncertainty are shown by the red line and the light red shaded band, respectively.}
    \label{fig:normfactor_mc}
\end{figure*}

Next, events with \chimodndf in the range $5 < \chimodndf < 20$ are used to check for possible dependence of the normalization factors on the VLQ candidate mass. Figure~\ref{fig:normalization} shows the \PQb-tag to \PQb-veto ratio as a function of \mvlq for events in the range $5 < \chimodndf < 20$ for all four leptonic channels. As a cross-check, the dependence of the \PQb-tag to \PQb-veto ratio as a function of \mvlq for simulated Drell--Yan events, which constitute the dominant background, is shown in Fig.~\ref{fig:normfactor_mc} over the full mass range, which verifies that the normalization factors do not have a significant dependence on the VLQ mass.

\begin{figure*}[hbtp]
    \centering
    \includegraphics[width=0.49\textwidth]{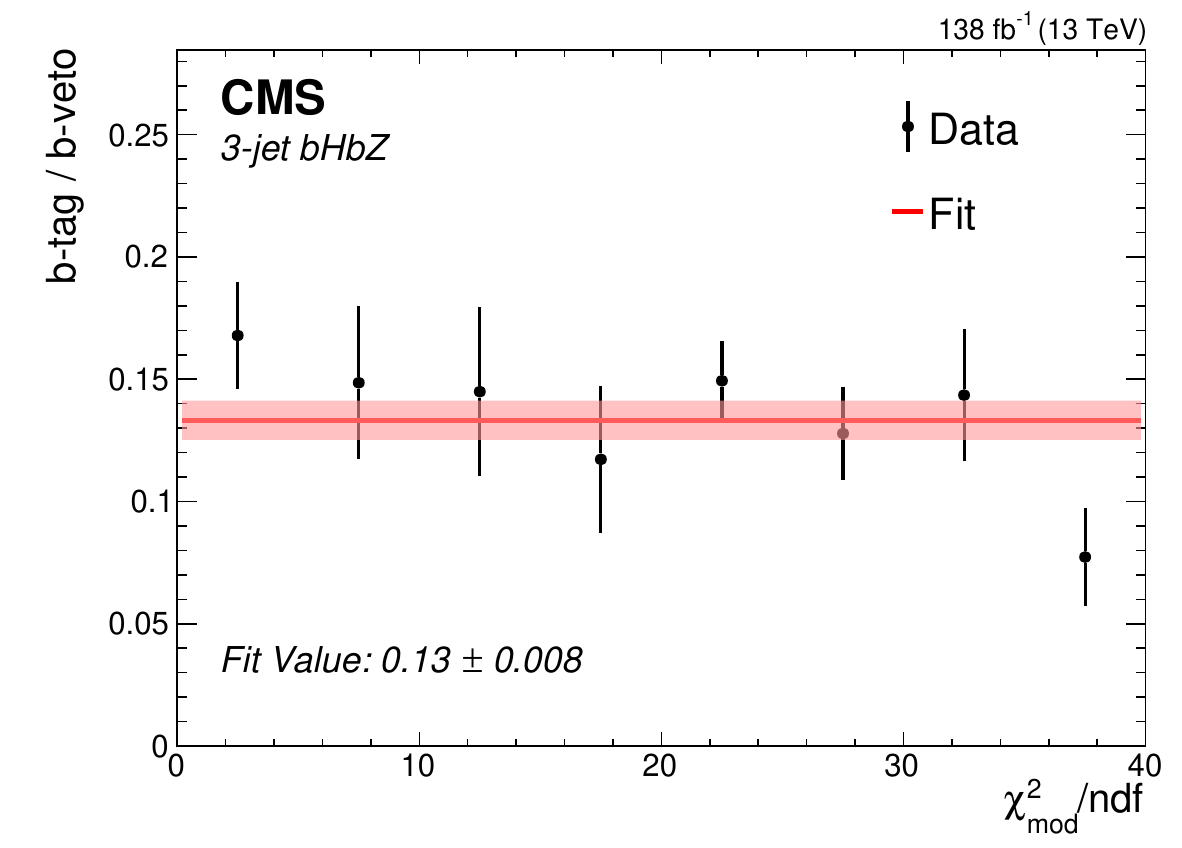}
    \includegraphics[width=0.49\textwidth]{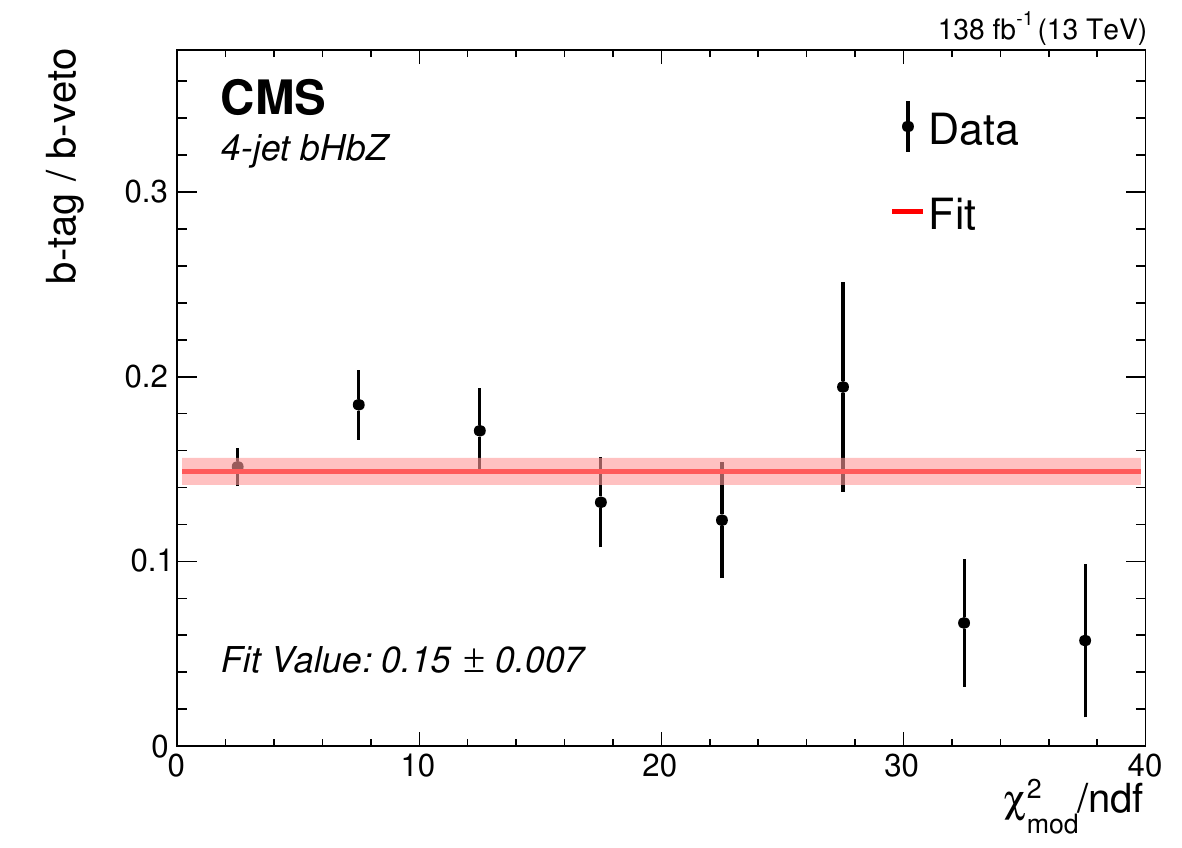}
    \includegraphics[width=0.49\textwidth]{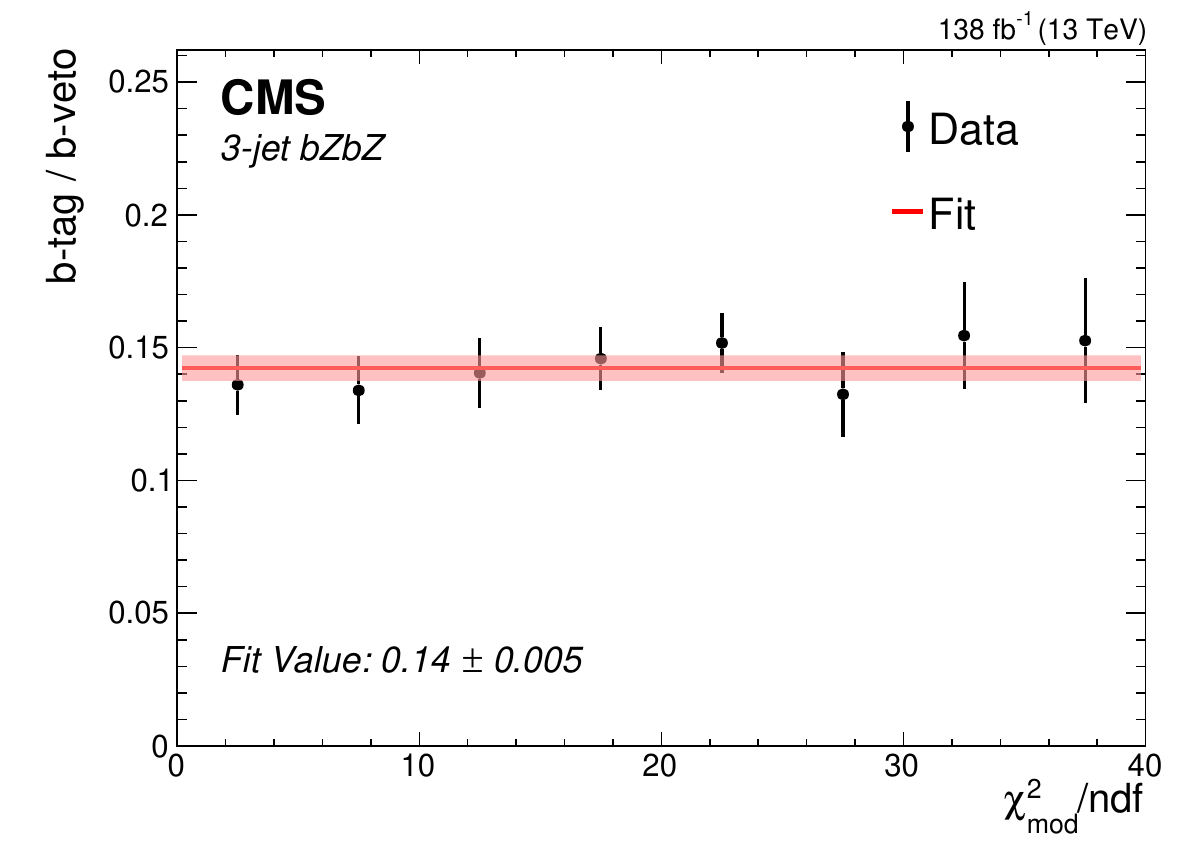}
    \includegraphics[width=0.49\textwidth]{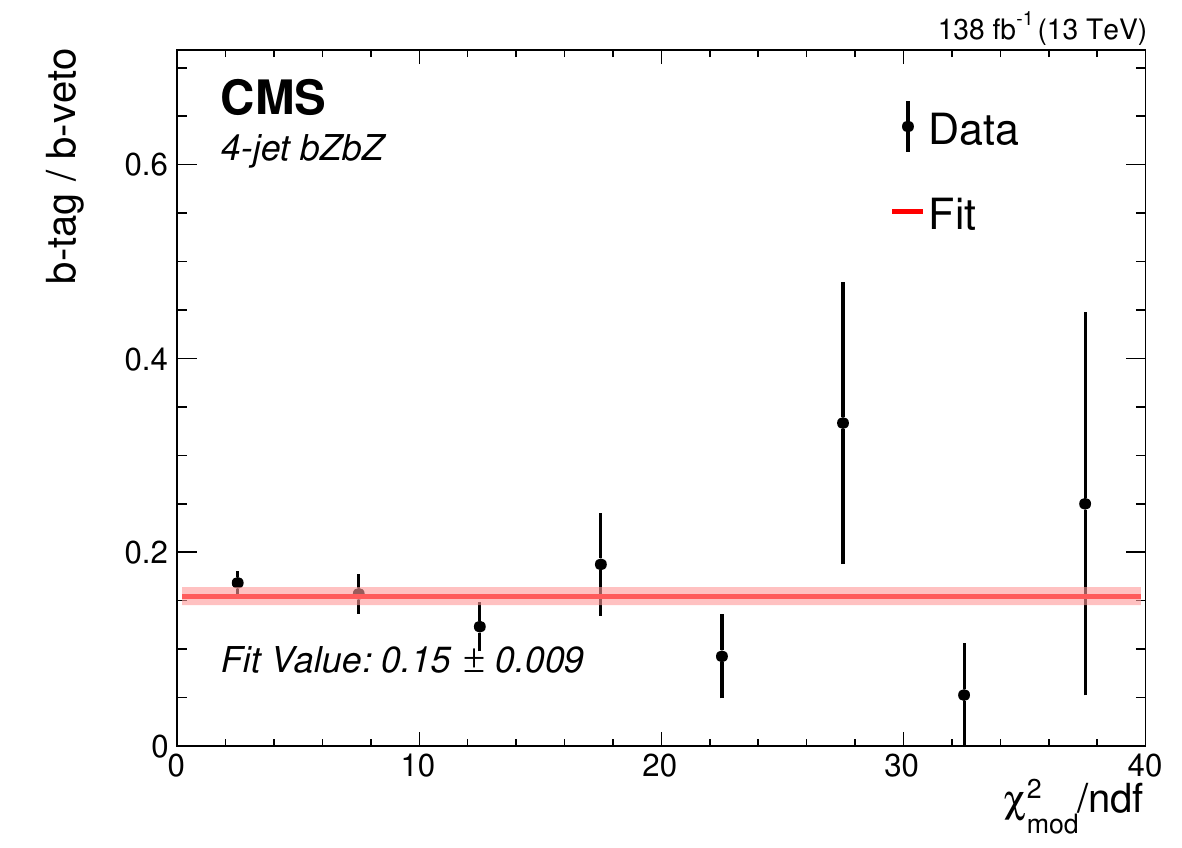}
    \caption{Normalization factor in the leptonic category as a function of \chimodndf. Upper row: 3-jet events in the \bHbZ (left) and \bZbZ (right) decay modes. Lower row: 4-jet events in the \bHbZ (left) and \bZbZ (right) decay modes. The fit to a constant value and its uncertainty are shown by the red line and the light red shaded band, respectively.}
    \label{fig:chireduction}
\end{figure*}

We next check that the \PQb-tag to \PQb-veto ratio in the range $5 < \chimodndf < 20$ correctly represents the ratio in the range $\chimodndf < 5$ by plotting the ratio as a function of \chimodndf, shown in Fig.~\ref{fig:chireduction}. No significant dependence over the \chimodndf region is observed. 

\begin{figure}[!htbp]\centering
\includegraphics[width=0.49\textwidth]{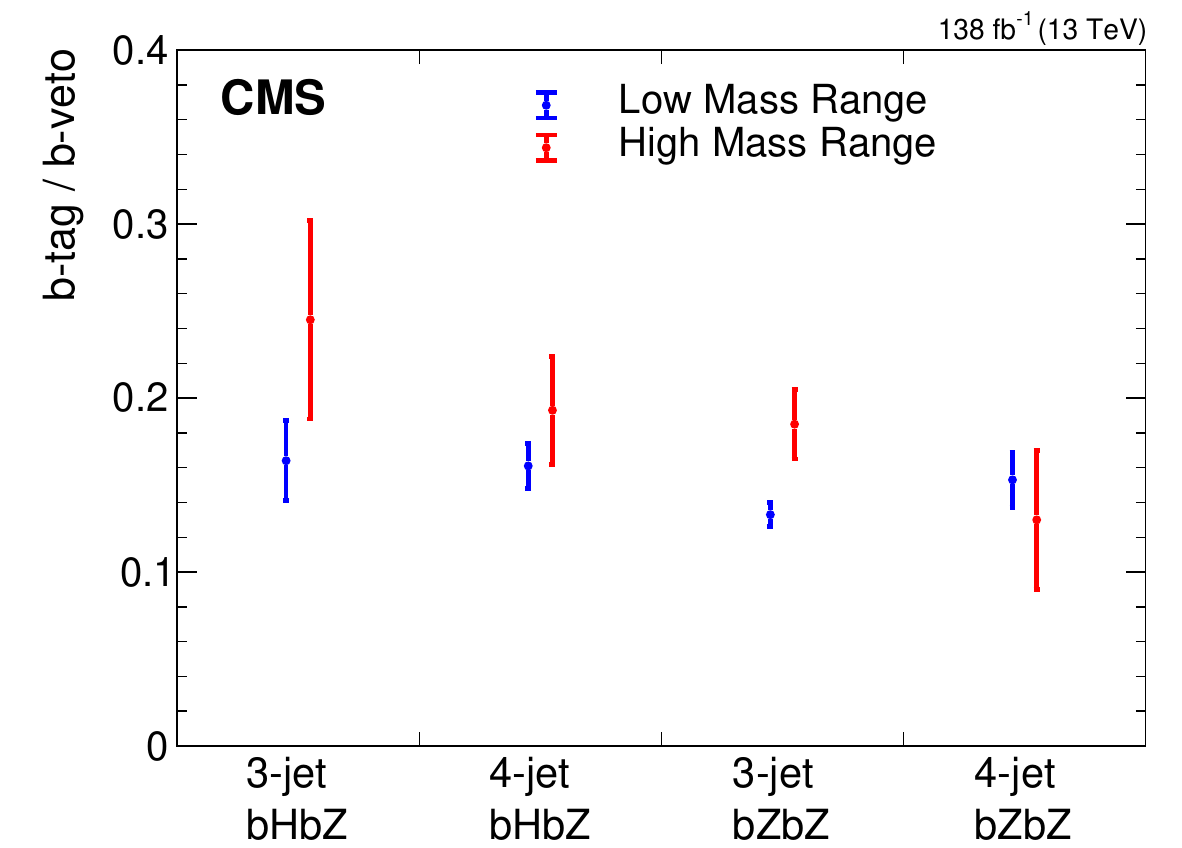}
\includegraphics[width=0.49\textwidth]{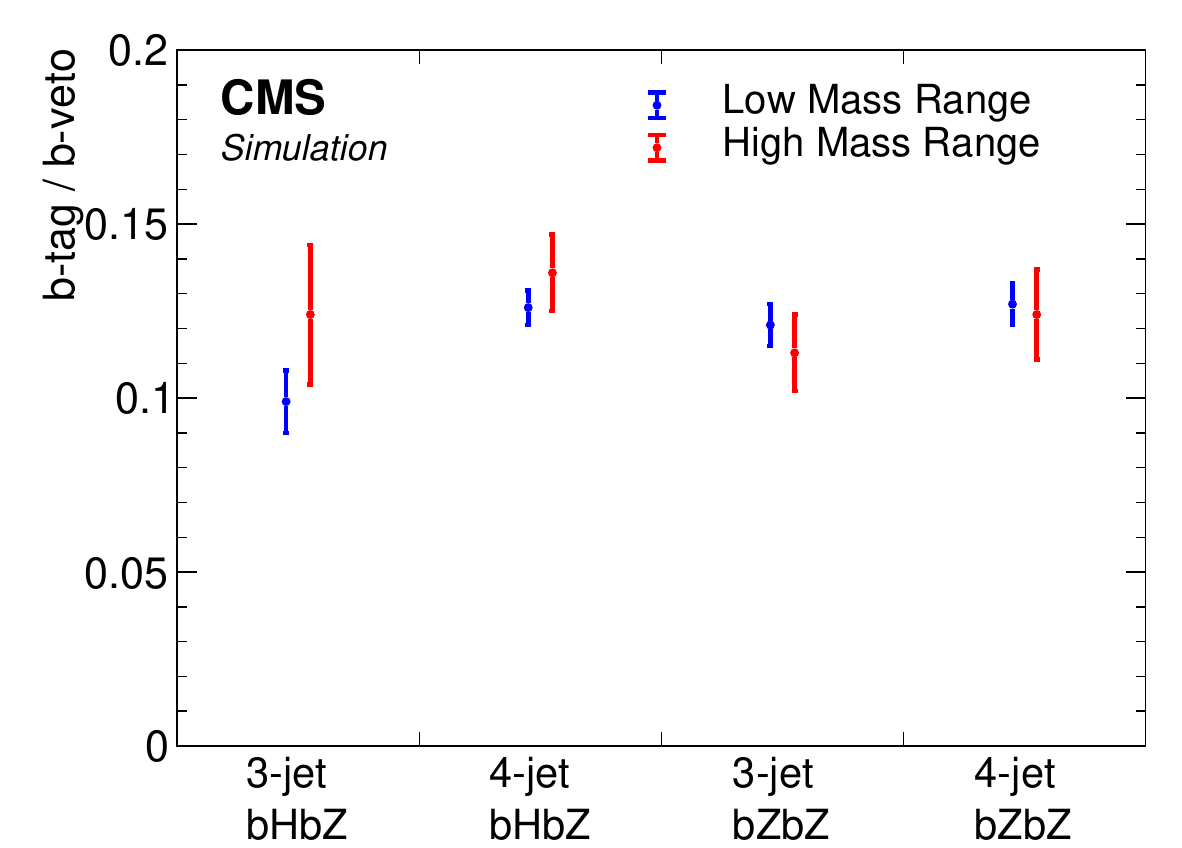}
\caption{\label{fig:BkgdEstimationComparison} Normalization factor in the low-mass region (450 to 900\GeV) and the high-mass region (800 to 2000\GeV) for events with $5 < \chimodndf < 20$ in data (\cmsLeft) and simulated Drell--Yan events with $\chimodndf < 5$ (\cmsRight). }
\end{figure}

The fit to a constant value is therefore sufficient for all channels, as there is no statistically significant evidence for a mass dependence of the \PQb-tag to \PQb-veto ratio. The normalization factor is thereby set to be the ratio of \PQb-tag events to  \PQb-veto events in the 450 to 900\GeV mass range using events with $\chimodndf < 5$, that is, the ratio of low-mass signal region events to low-mass control region events. 
Although there is no statistically significant evidence of a normalization factor mass dependence, steps are taken to estimate the uncertainty that such a dependence may create in this factor. For this, we compare the ratio of \PQb-tag events to \PQb-veto events in the low-mass range to the ratio  of \PQb-tag to \PQb-veto events in the high-mass range, using events in the \chimodndf range of $5 < \chimodndf < 20$. Figure~\ref{fig:BkgdEstimationComparison} (left) shows that the low-mass and high-mass ratios are consistent for all four channels. As a check on this procedure, the same comparison is performed for simulated Drell--Yan events using the optimally tuned \chimodndf. The results, displayed in Fig.~\ref{fig:BkgdEstimationComparison} (right), show that the low-mass and high-mass ratios are consistent, even after an upper bound is applied on the value of \chimodndf.

The estimate of the number of background events in the signal region, $n_{\mathrm{bkg}}$, as a function of \mvlq is given by the following expression:
\begin{linenomath}
\begin{equation}
\label{eq:background_l}
n_{\mathrm{bkg}}(\mvlq) = f(\mvlq) \frac{\int_{450\GeV}^{900\GeV}n_{\text{signal}}(m')\,\rd m'}{\int_{450\GeV}^{900\GeV}n_{\text{control}}(m')\,\rd m'},
\end{equation}
\end{linenomath}
where $f(\mvlq)$ is the fitted function for the number of candidates as a function of \mvlq in the control region, as shown in Fig.~\ref{fig:mvlq_distributions_l}, and the second term is the normalization factor, as determined by the ratio of the number of events in the signal region to the control region with $450 < \mvlq < 900\GeV$. For the final evaluation of the number of events in the sideband region, the \chimodndf requirement for each particular channel is applied.

\section{Systematic uncertainties}\label{section:systematics}

The systematic uncertainties are listed in Tables~\ref{table:all_uncertainties_a} and~\ref{table:all_uncertainties_c}. They include the following:

\begin{itemize}

    \item Integrated luminosity: The integrated luminosities for the 2016--2018 data-taking years have 1.2--2.5\% individual uncertainties~\cite{CMS-LUM-17-003,CMS-PAS-LUM-17-004,CMS-PAS-LUM-18-002}, while the overall uncertainty for the 2016--2018 period is 1.6\%.

    \item Trigger: The uncertainties in the trigger efficiencies are determined from the uncertainties in the fitting parameters from fits of the trigger efficiencies with respect to \HT, using constant functions for channels in the hadronic category and logistic functions for channels in the leptonic category. For 2016, 2017, and 2018 they are 2.0, 0.05, and 0.01\%, respectively, for the jet triggers used; 0.3, 0.2, and 0.2\% for the muon triggers, and 0.2, 0.3, and 0.2\% for the electron triggers.
    
    \item Dilepton \PZ boson efficiency:  The total dimuon and dielectron efficiencies are determined using simulated signal events. The net efficiency is calculated by multiplying the efficiencies for each step in the dilepton \PZ boson selection process: exactly two leptons, both leptons pass the identification and isolation requirements, and the dilepton \PZ boson invariant mass is between 80 and 102\GeV. The uncertainty in the total efficiency is found to be 0.3\% for both the dimuon and dielectron cases.
    
    \item Scale factors: Scale factors for jet energy, jet energy resolution, \PQb tagging efficiency, and lepton efficiency (including isolation, identification, and reconstruction for individual leptons) are needed to correct the distributions in simulation to match those in data. The resulting uncertainty in the simulated signal efficiency is determined by varying each scale factor up and down by 1 standard deviation. The uncertainties over each year are treated as uncorrelated. The uncertainties depend upon the event channel. Typical values are as follows: 0.5--4.0\% for jet energy, 0.5--2.5\% for jet resolution, 5.0--12\% for \PQb tagging, and 10--12\% for the overall lepton efficiency (where the larger value is due to the combination of isolation, identification, and reconstruction scale factors for the individual leptons).
    
    \item Background estimation in the fully hadronic category: There are two sources of uncertainty in the background estimate as a function of the average mass of the two VLQ candidates, $n_{\text{bkg}}(\mvlq)$ in Eq.~(\ref{eq:background_h}). The first is from the fit for $n(\mvlq)$, and it is determined from the uncertainties in the fit parameters $p_0$, $p_1$, and $p_2$.  The second is the uncertainty in the mass dependence of the BJTF factor, which is similarly determined from the uncertainties in the first-order polynomial fit parameters, $p_0$ and $p_1$, shown in the definition of $\varepsilon(m)$. For each of the fits, the nuisance parameters are determined by decorrelating the fit parameters using the covariance matrix of the fit. Additionally, we include an uncertainty to account for our choice of fit function (referred to as ``BJTF fit form'' in Table~\ref{table:all_uncertainties_a}).  

    \item Background estimation in the dileptonic category: Similarly, there are two sources of uncertainty in the background estimate as a function of the average mass of the two VLQ candidates, $n_{\text{bkg}}(\mvlq)$ in Eq.~(\ref{eq:background_l}). The first arises from the fit for $f(\mvlq)$, and it is determined from the uncertainties in the exponential fit parameters.  The second is the uncertainty in the mass dependence of the normalization factor, the \PQb-tag to \PQb-veto ratio, which is assigned to be one-half of the absolute value of the difference between this ratio in the low- and high-mass ranges, as shown in Fig.~\ref{fig:BkgdEstimationComparison} (left).

    \item PDFs: The uncertainty associated with the choice of PDF set is calculated by following the PDF4LHC procedure to generate a set of weights for each uncertainty in the NNPDF3.0 distribution~\cite{Butterworth:2015oua}. The uncertainty in the signal acceptance is then determined from the standard deviation of the distribution of the signal acceptance for each weight.

    \item QCD scale: The systematic uncertainty on the QCD scale is estimated by independently multiplying the default values of the renormalization and factorization scales, \muR and \muF, by factors of 2.0 and 0.5, and assigning the systematic error to be half of the resulting range of cross sections. In this procedure, following the results from \cite{Catani:2003zt}, the cases of varying one scale by 2.0 and the other by 0.5 are excluded since these correspond to unjustified values of $\log(\muR/\muF)$. The range over the remaining seven combinations is then used. 
    
    \item Pileup: The uncertainty due to the number of pileup events in the simulated events is evaluated by varying the $\Pp\Pp$ inelastic cross section~\cite{CMS:2018mlc} by $\pm$1 standard deviation and determining the resulting change in the event selection efficiency.

    \item Trigger prefiring: Because of detector effects, ECAL and muon triggers can appear in the bunch crossing preceding the actual collision. This effect is not accounted for in simulation, so corrections are applied to obtain the correct trigger efficiency. The uncertainties in these corrections are propagated to the event efficiency to obtain the resulting uncertainty.
\end{itemize}

Tables~\ref{table:all_uncertainties_a} and \ref{table:all_uncertainties_c} summarize these uncertainties for the case of 40\% \BrBbH, 40\% \BrBbZ, and 20\% \BrBtW branching fractions and a VLQ mass of 1400\GeV. For each uncertainty, it is noted whether it affects the signal efficiency or the background estimate and whether it affects the shape of the mass distribution or only the overall rate.

In general, the statistical uncertainty has the most significant impact on the results, in part because it also increases the systematic uncertainties related to the background fit functions. These uncertainties tend to be largest for the 4-jet channels, since these have the fewest events.

In addition, an uncertainty of 6\% is assigned to account for uncertainties in the VLQ pair production cross section~\cite{Sirunyan:2018omb}. This uncertainty is separate from the others described above, and it is used only for the error band on the theory curve shown in the exclusion limit plots in Fig.~\ref{fig:results_mass}. 

\begin{table*}[!hbtp]
    \centering
    \topcaption{Table of systematic uncertainties for the fully hadronic channels for a simulated signal mass of 1400\GeV and branching fractions of $\BrBbH = 40\%$, $\BrBbZ = 40\%$, and $\BrBtW = 20\%$. The only parameters in the fits that have significant uncertainties ($>$0.01\%) are the scaling parameters (denoted by $p_0$).}
    \label{table:all_uncertainties_a}
    \cmsTableSmall{\begin{scotch}{lccccc}
    Systematic & Signal/Background & Rate/Shape & 4 jets & 5 jets & 6 jets \\ \hline
Jet trigger efficiency & Signal & Rate & 0.6\% & 0.6\% & 0.6\% \\
Luminosity & Signal & Rate & 1.6\% & 1.6\% & 1.6\% \\
PDFs & Signal & Rate & 1.5\% & 1.5\% & 1.5\% \\
QCD \muR and \muF scales & Signal & Rate & 14.0\% & 14.0\% & 14.0\% \\
 [\cmsTabSkip]
\multicolumn{6}{c}{\textit{Fully hadronic \bHbH event mode}} \\ 
 [\cmsTabSkip]
Low-mass BJTF & Background & Rate & 70.7\% & 18.9\% & 13.5\% \\
Background fit $p_0$ & Background & Shape & 9.7\% & 9.6\% & 9.6\% \\
BJTF fit form & Background & Rate & 5.6\% & 21.4\% & 0.1\% \\
BJTF fit $p_0$ & Background & Shape & 9.7\% & 9.6\% & 9.6\% \\
Jet tag scale factors & Signal & Shape & 4.6\% & 5.8\% & 7.1\% \\
Jet energy resolution & Signal & Shape & 0.3\% & 1.4\% & 2.0\% \\
Jet energy scale & Signal & Shape & 1.3\% & 2.1\% & 3.4\% \\
Pileup & Signal & Shape & 0.3\% & 0.3\% & 1.5\% \\
 [\cmsTabSkip]
\multicolumn{6}{c}{\textit{Fully hadronic \bHbZ event mode}} \\ 
 [\cmsTabSkip]
Low-mass BJTF & Background & Rate & 17.2\% & 8.2\% & 14.6\% \\
Background fit $p_0$ & Background & Shape & 9.7\% & 9.6\% & 9.4\% \\
BJTF fit form & Background & Rate & 0.7\% & 1.5\% & 1.3\% \\
BJTF fit $p_0$ & Background & Shape & 9.7\% & 9.6\% & 9.4\% \\
Jet tag scale factors & Signal & Shape & 4.8\% & 5.7\% & 7.2\% \\
Jet energy resolution & Signal & Shape & 0.2\% & 1.0\% & 1.6\% \\
Jet energy scale & Signal & Shape & 1.2\% & 0.3\% & 1.4\% \\
Pileup & Signal & Shape & 0.1\% & 0.0\% & 1.1\% \\
 [\cmsTabSkip]
\multicolumn{6}{c}{\textit{Fully hadronic \bZbZ event mode}} \\ 
 [\cmsTabSkip]
Low-mass BJTF & Background & Rate & 20.9\% & 3.8\% & 4.5\% \\
Background fit $p_0$ & Background & Shape & 9.8\% & 9.8\% & 9.5\% \\
BJTF fit form & Background & Rate & 10.4\% & 0.0\% & 34.3\% \\
BJTF fit $p_0$ & Background & Shape & 9.8\% & 9.8\% & 9.5\% \\
Jet tag scale factors & Signal & Shape & 5.0\% & 5.0\% & 4.9\% \\
Jet energy resolution & Signal & Shape & 0.3\% & 0.5\% & 1.0\% \\
Jet energy scale & Signal & Shape & 1.0\% & 2.8\% & 3.9\% \\
Pileup & Signal & Shape & 0.4\% & 0.1\% & 0.2\% \\
 [\cmsTabSkip]
\multicolumn{5}{c}{\textit{Fully hadronic \bHtW event mode}} \\ 
 [\cmsTabSkip]
Low-mass BJTF & Background & Rate & \NA & 7.4\% & 3.2\% \\
Background fit $p_0$ & Background & Shape & \NA & 9.4\% & 10.2\% \\
BJTF fit form & Background & Rate & \NA & 0.3\% & 0.6\% \\
BJTF fit $p_0$ & Background & Shape & \NA & 9.4\% & 10.2\% \\
Jet tag scale factors & Signal & Shape & \NA & 4.1\% & 5.6\% \\
Jet energy resolution & Signal & Shape & \NA & 0.7\% & 0.4\% \\
Jet energy scale & Signal & Shape & \NA & 0.3\% & 1.5\% \\
Pileup & Signal & Shape & \NA & 0.1\% & 0.7\% \\
 [\cmsTabSkip]
\multicolumn{5}{c}{\textit{Fully hadronic \bZtW event mode}} \\ 
 [\cmsTabSkip]
Low-mass BJTF & Background & Rate & \NA & 8.0\% & 1.6\% \\
Background fit $p_0$ & Background & Shape & \NA & 9.7\% & 10.6\% \\
BJTF fit form & Background & Rate & \NA & 0.6\% & 1.9\% \\
BJTF fit $p_0$ & Background & Shape & \NA & 9.7\% & 10.6\% \\
Jet tag scale factors & Signal & Shape & \NA & 3.6\% & 4.1\% \\
Jet energy resolution & Signal & Shape & \NA & 1.0\% & 0.4\% \\
Jet energy scale & Signal & Shape & \NA & 0.5\% & 1.4\% \\
Pileup & Signal & Shape & \NA & 0.0\% & 0.6\% \\ 
\end{scotch}}
\end{table*}

 \begin{table*}[!hbtp]
    \centering
    \topcaption{Table of systematic uncertainties for the dileptonic \bHbZ and \bZbZ channels for a simulated signal mass of 1400\GeV and branching fractions of $\BrBbH = 40\%$, $\BrBbZ = 40\%$, and $\BrBtW = 20\%$.}
    \label{table:all_uncertainties_c}
    \cmsTable{\begin{scotch}{lcccc}
    Systematic & Signal/Background & Rate/Shape & 3 jets & 4 jets \\ \hline 
$\PZ \to \Pe\Pe$ efficiency & Signal & Rate & 0.8\% & 0.8\% \\
$\PZ \to \PGm\PGm$ efficiency & Signal & Rate & 0.6\% & 0.6\% \\
Electron trigger efficiency & Signal & Rate & 0.4\% & 0.4\% \\
Muon trigger efficiency & Signal & Rate & 0.1\% & 0.1\% \\
Luminosity & Signal & Rate & 1.6\% & 1.6\% \\
PDFs & Signal & Rate & 2.0\% & 2.0\% \\
QCD \muR and \muF scales & Signal & Rate & 15.1\% & 15.1\% \\
 [\cmsTabSkip]
\multicolumn{5}{c}{\textit{Dileptonic \bHbZ event mode}} \\ 
 [\cmsTabSkip]
Background exponential fit & Background & Shape & 24.4\% & 16.1\% \\
Jet tag scale factors & Signal & Shape & 6.9\% & 7.8\% \\
Normalization factor mass dependence & Background & Rate & 55.9\% & 7.1\% \\
Normalization factor (\PQb-tag/\PQb-veto) & Background & Rate & 14.4\% & 8.2\% \\
Jet energy resolution & Signal & Shape & 0.4\% & 1.3\% \\
Jet energy scale & Signal & Shape & 0.5\% & 2.5\% \\
Lepton scale factors & Signal & Shape & 12.4\% & 11.4\% \\
Pileup & Signal & Shape & 0.5\% & 1.1\% \\
Trigger prefiring & Signal & Shape & 0.3\% & 0.3\% \\
 [\cmsTabSkip]
\multicolumn{5}{c}{\textit{Dileptonic \bZbZ event mode}} \\ 
 [\cmsTabSkip]
Background exponential fit & Background & Shape & 26.0\% & 20.1\% \\
Jet tag scale factors & Signal & Shape & 7.1\% & 8.0\% \\
Normalization factor mass dependence & Background & Rate & 4.6\% & 10.6\% \\
Normalization factor (\PQb-tag/\PQb-veto) & Background & Rate & 14.6\% & 9.4\% \\
Jet energy resolution & Signal & Shape & 0.4\% & 0.8\% \\
Jet energy scale & Signal & Shape & 0.2\% & 0.6\% \\
Lepton scale factors & Signal & Shape & 11.7\% & 11.6\% \\
Pileup & Signal & Shape & 0.5\% & 0.4\% \\
Trigger prefiring & Signal & Shape & 0.3\% & 0.3\% \\
    \end{scotch}}
\end{table*}

\ifthenelse{\boolean{cms@external}}{}{\clearpage}

\section{Results}

\begin{figure*}[!htp]
\centering
\includegraphics[width=0.4\textwidth]{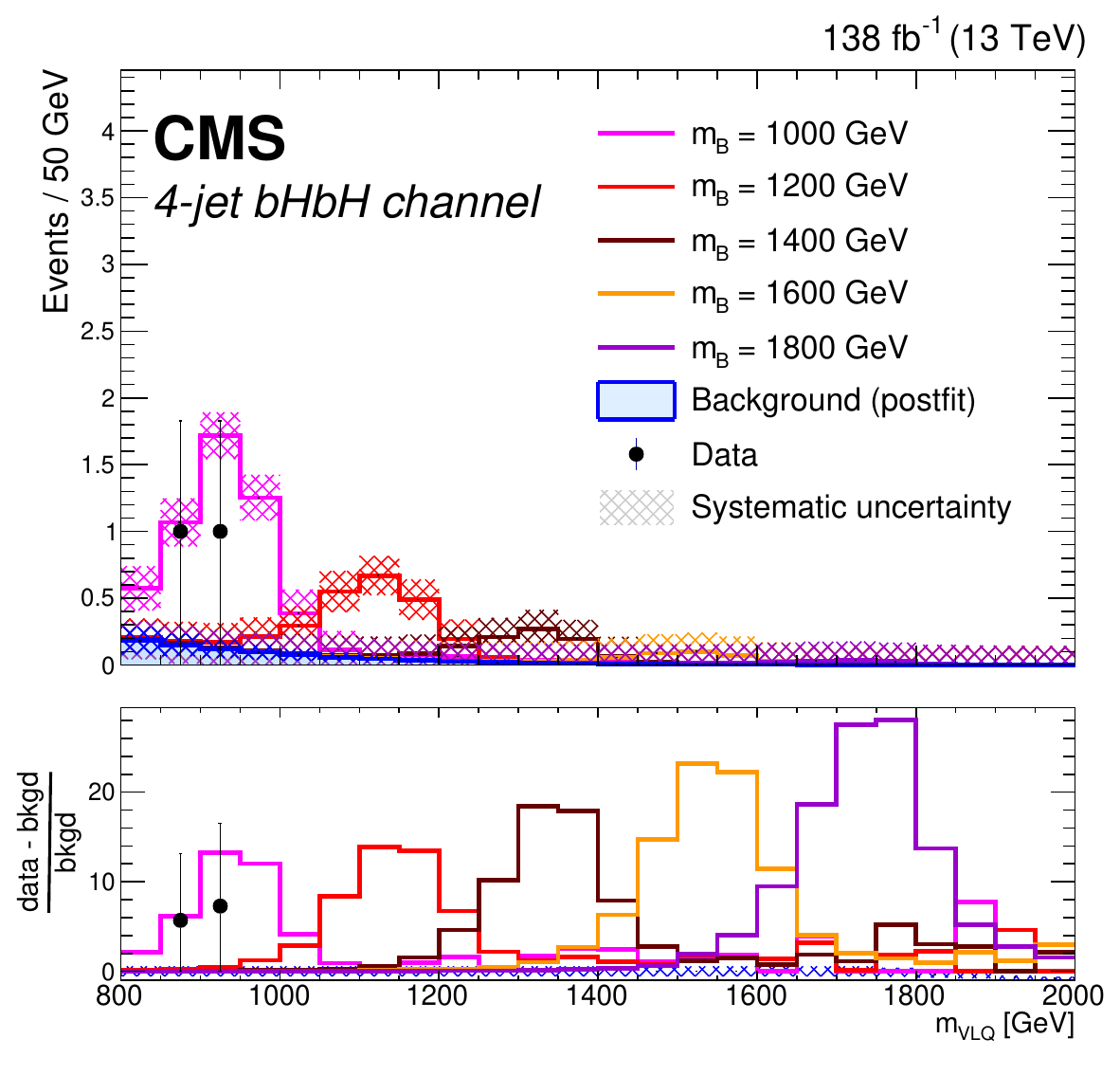}
\includegraphics[width=0.4\textwidth]{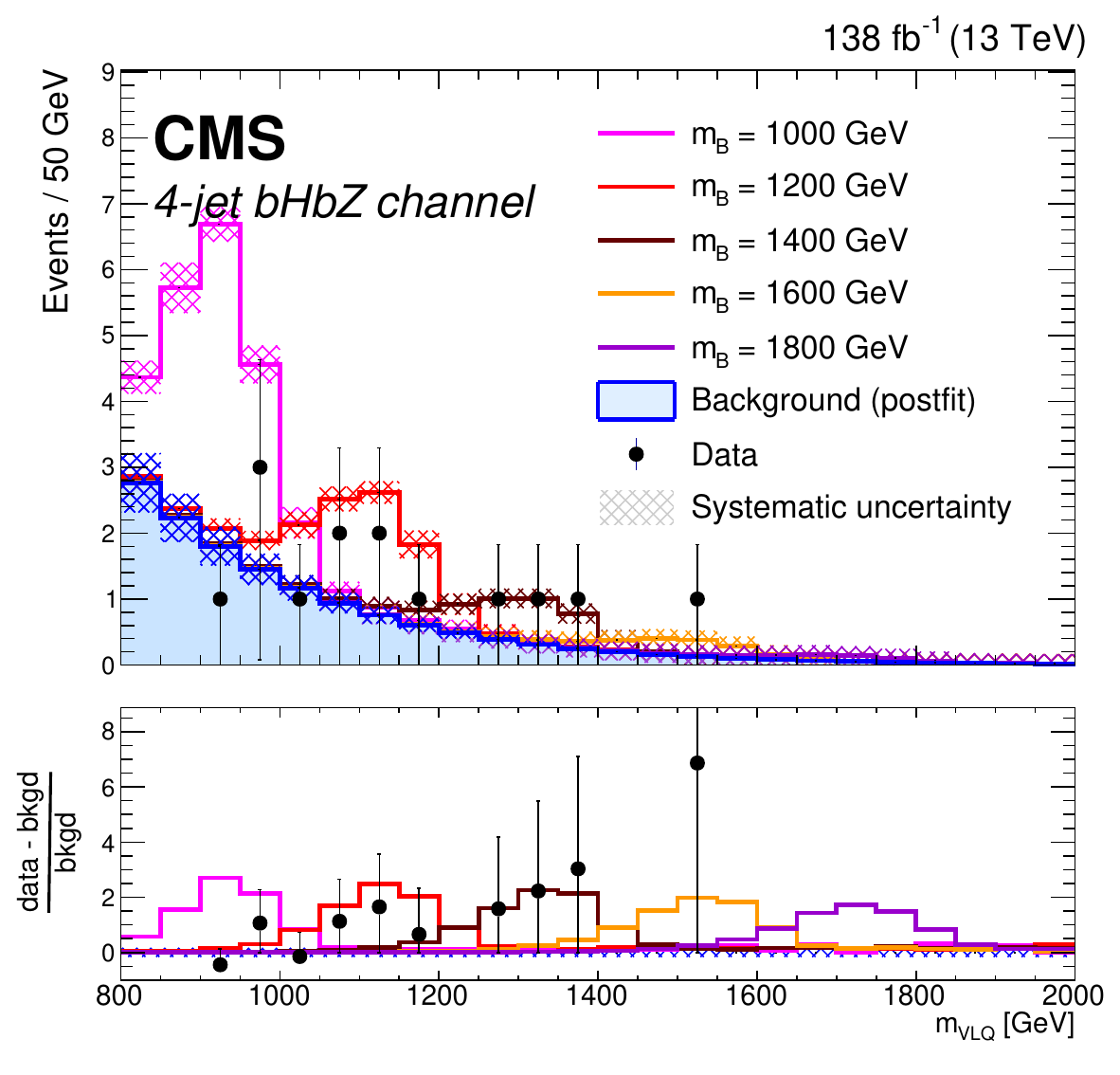}
\includegraphics[width=0.4\textwidth]{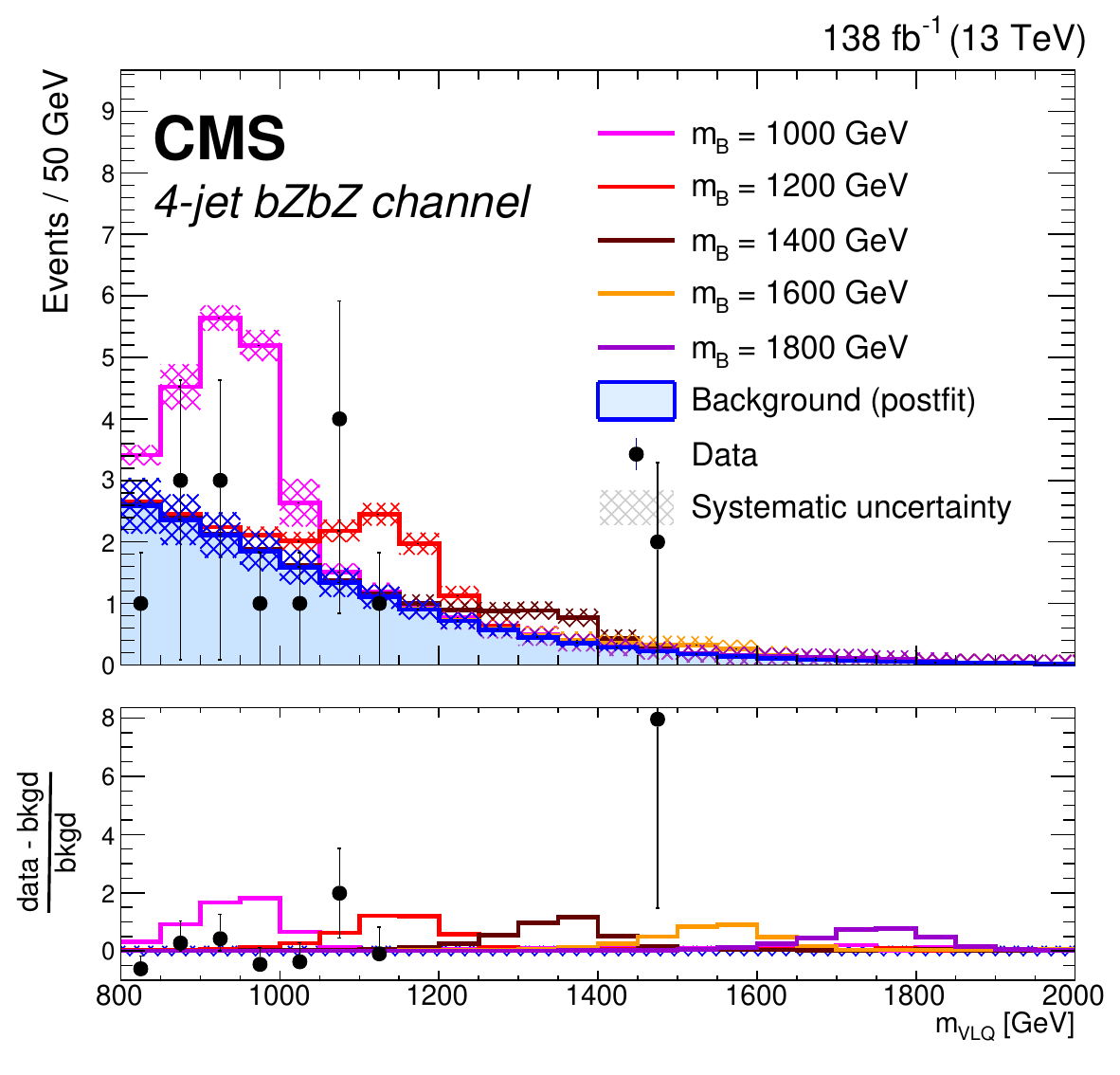}
\caption{\label{fig:sigbkgd_had} Distributions of reconstructed VLQ mass for expected postfit background (blue histogram), signal plus background (colored lines), and observed data (black points) for events in the hadronic category. The channels shown are 4-jet \bHbH (upper left), 4-jet \bHbZ (upper right), 4-jet \bZbZ (lower center). Five signal masses are shown: 1000 (magenta), 1200 (red), 1400 (maroon), 1600 (orange), and 1800\GeV (purple). The signal distributions are normalized to the number of events estimated from the expected VLQ production cross section. The assumed branching fractions are $\BrBbH = \BrBbZ = 50\%$, $\BrBtW = 0\%$. The background distribution is independent of the signal branching fractions. The hatched regions indicate the total systematic uncertainties in the background estimate.}
\end{figure*}

\begin{figure*}[!htp]
\centering
\includegraphics[width=0.4\textwidth]{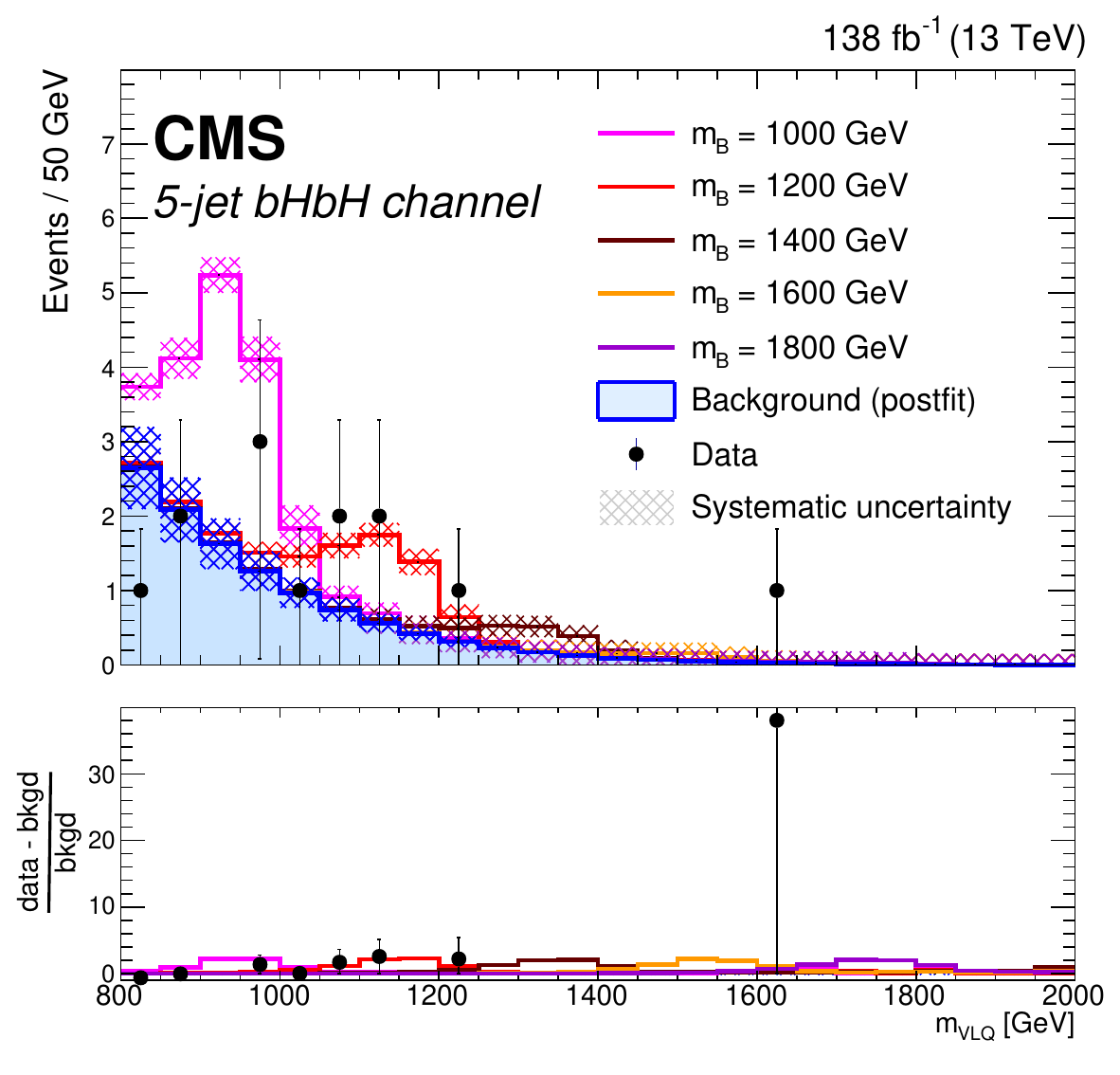}
\includegraphics[width=0.4\textwidth]{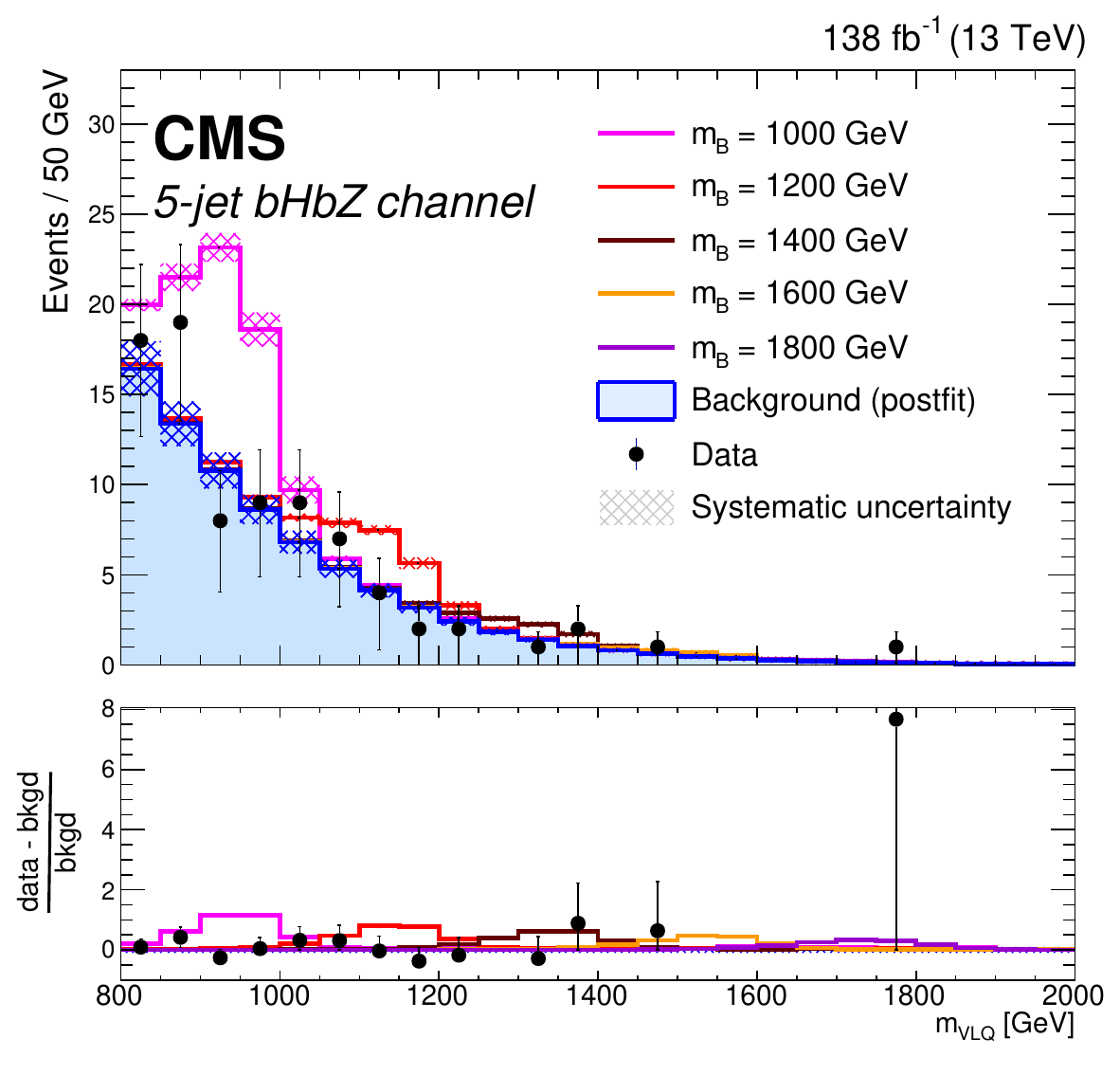}
\includegraphics[width=0.4\textwidth]{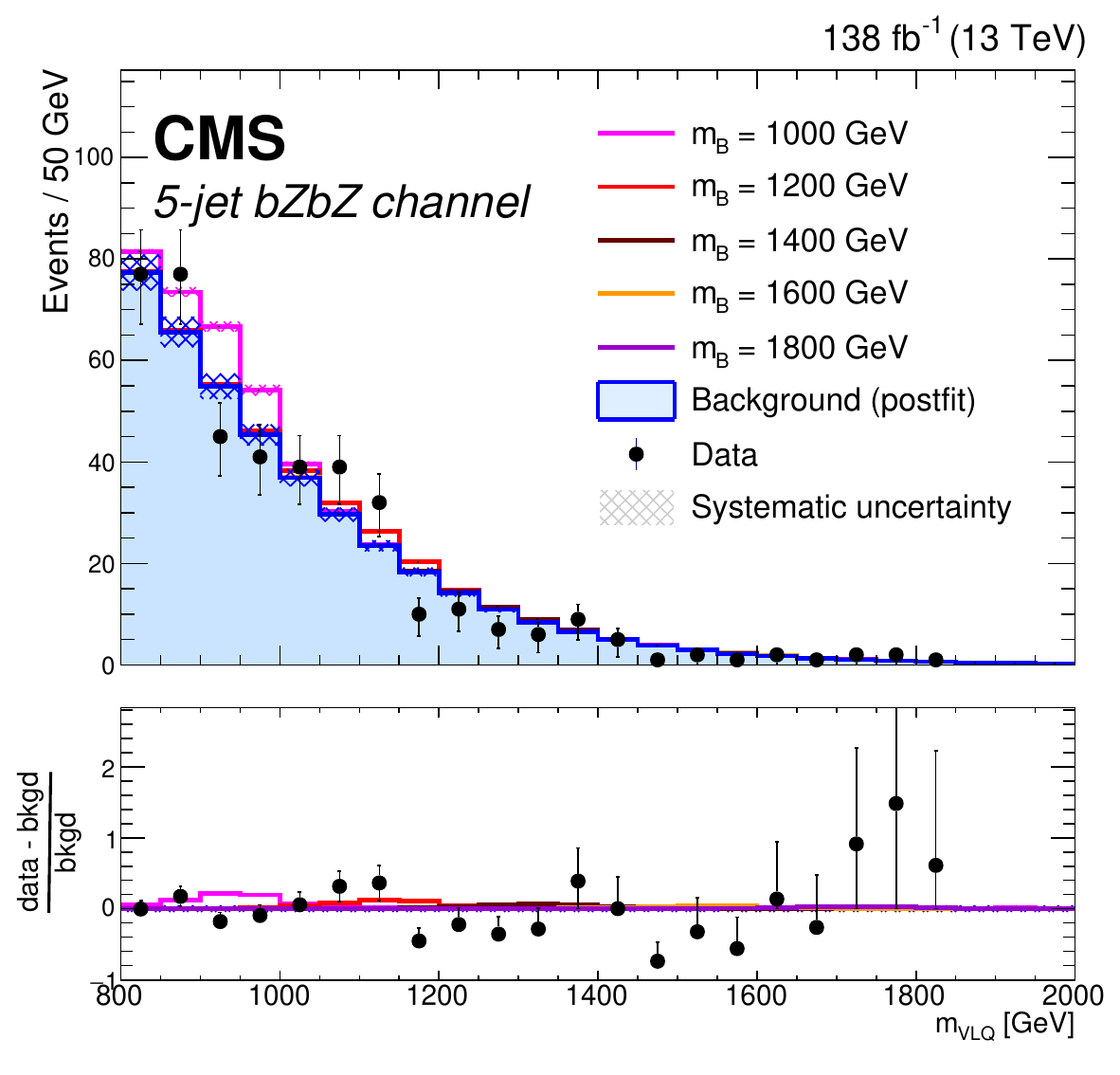}
\includegraphics[width=0.4\textwidth]{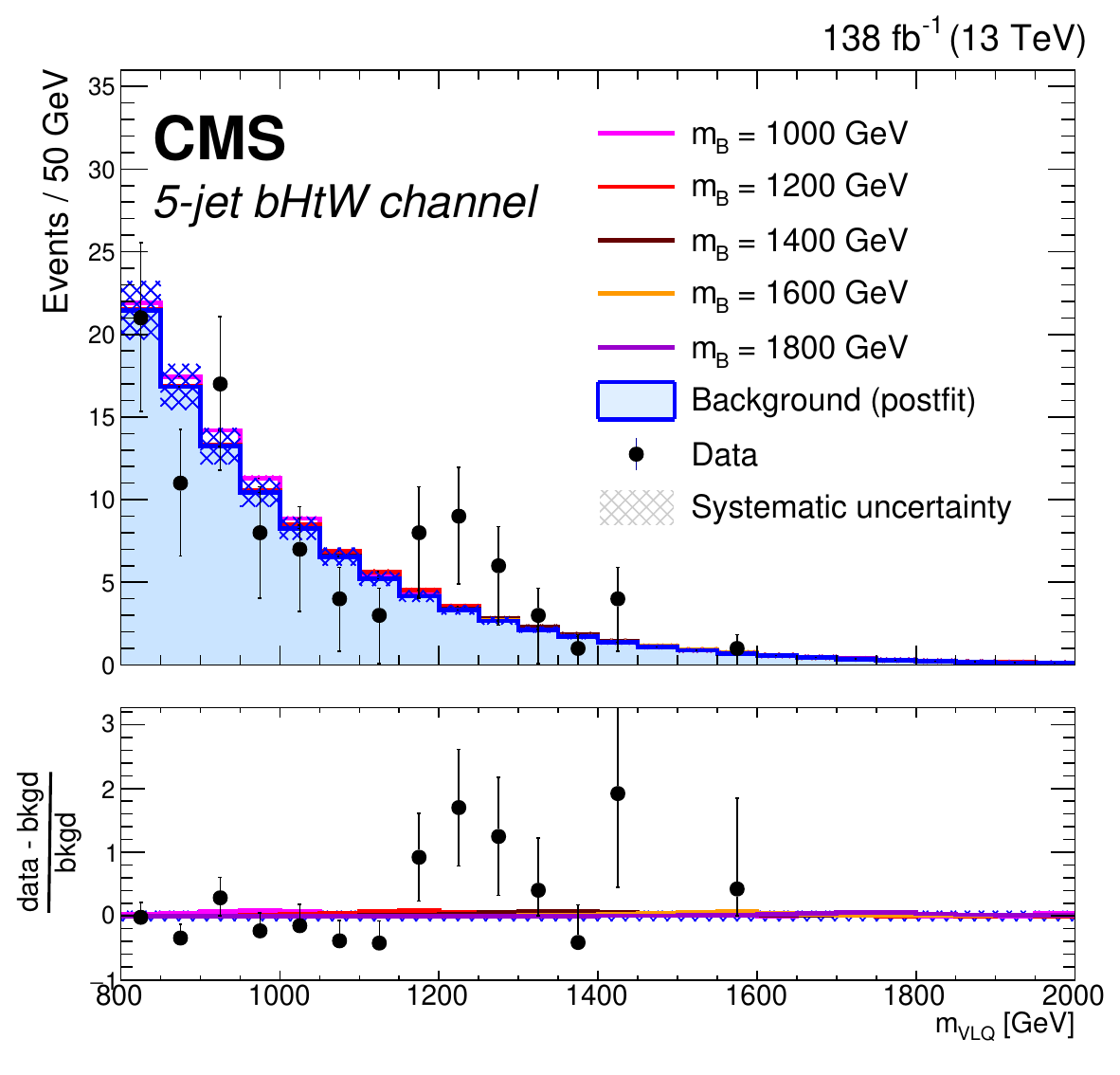}
\includegraphics[width=0.4\textwidth]{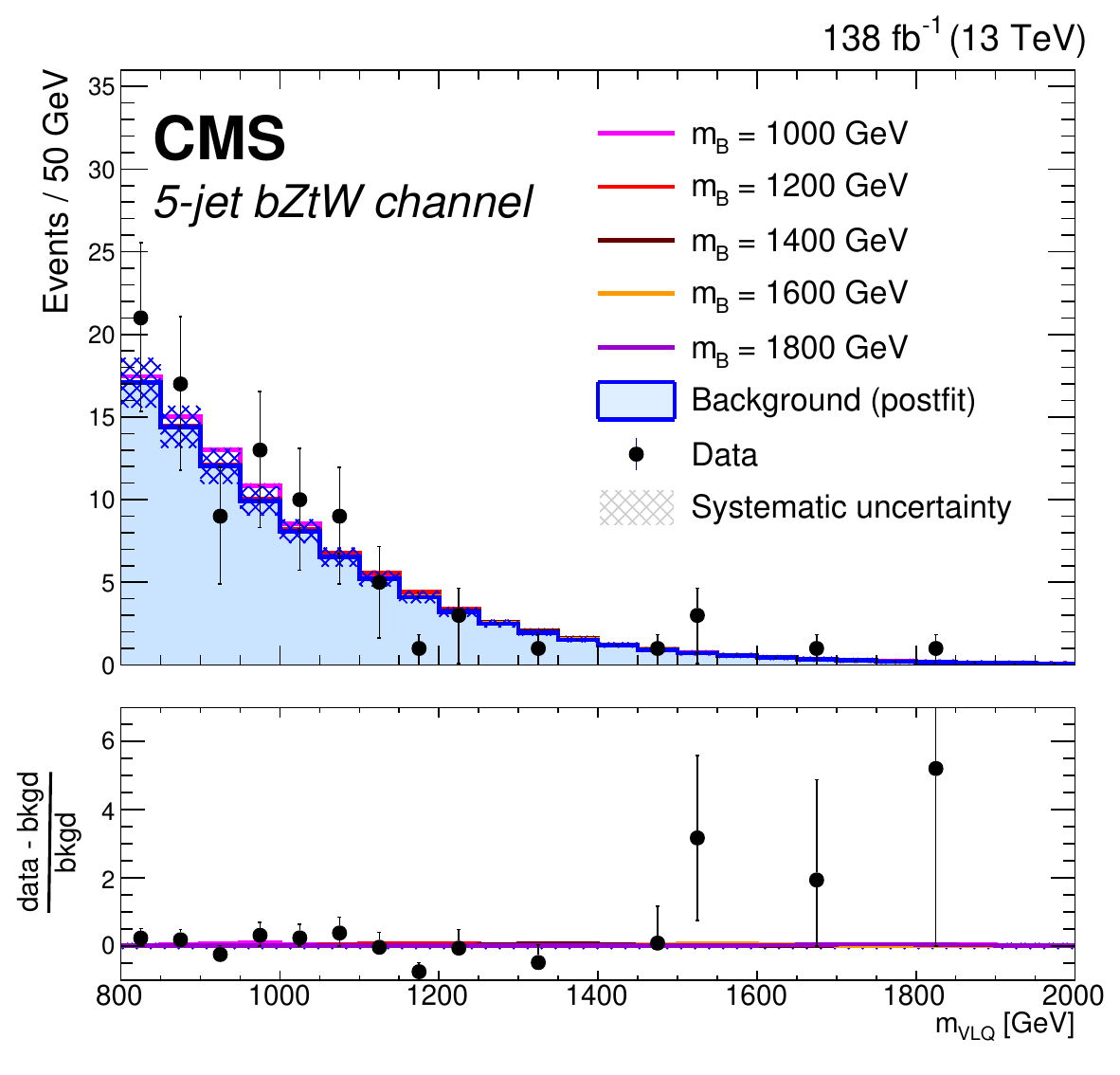}
\caption{\label{fig:sigbkgd_had_5} Distributions of reconstructed VLQ mass for expected postfit background (blue histogram), signal plus background (colored lines), and observed data (black points) for events in the hadronic category. The channels shown are 5-jet \bHbH (upper left), 5-jet \bHbZ (upper right), 5-jet \bZbZ (middle left), 5-jet \bHtW (middle right), and 5-jet \bZtW (lower center). Five signal masses are shown: 1000 (magenta), 1200 (red), 1400 (maroon), 1600 (orange), and 1800\GeV (purple). The signal distributions are normalized to the number of events estimated from the expected VLQ production cross section. The assumed branching fractions are $\BrBbH = \BrBbZ = 50\%$, $\BrBtW = 0\%$. The background distribution is independent of the signal branching fractions. The hatched regions indicate the total systematic uncertainties in the background estimate.}
\end{figure*}

\begin{figure*}[!htp]
\centering
\includegraphics[width=0.4\textwidth]{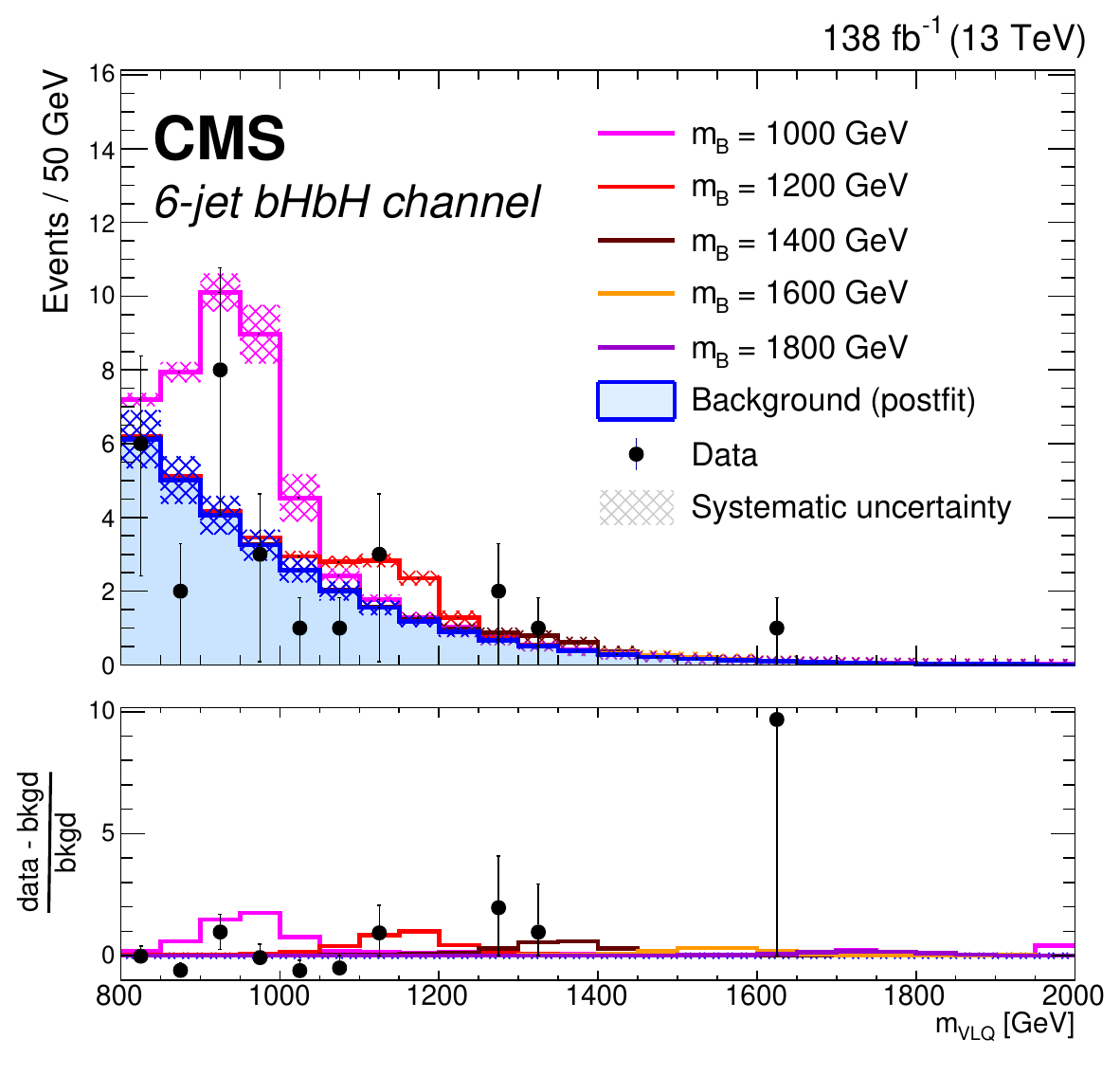}
\includegraphics[width=0.4\textwidth]{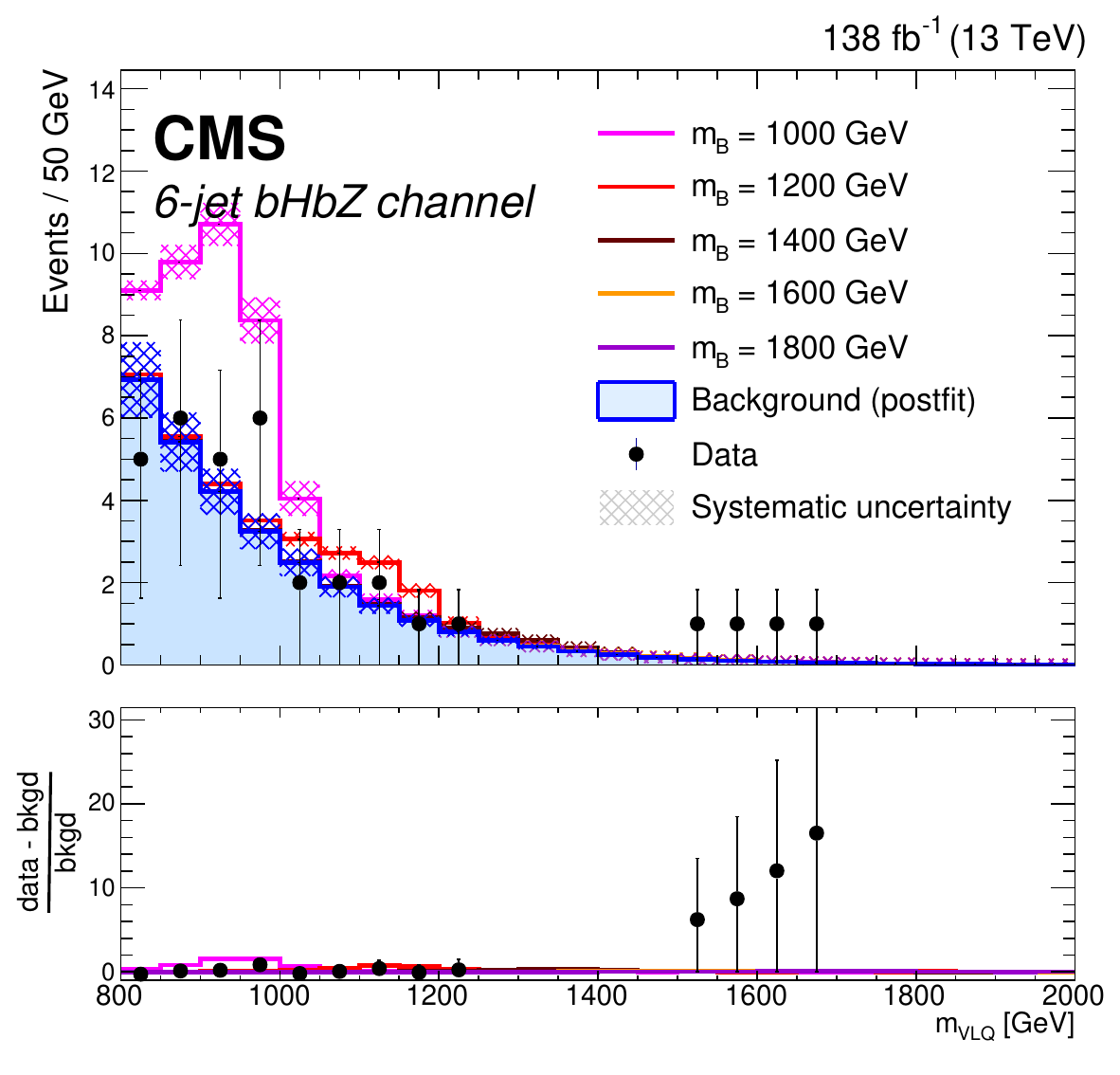}
\includegraphics[width=0.4\textwidth]{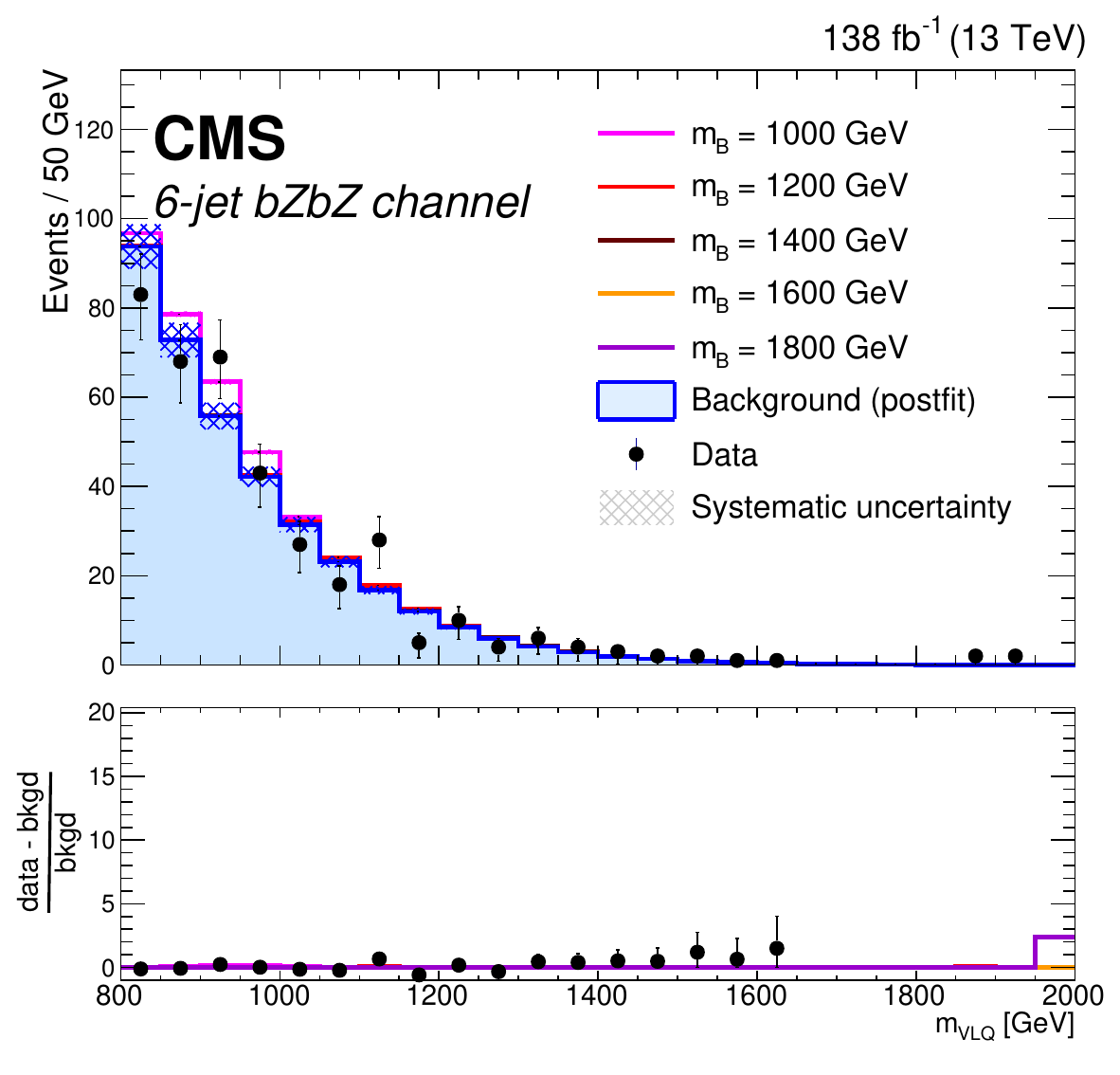}
\includegraphics[width=0.4\textwidth]{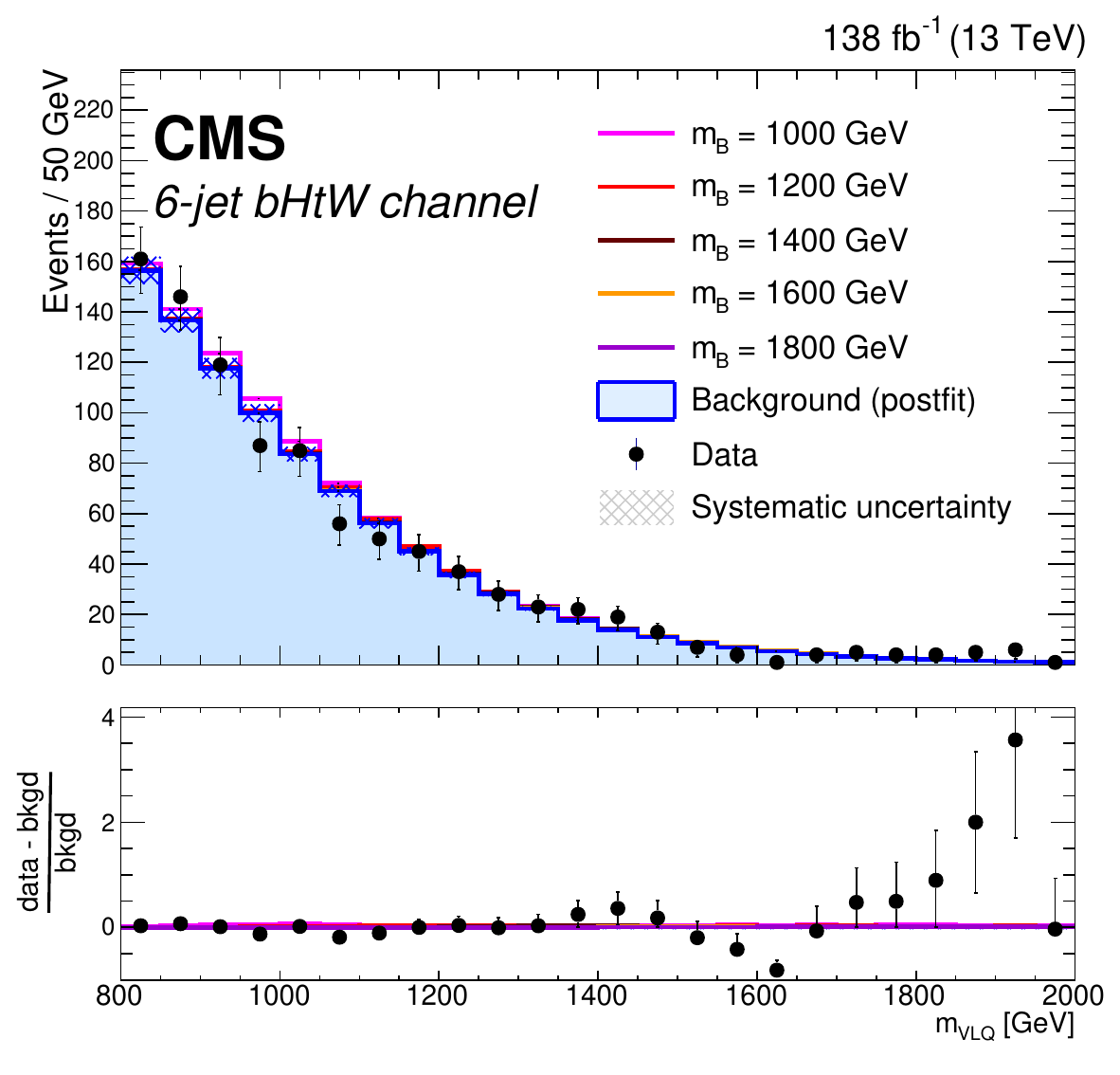}
\includegraphics[width=0.4\textwidth]{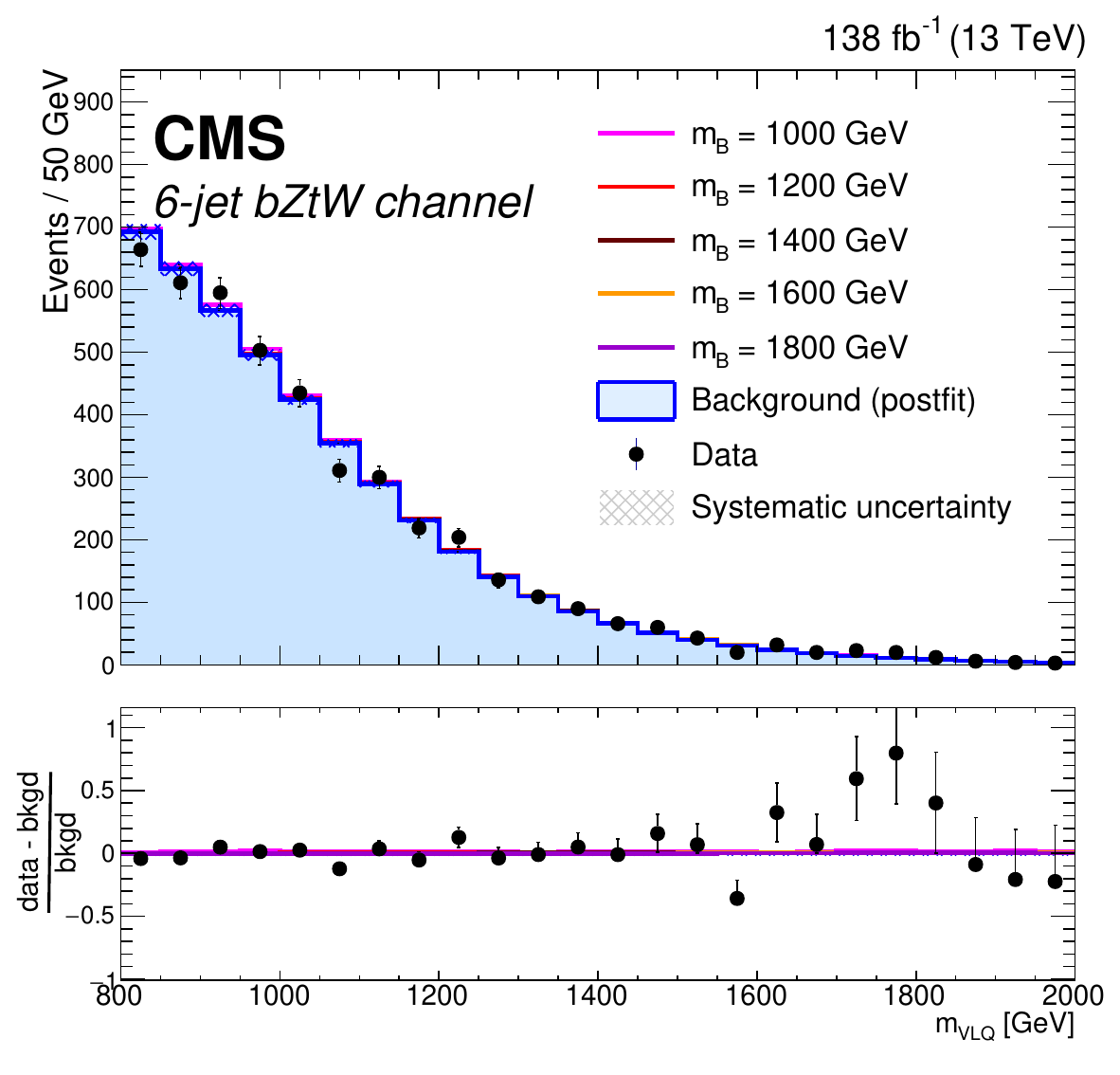}
\caption{\label{fig:sigbkgd_had_6} Distributions of reconstructed VLQ mass for expected postfit background (blue histogram), signal plus background (colored lines), and observed data (black points) for events in the hadronic category. The channels shown are 6-jet \bHbH (upper left), 6-jet \bHbZ (upper right), 6-jet \bZbZ (middle left), 6-jet \bHtW (middle right), and 6-jet \bZtW (lower center). Five signal masses are shown: 1000 (magenta), 1200 (red), 1400 (maroon), 1600 (orange), and 1800\GeV (purple). The signal distributions are normalized to the number of events estimated from the expected VLQ production cross section. The assumed branching fractions are $\BrBbH = \BrBbZ = 50\%$, $\BrBtW = 0\%$. The background distribution is independent of the signal branching fractions. The hatched regions indicate the total systematic uncertainties in the background estimate.}
\end{figure*}

\begin{figure*}[!htp]
\centering
\includegraphics[width=0.49\textwidth]{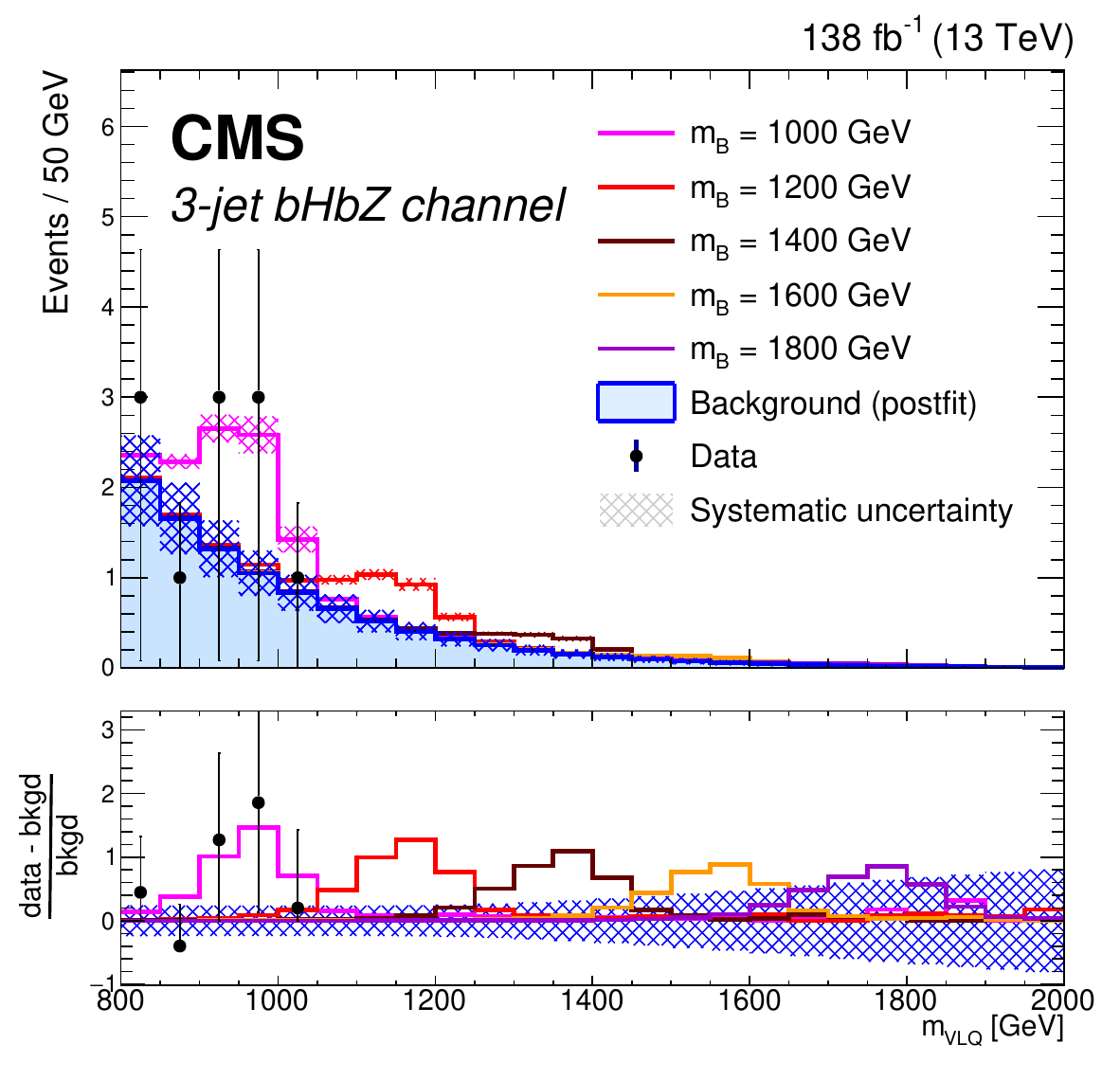}
\includegraphics[width=0.49\textwidth]{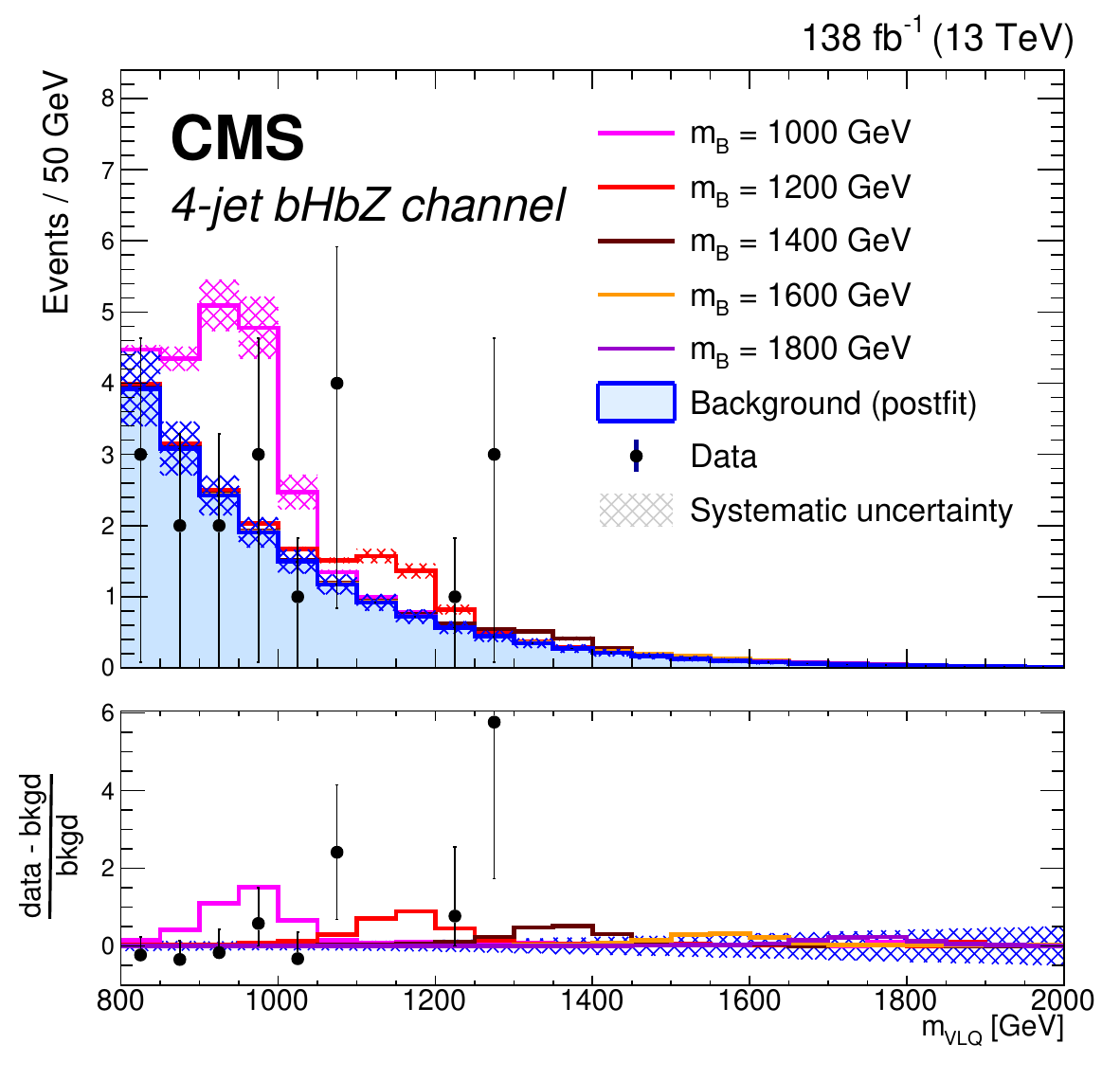}
\includegraphics[width=0.49\textwidth]{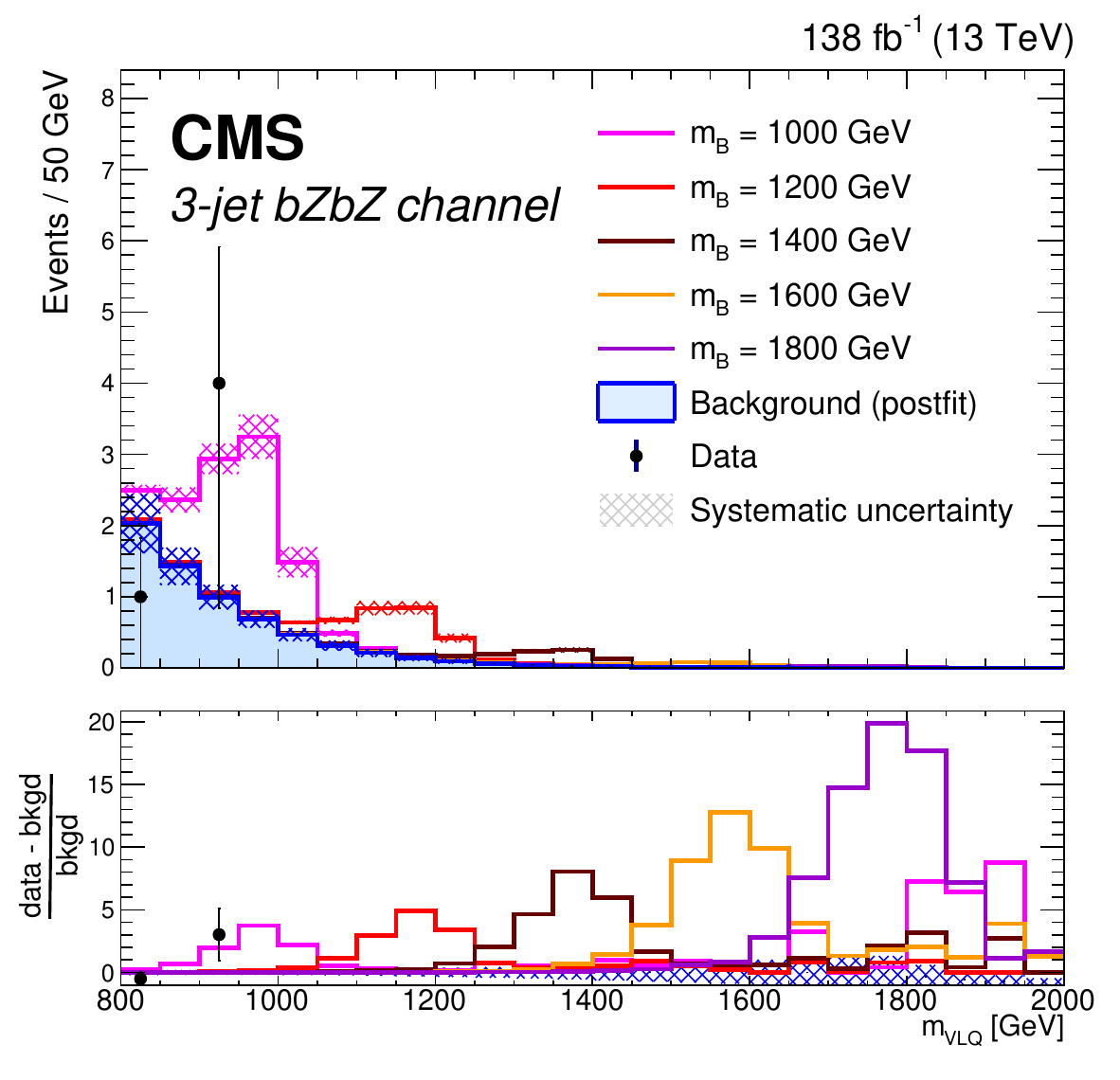}
\includegraphics[width=0.49\textwidth]{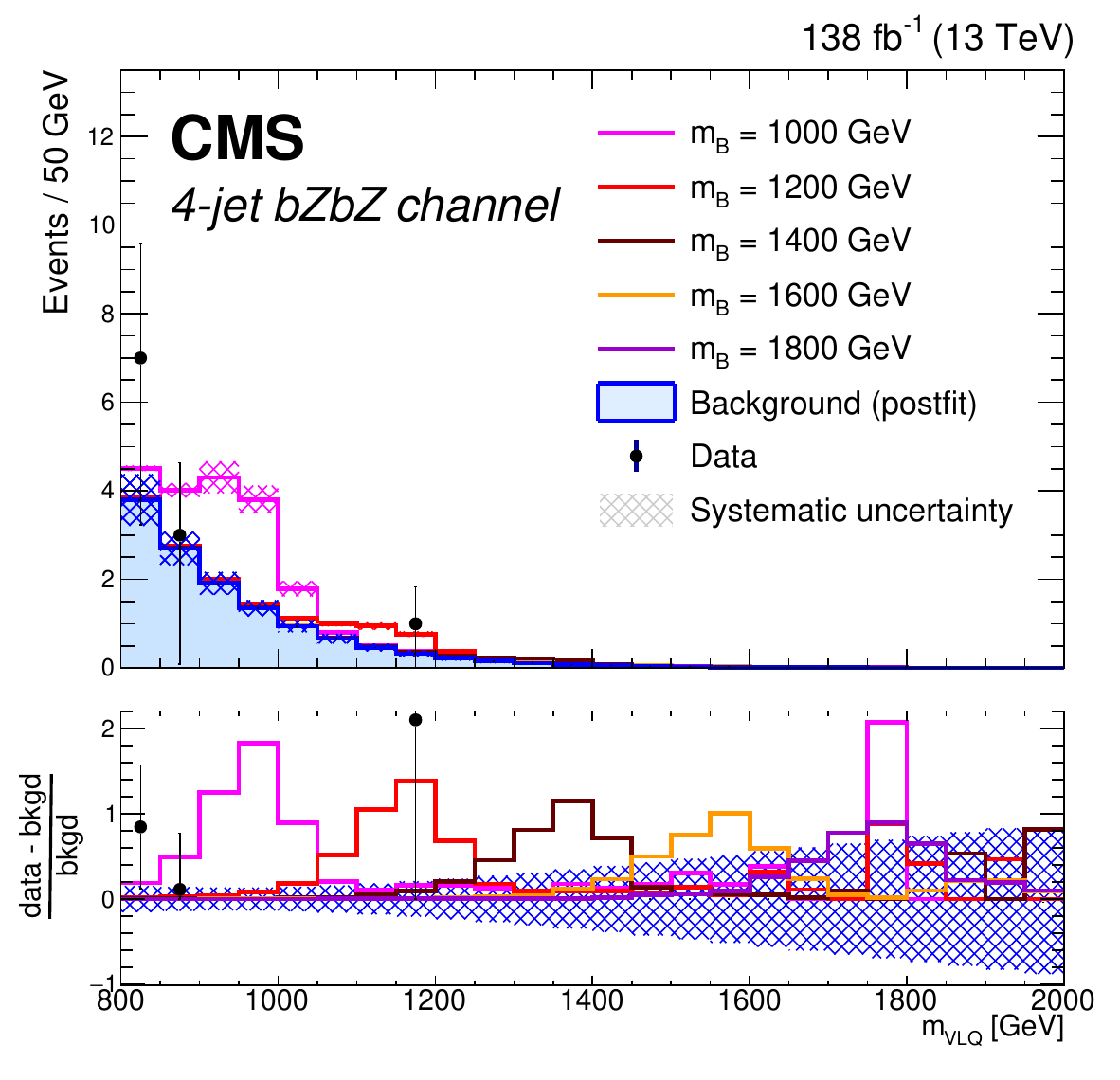}
\caption{\label{fig:sigbkgd_lep} Distributions of reconstructed VLQ mass for expected postfit background (blue histogram), signal plus background (colored lines), and observed data (black points) for events in the leptonic category. The channels shown are 3-jet \bHbZ (upper left), 4-jet \bHbZ (upper right), 3-jet \bZbZ (lower left), and 4-jet \bZbZ (lower right). Five signal masses are shown: 1000 (magenta), 1200 (red), 1400 (maroon), 1600 (orange), and 1800\GeV (purple). The signal distributions are normalized to the number of events estimated from the expected VLQ production cross section. The assumed branching fractions are $\BrBbH = \BrBbZ = 50\%$, $\BrBtW = 0\%$. The background distribution is independent of the signal branching fractions. The hatched regions indicate the total systematic uncertainties in the background estimate.}
\end{figure*}

\begin{figure*}[hbtp]
    \centering
    \includegraphics[width=0.45\textwidth]{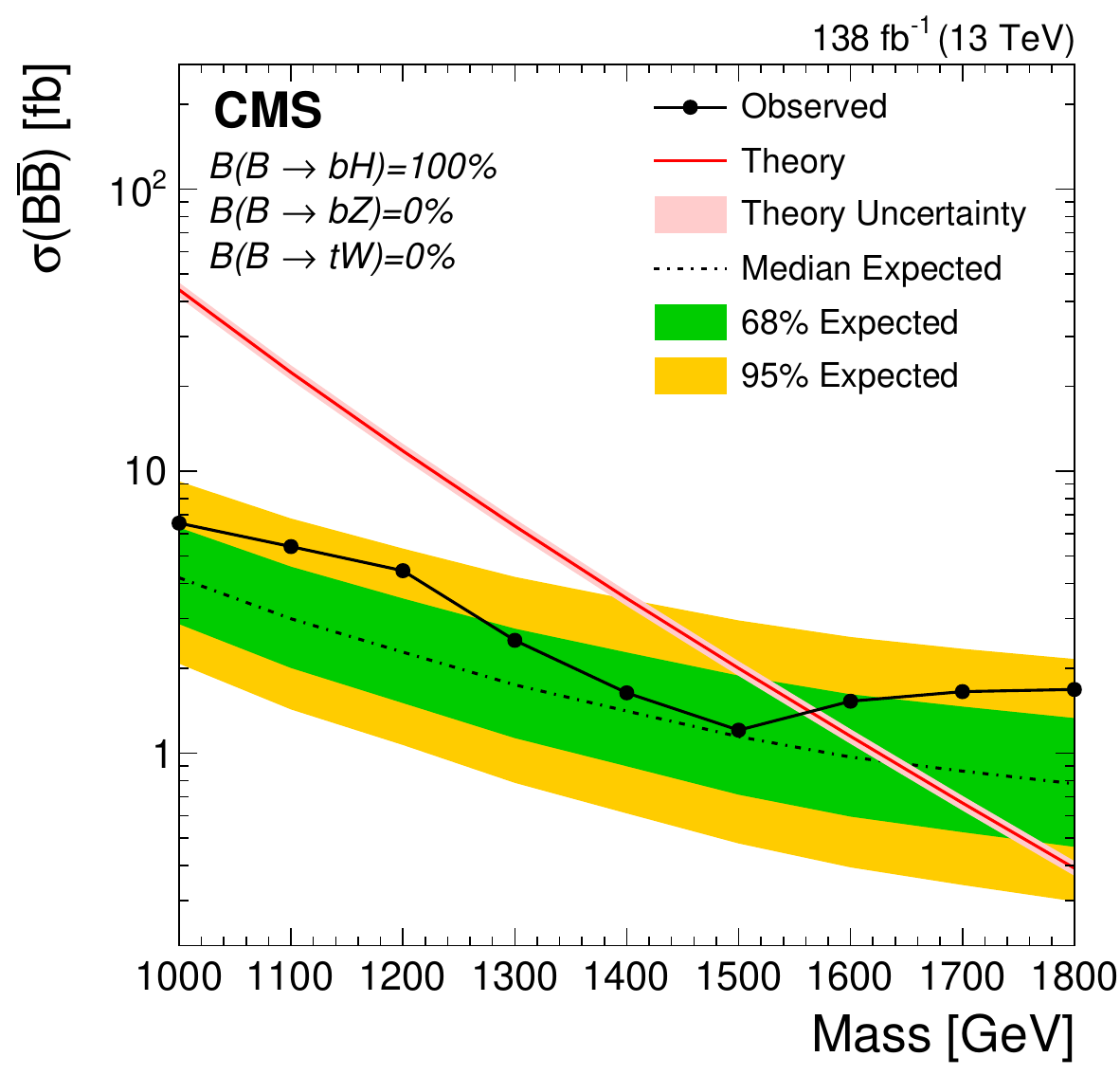}
    \includegraphics[width=0.45\textwidth]{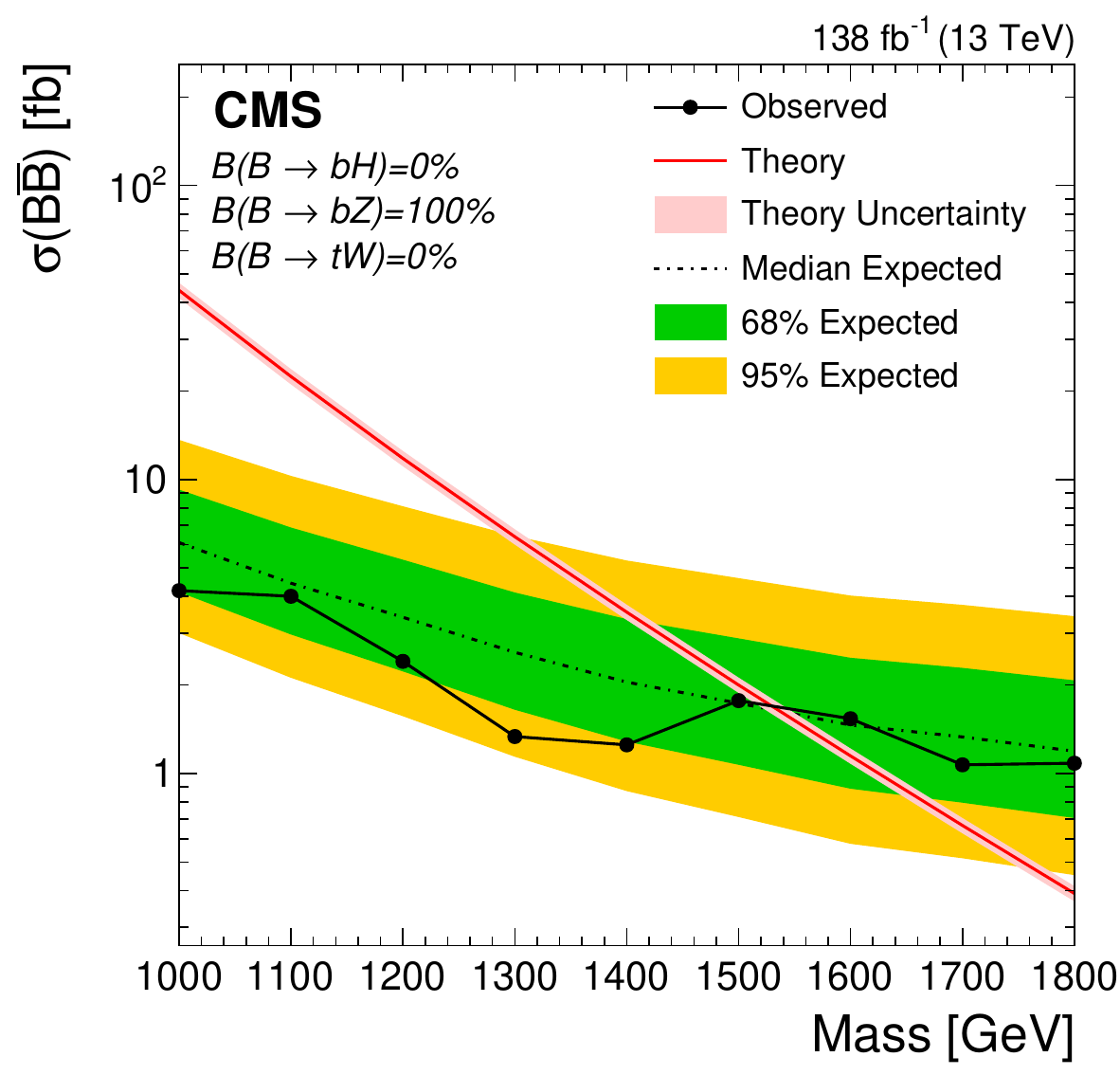}
    \includegraphics[width=0.45\textwidth]{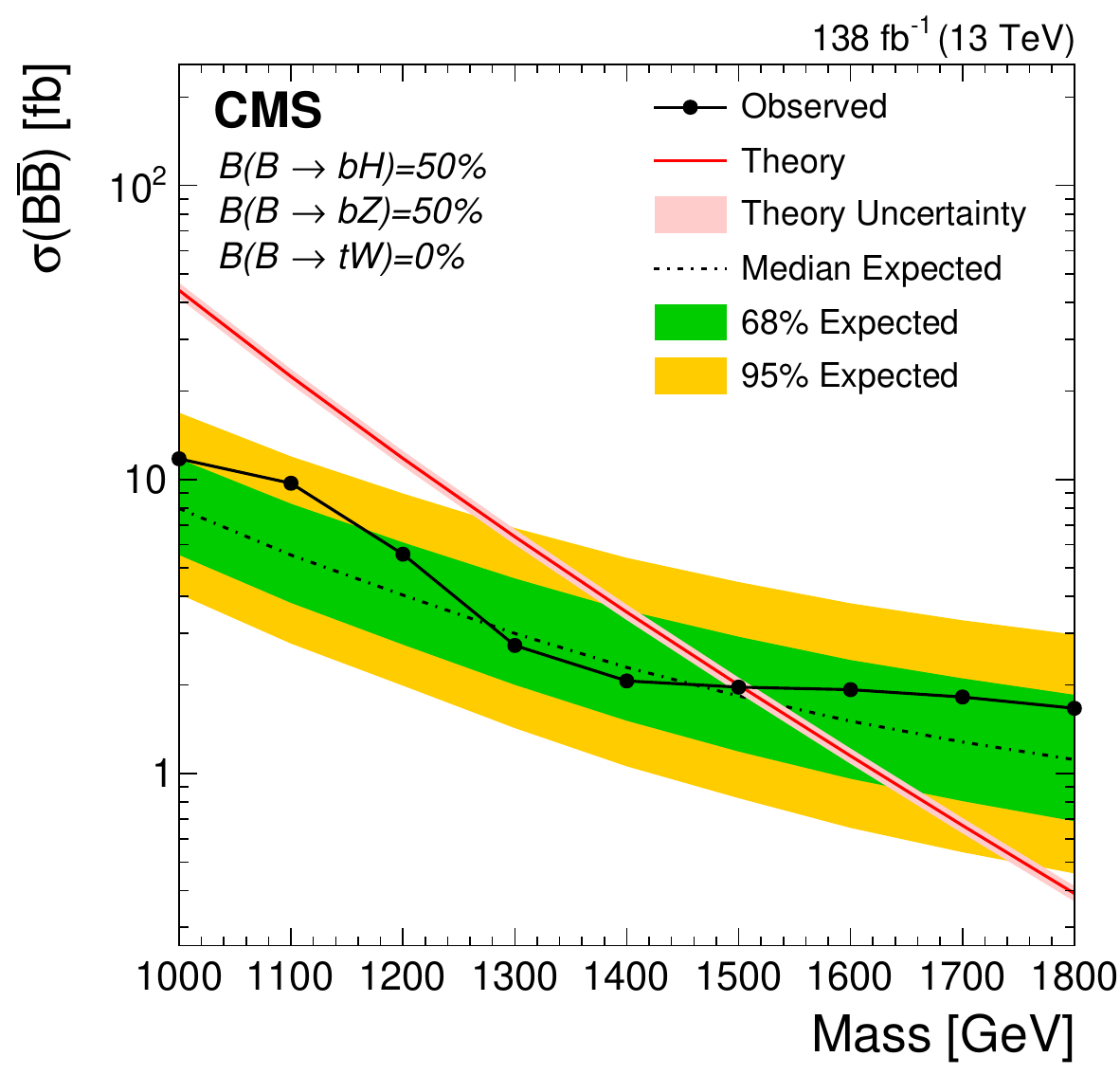}
    \includegraphics[width=0.45\textwidth]{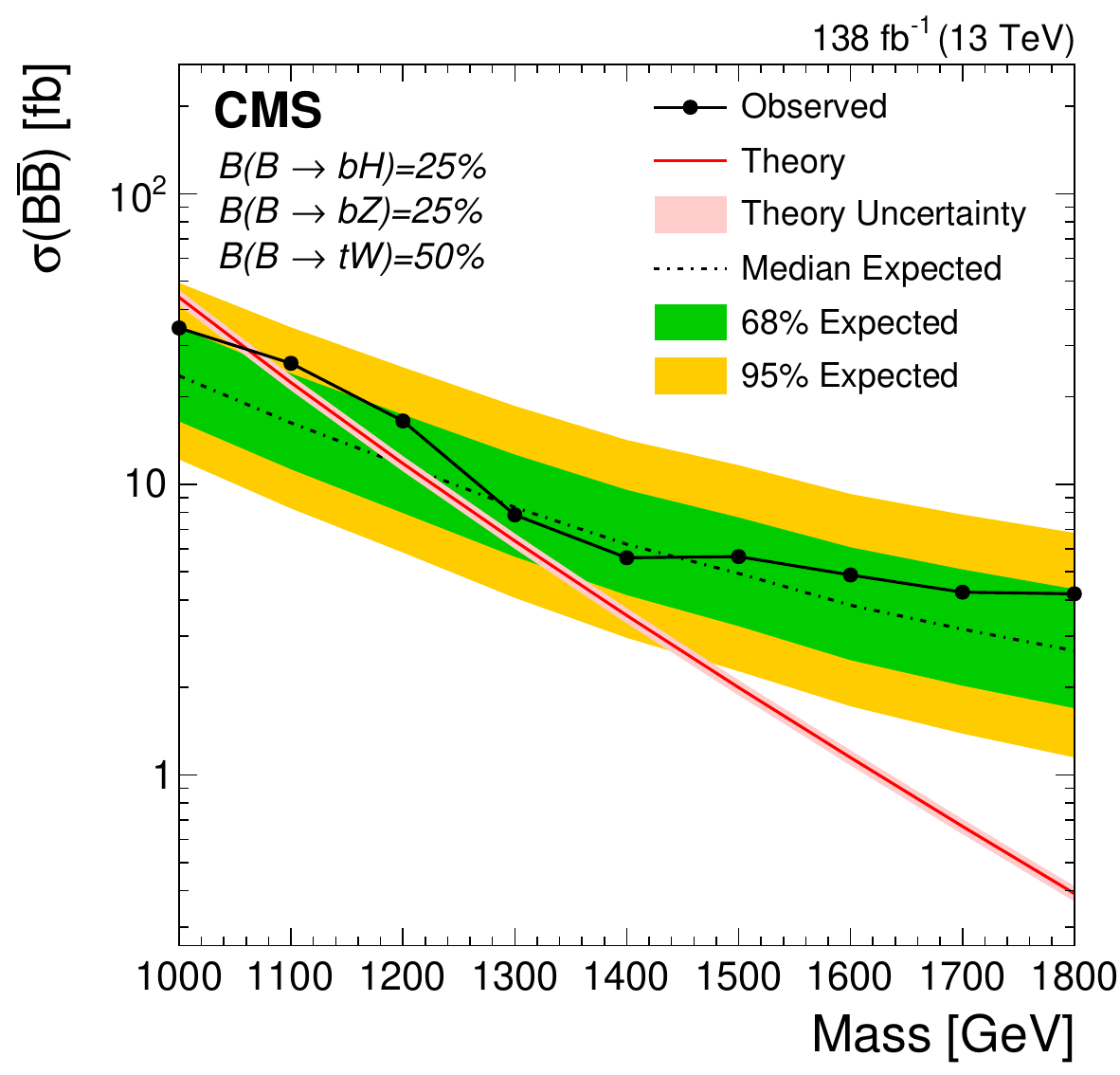}
    \caption{Limit at 95\% \CL on the cross section for VLQ pair production for four different branching fraction hypotheses: $\BrBbH = 100\%$ (upper left), $\BrBbZ = 100\%$ (upper right), $\BrBbH = \BrBbZ = 50\%$, corresponding to the $\VLQT\PB$ doublet model with no $\VLQT\PQt$ mixing and also to the large VLQ mass $\PX\VLQT\PB$ triplet model (lower left), and $\BrBbH = \BrBbZ = 25\%$, $\BrBtW = 50\%$, corresponding to the large VLQ mass $\VLQT\PB\VLQY$ triplet model (lower right). The expected limit is shown as the dashed line, with the 68 and 95\% uncertainties shown by the green (inner) and yellow (outer) bands, respectively. The theoretical cross section and its uncertainty are shown by the red line and light red band.}
    \label{fig:results_mass}    
\end{figure*}

\begin{figure}[hbtp]
    \centering
    \includegraphics[width=\cmsFigWidth]{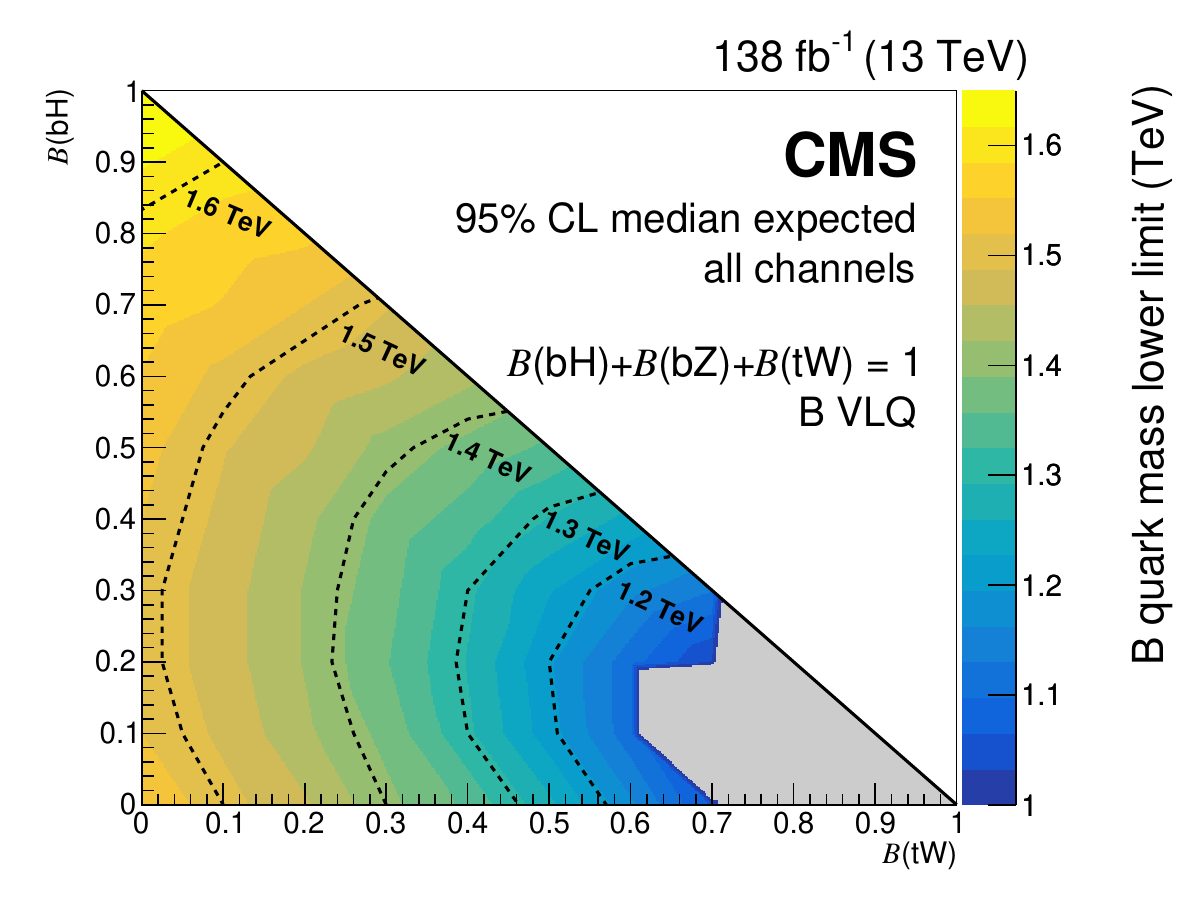}
    \caption{Median expected exclusion limits on the VLQ mass at 95\% \CL as a function of the branching fractions \BrBbH and \BrBtW, with $\BrBtW = 1 - \BrBbH - \BrBbZ$. The gray area corresponds to the region where the exclusion limit is less than 1000\GeV.}
   \label{fig:results_triangle_exp}
\end{figure}

\begin{figure}[hbtp]
    \centering
    \includegraphics[width=\cmsFigWidth]{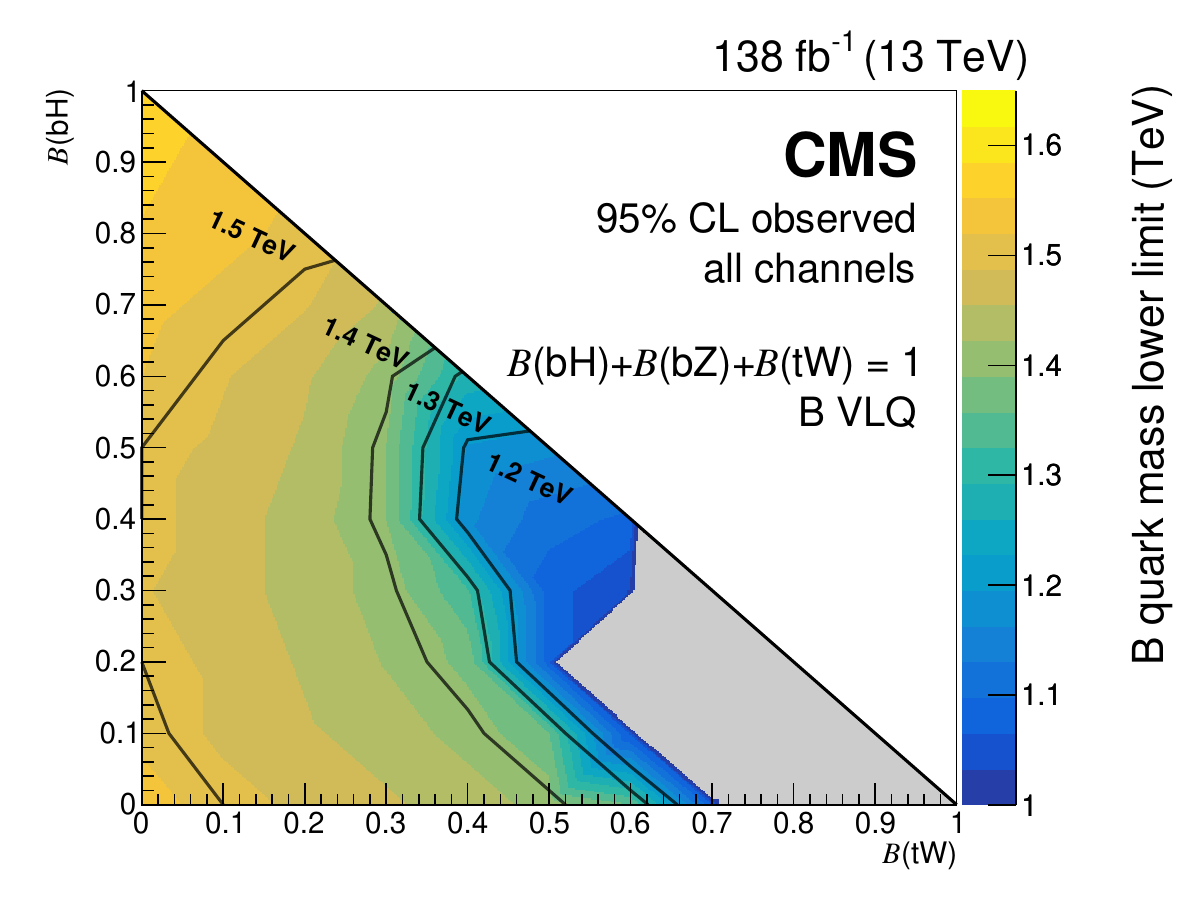}
    \caption{Observed exclusion limits on the VLQ mass at 95\% \CL as a function of the branching fractions \BrBbH and \BrBtW, with $\BrBtW = 1 - \BrBbH - \BrBbZ$. The gray area corresponds to the region where the exclusion limit is less than 1000\GeV.}
   \label{fig:results_triangle_obs}
\end{figure}

The data distributions in each channel are fit as a function of the reconstructed VLQ mass using \textsc{Combine}~\cite{CMS:2024onh}, including the relevant systematic uncertainties listed in Tables~\ref{table:all_uncertainties_a} and~\ref{table:all_uncertainties_c} as uncorrelated nuisance parameters. Figures~\ref{fig:sigbkgd_had}--\ref{fig:sigbkgd_had_6} show, for the hadronic category, the comparison of the distribution of the reconstructed VLQ mass, after the final optimized selection requirements have been applied, for data, expected background, and simulated signal events with VLQ masses of 1000, 1200, 1400, 1600, and 1800\GeV. Figure~\ref{fig:sigbkgd_lep} shows the same comparison for selected channels in the leptonic category. Branching fractions of $\BrBbH = 50\%$ and $\BrBbZ = 50\%$ are assumed. The data in Figs.~\ref{fig:sigbkgd_had}--\ref{fig:results_triangle_obs} are available in tabular form in the HEPData record for this analysis~\cite{hepdata}.

No statistically significant excess of data over the background expectation is observed in this mass range. We set exclusion limits on the VLQ mass as a function of the branching fractions \BrBbH, \BrBbZ, and \BrBtW. The results for the fully hadronic and leptonic categories, and for all jet multiplicities and event modes, are combined to set the limits. The signal extraction procedure is based on a binned maximum likelihood fit, where each systematic uncertainty is incorporated as a nuisance parameter into the fit, with the effect  of the systematic uncertainty included as a lognormal probability distribution per bin. The \CLs criterion~\cite{CLS2,CLS1} is used to obtain a limit at 95\% \CL using the profile likelihood test statistic in the asymptotic approximation~\cite{Cowan:2010js}.

 The exclusion limit on the VLQ mass is derived from the intersection of the limit curve with the predicted theoretical cross section as a function of the VLQ mass. Figure~\ref{fig:results_mass} shows the observed and expected limits at 95\% \CL on the cross section of VLQ pair production as functions of VLQ mass, for four representative branching fraction hypotheses: $\BrBbH = 100\%$; $\BrBbZ = 100\%$; $\BrBbH = \BrBbZ = 50\%$, corresponding to the $\VLQT\PB$ doublet model with no $\VLQT\PQt$ mixing and also to the large VLQ mass $\PX\VLQT\PB$ triplet model; and $\BrBbH = \BrBbZ = 25\%$, $\BrBtW = 50\%$, corresponding to the large VLQ mass $\VLQT\PB\VLQY$ triplet and \PB singlet models. 

Figures~\ref{fig:results_triangle_exp} and~\ref{fig:results_triangle_obs} show the expected and observed exclusion limits, respectively, on VLQ mass as a function of \BrBbH and \BrBtW, omitting points for which the exclusion limit is less than 1000\GeV. In the 100\% \BbZ model, the limits on the \PB VLQ mass have been improved by 120\GeV compared to the previous best limits, set by ATLAS~\cite{ATLAS:2022hnn}; in the $\VLQT\PB$ doublet model with no $\VLQT\PQt$ mixing, the limits have been improved by 50\GeV compared to the previous best limits, set by CMS~\cite{CMS:2020ttz}.

\ifthenelse{\boolean{cms@external}}{}{\clearpage}

\section{Summary}

A search for bottom-type vector-like quarks (\PB VLQs) has been presented, using data from proton-proton collisions collected by the CMS detector in 2016--2018 at $\sqrt{s} = 13\TeV$. Results are combined from the fully hadronic category, where each \PB VLQ decays into either a \PQb quark and a Higgs boson (\PH), a \PQb quark and a \PZ boson, or a \PQt quark and a \PW boson; and the leptonic category, where each \PB VLQ decays into a \PQb quark and either an \PH or a \PZ boson, and at least one decay includes a \PZ boson that decays into a pair of charged leptons. Since the dileptonic (\Pe and \PGm) branching fraction of the \PZ boson is 6.7\%, the rate of four-lepton events is an order of magnitude less than that of two-lepton events, which require only one dileptonic \PZ boson decay. Therefore, including the four-lepton signature would not provide a significant increase in sensitivity at the current integrated luminosity. To account for the fact that the two jets from an \PH, a \PZ, or a \PW boson decay may be reconstructed separately, or may be merged into a single reconstructed jet due to a high Lorentz boost, events are separated into different jet multiplicity categories and reconstructed appropriately. Backgrounds are estimated from data, and limits are set on the VLQ mass at 95\% confidence level as a function of the branching fractions \BrBbH, \BrBbZ, and \BrBtW. The most significant improvement over previous results is an increased sensitivity for scenarios with large \BrBbZ due to the inclusion of events with leptonic \PZ boson decays. The current results represent the most stringent limits on \PB VLQs to date.

\begin{acknowledgments}
We congratulate our colleagues in the CERN accelerator departments for the excellent performance of the LHC and thank the technical and administrative staffs at CERN and at other CMS institutes for their contributions to the success of the CMS effort. In addition, we gratefully acknowledge the computing centers and personnel of the Worldwide LHC Computing Grid and other centers for delivering so effectively the computing infrastructure essential to our analyses. Finally, we acknowledge the enduring support for the construction and operation of the LHC, the CMS detector, and the supporting computing infrastructure provided by the following funding agencies: SC (Armenia), BMBWF and FWF (Austria); FNRS and FWO (Belgium); CNPq, CAPES, FAPERJ, FAPERGS, and FAPESP (Brazil); MES and BNSF (Bulgaria); CERN; CAS, MoST, and NSFC (China); MINCIENCIAS (Colombia); MSES and CSF (Croatia); RIF (Cyprus); SENESCYT (Ecuador); ERC PRG, RVTT3 and MoER TK202 (Estonia); Academy of Finland, MEC, and HIP (Finland); CEA and CNRS/IN2P3 (France); SRNSF (Georgia); BMBF, DFG, and HGF (Germany); GSRI (Greece); NKFIH (Hungary); DAE and DST (India); IPM (Iran); SFI (Ireland); INFN (Italy); MSIP and NRF (Republic of Korea); MES (Latvia); LMTLT (Lithuania); MOE and UM (Malaysia); BUAP, CINVESTAV, CONACYT, LNS, SEP, and UASLP-FAI (Mexico); MOS (Montenegro); MBIE (New Zealand); PAEC (Pakistan); MES and NSC (Poland); FCT (Portugal); MESTD (Serbia); MCIN/AEI and PCTI (Spain); MOSTR (Sri Lanka); Swiss Funding Agencies (Switzerland); MST (Taipei); MHESI and NSTDA (Thailand); TUBITAK and TENMAK (Turkey); NASU (Ukraine); STFC (United Kingdom); DOE and NSF (USA).
    
\hyphenation{Rachada-pisek} Individuals have received support from the Marie-Curie programme and the European Research Council and Horizon 2020 Grant, contract Nos.\ 675440, 724704, 752730, 758316, 765710, 824093, 101115353, and COST Action CA16108 (European Union); the Leventis Foundation; the Alfred P.\ Sloan Foundation; the Alexander von Humboldt Foundation; the Science Committee, project no. 22rl-037 (Armenia); the Belgian Federal Science Policy Office; the Fonds pour la Formation \`a la Recherche dans l'Industrie et dans l'Agriculture (FRIA-Belgium); the Agentschap voor Innovatie door Wetenschap en Technologie (IWT-Belgium); the F.R.S.-FNRS and FWO (Belgium) under the ``Excellence of Science -- EOS" -- be.h project n.\ 30820817; the Beijing Municipal Science \& Technology Commission, No. Z191100007219010 and Fundamental Research Funds for the Central Universities (China); the Ministry of Education, Youth and Sports (MEYS) of the Czech Republic; the Shota Rustaveli National Science Foundation, grant FR-22-985 (Georgia); the Deutsche Forschungsgemeinschaft (DFG), under Germany's Excellence Strategy -- EXC 2121 ``Quantum Universe" -- 390833306, and under project number 400140256 - GRK2497; the Hellenic Foundation for Research and Innovation (HFRI), Project Number 2288 (Greece); the Hungarian Academy of Sciences, the New National Excellence Program - \'UNKP, the NKFIH research grants K 124845, K 124850, K 128713, K 128786, K 129058, K 131991, K 133046, K 138136, K 143460, K 143477, 2020-2.2.1-ED-2021-00181, and TKP2021-NKTA-64 (Hungary); the Council of Science and Industrial Research, India; ICSC -- National Research Center for High Performance Computing, Big Data and Quantum Computing, funded by the EU NexGeneration program (Italy); the Latvian Council of Science; the Ministry of Education and Science, project no. 2022/WK/14, and the National Science Center, contracts Opus 2021/41/B/ST2/01369 and 2021/43/B/ST2/01552 (Poland); the Funda\c{c}\~ao para a Ci\^encia e a Tecnologia, grant CEECIND/01334/2018 (Portugal); the National Priorities Research Program by Qatar National Research Fund; MCIN/AEI/10.13039/501100011033, ERDF ``a way of making Europe", and the Programa Estatal de Fomento de la Investigaci{\'o}n Cient{\'i}fica y T{\'e}cnica de Excelencia Mar\'{\i}a de Maeztu, grant MDM-2017-0765 and Programa Severo Ochoa del Principado de Asturias (Spain); the Chulalongkorn Academic into Its 2nd Century Project Advancement Project, and the National Science, Research and Innovation Fund via the Program Management Unit for Human Resources \& Institutional Development, Research and Innovation, grant B37G660013 (Thailand); the Kavli Foundation; the Nvidia Corporation; the SuperMicro Corporation; the Welch Foundation, contract C-1845; and the Weston Havens Foundation (USA).    
\end{acknowledgments}

\bibliography{auto_generated}
\cleardoublepage \appendix\section{The CMS Collaboration \label{app:collab}}\begin{sloppypar}\hyphenpenalty=5000\widowpenalty=500\clubpenalty=5000
\cmsinstitute{Yerevan Physics Institute, Yerevan, Armenia}
{\tolerance=6000
A.~Hayrapetyan, A.~Tumasyan\cmsAuthorMark{1}\cmsorcid{0009-0000-0684-6742}
\par}
\cmsinstitute{Institut f\"{u}r Hochenergiephysik, Vienna, Austria}
{\tolerance=6000
W.~Adam\cmsorcid{0000-0001-9099-4341}, J.W.~Andrejkovic, T.~Bergauer\cmsorcid{0000-0002-5786-0293}, S.~Chatterjee\cmsorcid{0000-0003-2660-0349}, K.~Damanakis\cmsorcid{0000-0001-5389-2872}, M.~Dragicevic\cmsorcid{0000-0003-1967-6783}, P.S.~Hussain\cmsorcid{0000-0002-4825-5278}, M.~Jeitler\cmsAuthorMark{2}\cmsorcid{0000-0002-5141-9560}, N.~Krammer\cmsorcid{0000-0002-0548-0985}, A.~Li\cmsorcid{0000-0002-4547-116X}, D.~Liko\cmsorcid{0000-0002-3380-473X}, I.~Mikulec\cmsorcid{0000-0003-0385-2746}, J.~Schieck\cmsAuthorMark{2}\cmsorcid{0000-0002-1058-8093}, R.~Sch\"{o}fbeck\cmsorcid{0000-0002-2332-8784}, D.~Schwarz\cmsorcid{0000-0002-3821-7331}, M.~Sonawane\cmsorcid{0000-0003-0510-7010}, S.~Templ\cmsorcid{0000-0003-3137-5692}, W.~Waltenberger\cmsorcid{0000-0002-6215-7228}, C.-E.~Wulz\cmsAuthorMark{2}\cmsorcid{0000-0001-9226-5812}
% ...
\par}
\vskip\cmsinstskip
\dag:~Deceased\\
$^{1}$Also at Yerevan State University, Yerevan, Armenia\\
$^{2}$Also at TU Wien, Vienna, Austria\\
% ...
\end{sloppypar}
%%% END EDITABLE REGION %%%
% skeleton_end
\end{document}